\documentclass[aps, twocolumn,nofootinbib, amsmath,amssymb, aps, floatfix]{revtex4-1}

\usepackage{aas_macros} 
\usepackage[colorlinks=true,citecolor=blue,linkcolor=blue,urlcolor=blue, backref=false,pdfborder={0 0 0}]{hyperref}

\usepackage[utf8]{inputenc}

\usepackage{float}

\newcommand{\p}[2]{\frac{\partial #1}{\partial #2}}

\newcommand{\deriv}[2]{\frac{\textrm{d} #1}{\textrm{d} #2}}
\newcommand{\td}{\,\textrm{d}}

\newcommand{\beq}{\begin{equation}}
\newcommand{\eeq}{\end{equation}}
\newcommand{\barr}{\begin{eqnarray}}
\newcommand{\earr}{\end{eqnarray}}
\newcommand{\ball}{\begin{align}}
\newcommand{\eall}{\end{align}}
\newcommand{\bs}{\boldsymbol}
\newcommand{\rme}{\textrm{e}}

\newcommand{\rev}[1]{{{#1}}}

\usepackage{enumitem}
\usepackage{cancel}
\usepackage{graphicx}
\usepackage{dcolumn}
\usepackage{bm}
\usepackage{braket}
\usepackage{placeins}
\usepackage{color,soul}
\usepackage[toc,page]{appendix}
\begin{document}

\title{Perturbed recombination from inhomogeneous photon injection \texorpdfstring{\\}{} 
and application to accreting primordial black holes}
\author{Trey W. Jensen}
\author{Yacine Ali-Ha{\"i}moud}
\affiliation{Center for Cosmology and Particle Physics \\
Department of Physics, New York University \\
New York, NY 10003, USA}

\date{\today}

\begin{abstract}
Exotic electromagnetic energy injection in the early Universe may alter cosmological recombination, and ultimately cosmic microwave background (CMB) anisotropies. Moreover, if energy injection is inhomogeneous, it may induce a spatially varying ionization fraction, and non-Gaussianity in the CMB. The observability of these signals, however, is contingent upon how far the injected particles propagate and deposit their energy into the primordial plasma, relative to the characteristic scale of energy injection fluctuations. In this study we inspect the spatial properties of energy deposition and perturbed recombination resulting from an inhomogeneous energy injection of sub-10 MeV photons, relevant to accreting primordial black holes (PBHs). We develop a novel Monte Carlo radiation transport code accounting for all relevant photon interactions in this energy range, and including secondary electron energy deposition efficiency through a new analytic approximation. For a specified injected photon spectrum, the code outputs an injection-to-deposition Green’s function depending on time and distance from the injection point. Combining this output with a linearized solution of the perturbed recombination problem, we derive time- and scale-dependent deposition-to-ionization Green’s functions. We apply this general framework to accreting PBHs, whose luminosity is strongly spatially modulated by supersonic relative velocities between cold dark matter and baryons. We find that the resulting spatial fluctuations of the free-electron fraction are of the same magnitude as its mean deviation from standard recombination, from which current CMB power spectra constraints are derived. This work suggests that the sensitivity to accreting PBHs might be substantially improved by propagating these inhomogeneities to CMB anisotropy power spectra and non-Gaussian statistics, which we study in subsequent papers. 

\end{abstract}

\maketitle

\section{Introduction}\label{sec:intro}

An alluring feature of dark matter (DM) candidates is their possible injection of electromagnetically interacting particles. Part of this injected energy is eventually deposited in the form of extra heating, excitation, and importantly, ionization of the primordial plasma around cosmological recombination \cite{adams98a,chen04a}. This changes the Thomson visibility function and diffusion damping scale, and ultimately the angular power spectra of cosmic microwave background (CMB) anisotropies observed today. This effect is at the basis of CMB anisotropy constraints on annihilating or decaying DM particles \cite{planck20b}, as well as evaporating \cite{Poulin_17b, Poulter_19} or accreting primordial black holes (PBHs) \cite{ricotti08a, yacine17a, poulin17a}. 

\begin{figure*}[hbt!]
\includegraphics[width=0.75\columnwidth]{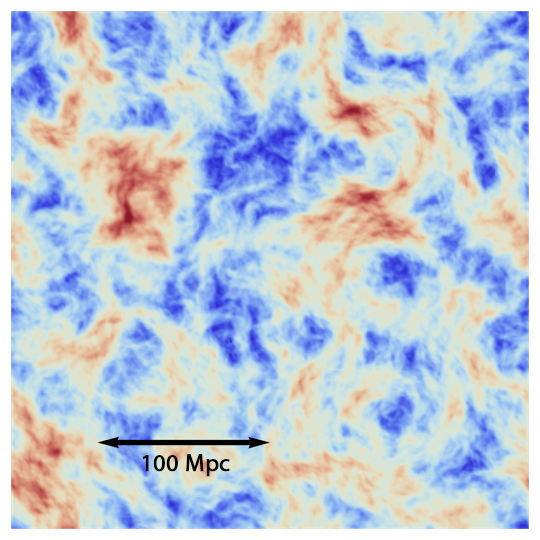}
\includegraphics[width=0.2\columnwidth]{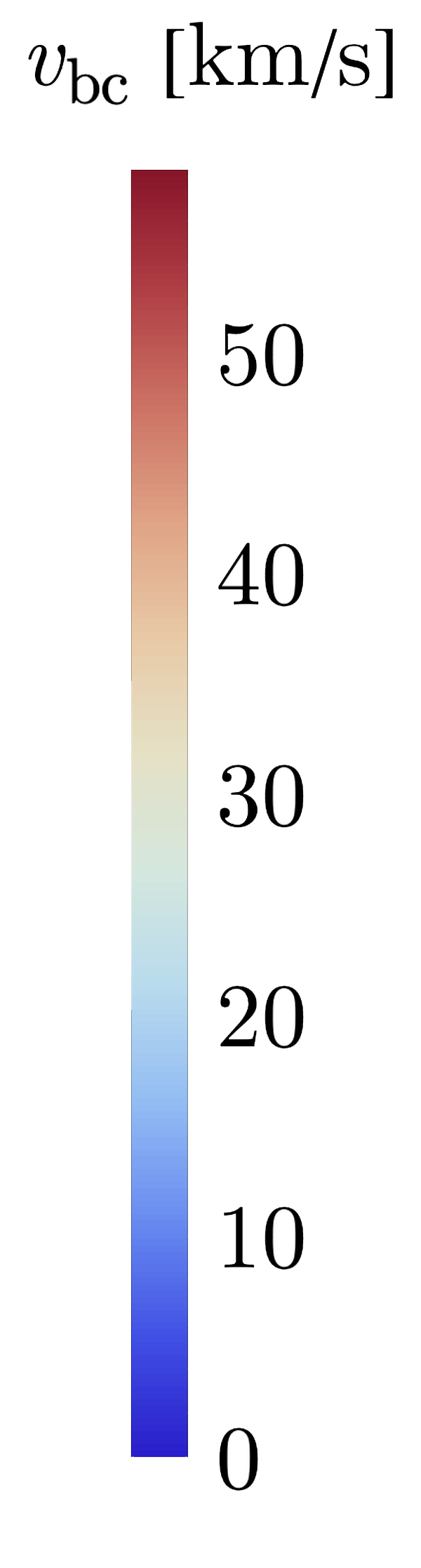}
\includegraphics[width=0.75\columnwidth]{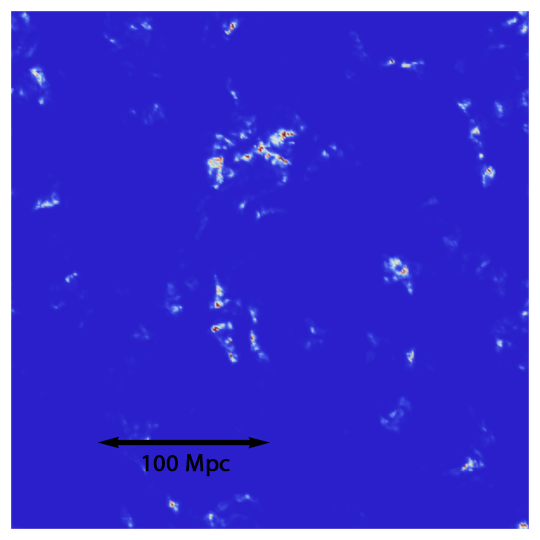}
\includegraphics[width=0.144\columnwidth]{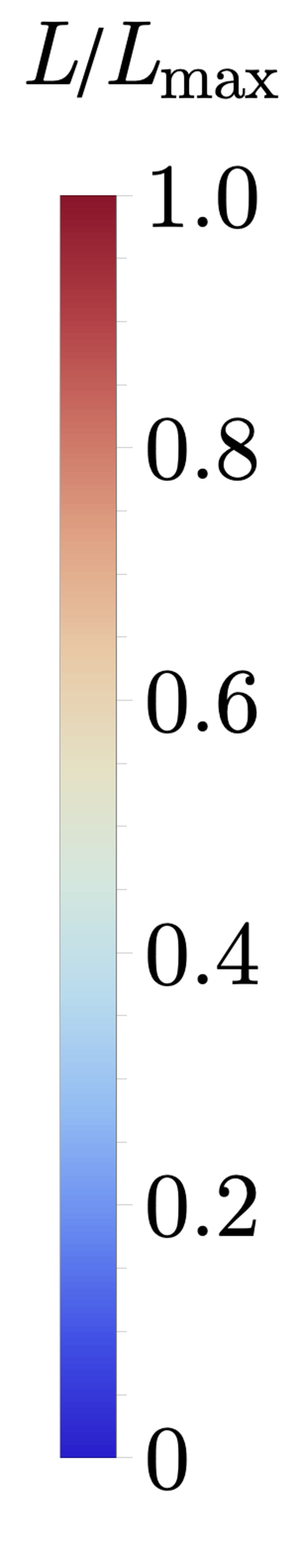}
\caption{\label{fig:relv_L} 300 Mpc $\times$ 300 Mpc slice of a realization of the relative velocity field $v_{\rm bc}$ (left), and the corresponding estimated PBH luminosity normalized by its maximum value (right), at $z = 1060$. At this redshift, the rms relative velocity is about five times the speed of sound $c_s$. As a consequence, the PBH luminosity $L \propto (c_s^2 + v_{\rm bc}^2)^{-3}$ is highly inhomogeneous, with most of the radiation concentrated in small regions with subsonic relative velocities. Figure credit: Julian Mu\~{n}oz.} 
\end{figure*}

A key step of the underlying calculation is to convert energy \emph{injection} into energy \emph{deposition}. This has been the subject of extensive studies in the context of \emph{homogeneous} energy injection \citep{slatyer09a, galli11a, hutsi11a, finkbeiner12a, slatyer16a, hongwan20a}, culminating in publicly available code packages \cite{slatyer16a, hongwan20a} that project a broad class of homogeneous energy injections into modified cosmological ionization histories. Energy injection, however, need not be spatially uniform. For instance, DM density perturbations around recombination would imply an inhomogeneous energy injection rate if DM annihilates or decays \cite{dvorkin13a}, or if part of it is made of evaporating or accreting PBHs. Another example, which provided the motivation for this work, is the highly nonuniform energy injection from accreting PBHs. As we describe in more detail below, this is due to the modulation of their accretion rate by supersonic relative velocities \cite{ferraro12a}, as illustrated vividly in Fig.~\ref{fig:relv_L}. 
Beyond these specific examples, there is no reason to expect that exotic energy injection in the early Universe should be spatially uniform in general. 

An interesting consequence of inhomogeneous energy injection is that it could imply spatial perturbations in the ionization history. In turn, inhomogeneous recombination gives rise to non-Gaussian signatures in CMB anisotropies \citep{senatore09a,khatri10a, dvorkin13a}, which are \emph{qualitatively} different from the change of the CMB power spectra resulting from homogeneous perturbations to recombination, and could be significantly more constraining. In order to quantify these effects accurately, the first step is to understand the spatial aspect of energy deposition: if it is highly nonlocal, it may partially smear out inhomogeneities in the injected power, thus partially wash out non-Gaussian signatures in CMB anisotropies (see the appendix of Ref.~\cite{dvorkin13a} for a discussion). A detailed inspection on how electromagnetically interacting particles deposit their energy spatially on cosmological scales has, to the authors' knowledge, yet to be published. 
This work lays out the initial steps in such a study, with the eventual goal of translating arbitrary spatial variations of energy injection into an inhomogeneous recombination history, and, ultimately, non-Gaussian CMB anisotropies.

As a first step in this program, we tackle the problem of injection of sub-10 MeV photons, which, among other possible applications, is relevant to accreting PBHs. In this energy regime, photons are subject only to Compton scattering off bound and unbound electrons and photoionization of neutral hydrogen and helium. The timescales for these photon interactions can be comparable to or longer than a Hubble time \cite{slatyer09a}, and we thus follow photons with a temporally and spatially dependent radiative transport code. In contrast, the energetic electrons resulting from these interactions lose their energy on a short timescale through heating, collisional excitations, ionizations, and inverse Compton scattering (ICS) of CMB photons \citep{slatyer16a,hongwan20a}. For sub-10 MeV secondary electrons, ICS results in upscattered photons with low enough energies that they very quickly deposit it into the plasma, unless they are less energetic than 10.2 eV. In the latter case, their energy is effectively lost, only
contributing to spectral distortions of the CMB \cite{slatyer16a}. 
We develop a new and highly accurate analytic estimate of this loss fraction, allowing us to simply account for energy deposition by secondary electrons, and include it in our radiative transport code. Specializing to sub-10 MeV injected photons therefore involves a relatively simple algorithmic structure (e.g.~it is not necessary to re-inject photons resulting from ICS events back into the radiative transfer code), allowing us to focus our efforts on checking the robustness of the code, from which we extract novel spatial signatures of energy deposition. In particular, we will see that for photon energies $E \gtrsim $ MeV, the spatial dependence of energy deposition is very different from the Gaussian distribution one may expect from a simple diffusion length estimate \cite{dvorkin13a}. 

\rev{As long as the effect of non-standard energy injection on the thermal and ionization history is sufficiently small, the energy deposition rate is linearly related to the energy injection rate. Mathematically, these rates are connected through a time- and scale-dependent injection-to-deposition \emph{Green's function}, which we extract from our radiative transport simulations for a given photon injection spectrum. Under the same perturbative assumption, the change in ionization fraction is linearly related to the energy injection rate, which is described mathematically by a deposition-to-ionization Green's function. The convolution of these two functions leads to the injection-to-ionization Green's function. This tool is one of the main outcomes of this work: for a given injected spectrum, it serves to compute the time and scale dependence of ionization perturbations in response to any time- and scale-dependent energy injection history.}

We apply this formalism to the specific scenario of energy injection by accreting PBHs, which are expected to radiate photons up to energies $\sim 10$ MeV \cite{Shapiro_73, yacine17a}. In the mass range of $\sim 1-10^4$ $M_{\odot}$, the luminosity of accreting PBHs is large enough that it would leave observable signatures on the thermal history of the Universe, even if PBHs make a subdominant fraction of the DM. In fact, one of the strongest constraints on PBH abundance in this mass range results from their effect on the mean ionization history, thus CMB anisotropy power spectra \citep{ricotti08a,yacine17a,poulin17a, carr20b}. In the simple Bondi accretion model \cite{bondi52a}, the PBH accretion rate has a strong dependence on supersonic relative velocities $v_{\rm bc}$ of baryons and DM, $\dot{M} \propto (v_{\rm bc}^2 +  c_s^2)^{-3/2}$, where $c_s$ is the sound speed. This dependence propagates to the PBH luminosity, thus to the energy injection rate. This implies that, on top of the spatially averaged effect, from which the most conservative CMB anisotropy limits are derived \cite{yacine17a}, there ought to be \emph{order-unity} inhomogeneities in the energy injection rate. This is to be contrasted with the small fluctuations in DM density modulating the energy injected from their annihilation products, studied in Ref.~\cite{dvorkin13a}. The characteristic length scale of PBH luminosity inhomogeneities is set by the scale over which baryon-DM relative velocity fluctuate, of order $\sim 10^2$ Mpc \citep{Tseliakhovich_10}. Convolving the injected power with our injection-to-ionization Green’s function, we are able to compute the spatial perturbations to the recombination history. Importantly, we find that the finite spatial extent of energy deposition only partially washes out inhomogeneities in the modification to the ionization fraction, which retains order-unity relative spatial fluctuations. This result bodes well for non-Gaussian signatures in CMB anisotropies, which we calculate in follow-up publications.


The remainder of this paper is organized as follows. \rev{In Sec.~\ref{sec:dep}, we start by introducing the Green's function formalism used throughout this paper. We then review the physical processes relevant to sub-10 MeV injected photons, and describe our Monte Carlo radiation transport simulations}. In Sec.~\ref{sec:xe} we convert the energy deposited into an inhomogeneous recombination history. We apply our results to accreting PBHs in Sec.~\ref{sec:PBHs}. We conclude and outline future work in Sec.~\ref{sec:conc}. For completeness, we explicitly list all the cross sections relevant to this work in Appendix \ref{app:cross_sections}. Appendix \ref{app:Greens} describes a simple semi-analytic approximation for the energy deposition Green's function, which we use as a check for our simulations. We inspect whether the perturbed free-electron response due to accreting PBHs is linear in Appendix \ref{app:xe-checks}. Throughout this paper we adopt geometric units $G_{\rm Newt} = c = 1$.

\section{Energy deposition from injected sub-10 MeV photons}\label{sec:dep}

\subsection{Injection-to-deposition Green's function}\label{subsec:green}
We denote the rate of energy injection per baryon (specifically, per hydrogen nucleus), per logarithmic scale factor interval, per photon energy interval, by 
\beq
\frac{d \epsilon_{\rm inj}}{dE_\gamma} (a, \bs{r}) = \epsilon_{\rm inj}(a, \bs{r}) \Psi(E_\gamma, a, \bs{r}),
\eeq
where $\epsilon_{\rm inj}(a, \bs{r})$ is the total energy injected per baryon per logarithmic scale factor interval, and $\Psi(E_\gamma, a, \bs{r})$ is the injected photon spectrum, normalized such that $\int dE_\gamma~ \Psi(E_\gamma) = 1$.
Similarly, we denote by $\epsilon_{\rm dep}(a, \bs{r})$ the total energy deposited into the plasma, per baryon, per logarithmic scale factor interval. The quantities $\epsilon_{\rm inj, dep}$ are related to the volumetric rates of energy injection/deposition $\dot{\rho}_{\rm inj, dep}$ through $\epsilon_{\rm inj, dep} = \dot{\rho}_{\rm inj, dep}/H n_{\rm H}$, where $n_{\rm H}$ is the total number density of hydrogen (both neutral and ionized).

Assuming the thermal and ionization state of the gas is close to its standard history, the deposited power is linearly related to the injected power through a Green's function. Specifically, we define the energy-dependent dimensionless injection-to-deposition Green's function $G_{E_\gamma}(a_d, a_i, r)$ such that
\barr
\epsilon_{\rm dep}(a_d, \bs{r}') &=& \iiint d \ln a_i ~\frac{d^3 r}{4 \pi r^3} d E_\gamma ~  \nonumber\\
&&\times  G_{E_\gamma}(a_d, a_i, r) \frac{d\epsilon_{\rm inj}}{d E_\gamma}(a_i, \bs{r}' + \bs{r}).~\label{eq:genG}
\earr
Note that homogeneity and isotropy of the background plasma ensures that the Green's function only depends on the comoving distance $r$ from the injection point (rather than on the vector $\bs{r})$. 

In the case where the photon spectrum is spatially uniform, we may integrate out the energy dependence, and define a Green's function for the specific (time dependent) spectrum $\Psi(E_\gamma, a)$, 
\beq
G^{\rm inj}_{\rm dep}(a_d, a_i, r| \Psi) \equiv \int dE_\gamma \Psi(E_\gamma, a_i)  G_{E_\gamma}(a_d, a_i, r),
\eeq
such that
\barr
\epsilon_{\rm dep}(a_d, \bs{r}') &=& \iint d \ln a_i ~\frac{d^3 r}{4 \pi r^3} \nonumber\\
&& \times G^{\rm inj}_{\rm dep}(a_d, a_i, r| \Psi)~ \epsilon_{\rm inj}(a_i, \bs{r}' + \bs{r}).\label{eq:GPsi}
\earr
We define the spatially averaged Green's function,
\beq
\overline{G}_{\rm dep}^{\rm inj}(a_d, a_i|\Psi) \equiv  \int d \ln r~ G_{\rm dep}^{\rm inj}(a_d, a_i, r|\Psi),
\eeq
and similarly define $\overline{G}_{E_\gamma}(a_d, a_i)$. This Green's function connects the spatial average of the energy deposition rate $\overline{\epsilon}_{\rm dep}$ to the spatial average of the energy injection rate $\overline{\epsilon}_{\rm inj}$: in the case of a homogeneous injection spectrum, we get
\beq
\overline{\epsilon}_{\rm dep}(a_d) = \int d \ln a_i ~ \overline{G}_{\rm dep}^{\rm inj}(a_d, a_i|\Psi) ~\overline{\epsilon}_{\rm inj}(a_i).\label{eq:G-av}
\eeq
Lastly, we define the dimensionless Fourier transform of the Green's function,
\begin{align}
G_{\rm dep}^{\rm inj}(a_d, a_i, k|\Psi) &\equiv \int d^3 r ~ \frac{G_{\rm dep}^{\rm inj}(a_d, a_i, r|\Psi)}{4 \pi r^3} \frac{\sin (kr)}{kr}\nonumber\\
&= \int d \ln r ~ G_{\rm dep}^{\rm inj}(a_d, a_i, r|\Psi)\frac{\sin (kr)}{kr},
\end{align}
and similarly for $G_{E_\gamma}(a_d, a_i, k)$. This allows us to connect the Fourier components of the energy deposition rate to those of the energy injection rate; for a spatially uniform spectrum $\Psi$ (but inhomogeneous total injection rate $\epsilon_{\rm inj}$), we have
 \beq
{\epsilon}_{\rm dep}(a_d, \bs{k}) = \int \td \ln a_i ~ G^{\rm inj}_{\rm dep}(a_d, a_i, k|\Psi){\epsilon}_{\rm inj} (a_i, \bs{k}).\label{eq:genG_fourier}
\eeq

\subsection{Interactions processes for sub-10 MeV photons and electrons}

As photons propagate in the expanding Universe, they lose energy through redshifting and interacting with the plasma. At redshifts of interest, sub-10 MeV photons are only subject to two interactions: Compton scattering and photoionization \cite{slatyer09a}.

Sufficiently high energy photons Compton scatter with both free and bound electrons. At energies $E_\gamma \lesssim \hbar/a_0 = \alpha m_e \approx 4$ keV, where $a_0$ is the Bohr radius, Compton scattering with bound electrons is suppressed (see e.g.~Ref.~\cite{Hubbell_75}). Given that at these low energies, photoionization is the dominant source of energy loss for photons, for simplicity, and at no loss of accuracy, we may assume that photons scatter with bound and free electrons at all energies, with the unsuppressed Compton cross section. 

In addition, photons may photoionize hydrogen atoms if their energy is above $E_I = 13.6$ eV and helium atoms if their energy is above $24.6$ eV. We consider redshifts $z\lesssim 1800$, at which helium is fully recombined \cite{Switzer_08}, and thus need not account for photoionization of singly ionized helium. 

Upon Compton scattering, a photon with initial energy $E_\gamma \lesssim 10$ MeV transfers part of its energy to a secondary electron, with energy $E_e \lesssim 10$ MeV. This energy is typically much greater than atomic binding energies, thus results in an ionization event if the electron was initially bound. In the case of photoionization events, the photon terminates and deposits essentially all of its energy (minus the binding energy) into the freed electron. In both instances, the secondary electron then deposits all of its energy on a short timescale, as we describe below. Given that atomic binding energies are much less than the typical energies of secondary electrons, we can neglect the small amount of energy directly deposited by the initial photon in photoionization and Compton scattering events off bound electrons. 

Energetic electrons are subject to four possible interactions in the early Universe: they may collisionally ionize or excite a neutral hydrogen or helium atom, transfer part of their kinetic energy to (i.e.~heat up) another electron, or inverse Compton scatter (ICS) a CMB photon. Following either one of these interactions, the outgoing electron (or electrons, in the case of collisional ionization) promptly interacts again through either one of the four channels, and so on, until all of the initial electron's kinetic energy is used up. The timescales for these interactions are several orders of magnitude shorter than the Hubble timescale at the epochs of interest, so that the loss of energy of the initial electron can be approximated as effectively instantaneous \citep{slatyer16a,hongwan20a}, as well as spatially local. The end result is that a fraction of the initial electron's energy is eventually deposited into ionization, excitation and heating, and the remainder ends up in photons produced by ICS.

In principle, the photons resulting from ICS should be added to and evolved alongside the primary photon spectrum. However, for the electron energies of interest $E_e \lesssim 10$ MeV, the photons produced in ICS have energy $E_{\gamma}' \lesssim 4(E_e/m_e)^2 T_\gamma \sim 500~\textrm{eV}(z/10^3) (E_e/10~\textrm{MeV})^2$, much lower than the primary photons' energies. At these low energies, the fate of upscattered photons is simple to determine. Photons with energies $10.2 ~\textrm{eV} < E_\gamma \lesssim 500$ eV interact with the plasma on a timescale much shorter than the Hubble time \cite{slatyer09a}, by photoionizing or exciting a neutral atom; the subsequently produced electrons can themselves ionize, excite or heat the plasma, but have too little energy to efficiently ICS CMB photons. Therefore, upscattered photons with energies $10.2 ~\textrm{eV} < E_\gamma \lesssim 500$ eV promptly deposit their energy in the form of ionization, excitation and heating. In contrast, photons with energies $E_\gamma < 10.2$ eV no longer interact with the plasma.
This last channel is thus taken effectively as an energy sink, as far as the ionization and thermal history is concerned (but it can lead to CMB spectral distortions, which we do not consider in this work) \cite{slatyer16a}. 

In summary, we see that an electron with initial energy $E_e\lesssim 10$ MeV quickly deposits its energy into four channels: ionization, excitation, heating and non-interacting sub-10.2 eV photons, which constitutes a sink. Given an initial electron energy $E_e$, we define $F_{\rm sink}(E_e)$ to be the fraction of energy that goes into the latter channel. Its complement $F_{\rm dep}(E_e) \equiv 1 - F_{\rm sink}(E_e)$ is therefore the fraction of energy that is efficiently deposited into the plasma.

\subsection{Sub-10.2 eV ICS energy sink fraction for secondary electrons}\label{subsec:elec}

We now turn to computing the fraction $F_{\rm sink}(E_e)$ of an electron's energy that is lost to sub-10.2 eV ICS photons. Here we derive a simple yet remarkably accurate analytic solution, matching the numerical results of Ref.~\cite{hongwan20a}.

If an electron with energy $E_e$ collisionally ionizes a neutral atom, the end state consists of a free proton and two free electrons with energies $E_e'\geq E_e''$, such that $E_e' + E_e'' = E_e - E_I$. In principle one should keep track of both electrons following an ionization event. However, the differential ionization cross section $d\sigma_{\rm ion}/dE_e'$ is peaked at $|E_e - E_e'| \sim E_I$, i.e.~$E_e'' \sim E_I \ll E_e'$ (see Appendix \ref{app:cross_sections}). Therefore, in practice we may neglect the lower-energy electron. The differential rate of ionization events per final electron energy interval is therefore 
\beq
\frac{d \Gamma_{\rm ion}(E_e)}{dE_e'} = n_{\rm at} \frac{d\sigma_{\rm ion}(E_e)}{ d E_e'},
\eeq
where $n_{\rm at}$ is the abundance of the relevant atomic species: $n_{\rm at} = (1 - x_e) n_{\rm H}$ for hydrogen, where $x_e$ is the ionization fraction, and $n_{\rm at} = n_{\rm He}$ for neutral helium.

If an electron has energy greater than the atomic excitation energy $E_{\rm exc}$, it may collisionally excite a neutral hydrogen or helium atom. For simplicity, we only consider collisional excitation from the ground state to the first excited state, with $E_{\rm exc} = 10.2$ eV and 21.2 eV for hydrogen and helium, respectively. The corresponding differential rate takes the form 
\beq
\frac{d \Gamma_{\rm exc}(E_e)}{dE_e'} = n_{\rm at} \sigma_{\rm exc}(E_e) \delta(E_e' + E_{\rm exc} - E_e),
\eeq
where $\sigma_{\rm exc}(E_e)$ is the collisional excitation cross section, provided explicitly in Appendix~\ref{app:cross_sections}.

Similarly, we denote by $d \Gamma_{\rm heat}(E_e)/dE_e'$ the differential rate at which an electron with energy $E_e$ interacts with the plasma in ``heating events", producing a final electron with energy $E_e'$. We will see shortly that the relevant quantity of interest is the heating rate, given in Appendix \ref{app:cross_sections}, rather than this differential interaction rate. We only need the latter for intermediate calculations, and will assume that it is sharply peaked at $E_e' \approx E_e$ \citep{furlanetto10a}.

Lastly, let us consider ICS of CMB photons. The final state of an ICS event is an electron with energy $E_e'$ and an upscattered photon with energy $E_\gamma'$, such that $E_e' + E_\gamma' = E_e + E_{\gamma, \rm cmb} > E_e$. The total differential rate of ICS events is given by converting the doubly differential rate provided in Appendix \ref{app:cross_sections},
\beq
\frac{d \Gamma_{\rm ICS}(E_e)}{dE_e'} = \int dE_{\gamma}' \frac{d^2 \Gamma_{\rm ICS}(E_e)}{dE_e' dE_{\gamma}'}.
\eeq
As discussed earlier, upscattered photons with energy $E_\gamma' < E_{\rm exc} = 10.2$ eV no longer interact and are effectively an energy sink. We may define the corresponding differential rate by
\beq
\frac{d \Gamma_{\rm sink}(E_e)}{dE_e'} = \int^{E_{\rm exc}} d E_\gamma' \frac{d^2 \Gamma_{\rm ICS}(E_e)}{dE_e' dE_{\gamma}'},
\eeq
which is nonzero if $E_e'>E_e-E_{\rm exc}$.

We denote by $\Gamma_{\rm tot}(E_e)$ the total rate of interaction of the original electron:
\begin{align}
\Gamma_{\rm tot}(E_e) &= \int dE_e' \left( \frac{d \Gamma_{\rm ion}(E_e)}{d E_e'} + \frac{d \Gamma_{\rm exc}(E_e)}{d E_e'} \right.\nonumber\\
& ~~~~~~~~~~~~~ \left. + \frac{d \Gamma_{\rm heat}(E_e)}{d E_e'} + \frac{d \Gamma_{\rm ICS}(E_e)}{d E_e'}\right).
\end{align}
We are now in a position to compute the fraction $F_{\rm sink}(E_e)$ of the initial electron energy $E_e$ going into sub-10.2 eV photons. Assuming all the processes at play occur on a timescale much shorter than the expansion time, $F_{\rm sink}(E_e)$ satisfies the following integral equation:
\barr
 F_{\rm sink}(E_e) = \frac1{\Gamma_{\rm tot}(E_e)}\int dE_e' \frac{d \Gamma_{\rm sink}(E_e)}{dE_e'} \frac{E_e- E_e'}{E_e} \nonumber\\
+ \frac1{\Gamma_{\rm tot}(E_e)} \sum_{c} \int dE_e' \frac{d \Gamma_{c}(E_e)}{dE_e'} \frac{E_e'}{E_e} F_{\rm sink}(E_e'). \label{eq:Fc}
\earr
The first term in this equation accounts for the fraction of the electron's energy that directly goes into the sink channel, and the second term accounts for the indirect sink deposition, after first interacting through any of the channels $c$ = ionization, excitation, heat, and ICS. 

The discretized version of Eq.~\eqref{eq:Fc} was solved numerically in Ref.~\cite{hongwan20a}. Here we propose a simple approximation that dramatically simplifies the evaluation of $F_{\rm sink}(E_e)$, yet produces very accurate results. We consider electrons with energies $E_e \gg E_I$. In that case, in all the processes considered, electrons only lose a small fraction of their energy upon interacting. Mathematically, the rates $d \Gamma_c(E_e)/d E_e'$ are all peaked at $E_e'\approx E_e$, with a width much smaller than $E_e$. Assuming that $F_{\rm sink}(E_e)$ is a smooth function of $E_e$ (which we confirm a posteriori), inside the integral we may approximate  
\beq
E_e' F_{\rm sink}(E_e') \approx E_e F_{\rm sink}(E_e) - (E_e-E_e') \frac{d}{dE_e}(E_e F_{\rm sink}(E_e)).
\eeq
With this approximation, Eq.~\eqref{eq:Fc} becomes a simple first-order ordinary differential equation, 
\beq
\frac{d}{dE_e}(E_e F_{\rm sink}(E_e)) \approx \frac{\dot{\mathcal{E}}_{\rm sink}(E_e)}{\sum_{c}\dot{\mathcal{E}}_{c}(E_e)}, \label{eq:F-ode}
\eeq
where $\dot{\mathcal{E}}_c(E_e)$ is the rate of direct energy deposition through channel $c$:
\beq
\dot{\mathcal{E}}_c(E_e) \equiv \int dE_e' \frac{d \Gamma_c(E_e)}{dE_e'} (E_e - E_e'). \label{eq:dotE_c}
\eeq
Equation \eqref{eq:F-ode} has the explicit integral solution
\beq
F_{\rm sink}(E_e) = \frac1{E_e} \int_0^{E_e} dE_e' \frac{\dot{\mathcal{E}}_{\rm sink}(E_e')}{\sum_{c}\dot{\mathcal{E}}_{c}(E_e')}.\label{eq:elec_app}
\eeq

\begin{figure}[htbp]
\includegraphics[width=\columnwidth]{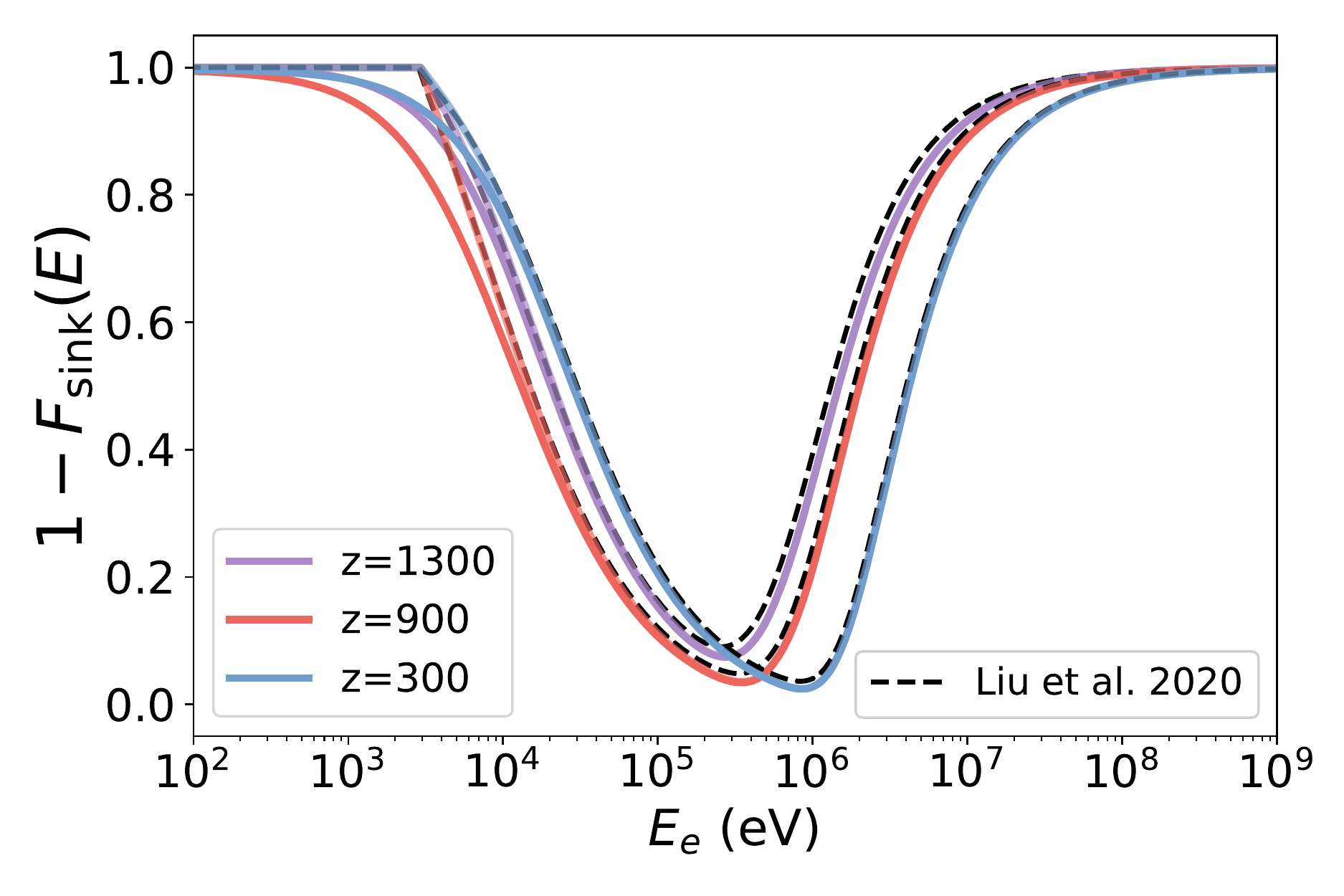}
\caption{\label{fig:rat} Total fraction of an electron's energy $E_e$ that does not end in sub-10.2 eV photons. For electron energies $E_e \lesssim 10$ MeV of interest in this paper, this corresponds to the fraction of energy efficiently deposited into the plasma in the form of ionizations, heating and excitations. Solid lines show our analytic approximation Eq.~\eqref{eq:elec_app} and dashed black lines are the numerical results from Ref.~\cite{hongwan20a}. The semitransparent lines are obtained with our approximation, but imposing a sharp boundary condition $F_{\rm sink}(E_e < 3$ keV) = 0, as is done in Ref.~\cite{hongwan20a}, and showing that this sharp cutoff is the main source of difference between our result and that of Ref.~\cite{hongwan20a} at $E_e \lesssim 10$ keV. Otherwise, we find a remarkable agreement at all energies and redshifts. We use the solid lines with no energy cutoff in this paper.}
\end{figure}

We show our analytic solution for $1 - F_{\rm sink}(E_e)$ in Fig.~\ref{fig:rat}, and compare it to the numerical solution of Ref.~\cite{hongwan20a}. The sharp feature in Ref.~\cite{hongwan20a}'s result is due to their assumption that electrons lose all their energy via atomic processes at $E_e < 3$ keV (see also \cite{slatyer16a}), implying $F_{\rm sink} = 0$ for $E_e < 3$ keV. When imposing this boundary condition for comparison, we find that our approximation agrees remarkably well with the results of Ref.~\cite{hongwan20a} across all redshifts and electron energies. For our computations we make no such cutoff and simply use Eq.~\eqref{eq:elec_app}, which gives a smooth transition $F_{\rm sink} (E_e) \rightarrow 0$ at low energies, differing somewhat from (and likely more accurate than) that of Ref.~\cite{hongwan20a} at electron energies $E_e \lesssim 10$ keV. 

Let us note that our calculation for the fraction $F_{\rm sink}(E_e)$ of electron energy deposited into sub-10.2 photon is accurate across all energies $E_e \gg E_I$ (as can be seen from the agreement with the numerical results of Ref.~\cite{hongwan20a} up to $E_e$ = GeV). However, its complement $1- F_{\rm sink}(E_e)$ can be interpreted as the fraction of energy efficiently deposited into ionizations, excitations and heating only for $E_e \lesssim 10$ MeV. For higher electron energies, part of the energy not ending in the sub-10.2 eV sink goes into higher-energy upscattered photons which do not necessarily interact immediately with the plasma, and would have to be followed numerically alongside the primary injected photons.

Let us also remark that our simple analytic approximation for $F_{\rm sink}(E_e)$ can be easily generalized to the fraction of electron energy deposited into ionizations, excitations, heating and ICS photons. For the former three channels, the relevant fractions $F_c(E_e)$ satisfy Eq.~\eqref{eq:elec_app}, with the substitution $\dot{\mathcal{E}}_{\rm sink}(E_e') \rightarrow \dot{\mathcal{E}}_c(E_e')$ in the numerator. For ICS photons, one can define a differential fraction of energy deposited into photons, per photon energy interval, $dF_{\rm ICS}(E_e)/d E_\gamma'$, satisfying Eq.~\eqref{eq:elec_app}, with the substitution $\dot{\mathcal{E}}_{\rm sink}(E_e') \rightarrow d\dot{\mathcal{E}}_{\rm ICS}(E_e')/dE_\gamma'$ in the numerator, where the latter is defined as in Eq.~\eqref{eq:dotE_c}, with $d \Gamma_c(E_e)/dE_e' \rightarrow d^2 \Gamma_{\rm ICS}(E_e)/dE_e'dE_\gamma'$.

\subsection{Radiation transport simulations}\label{subsec:radsim}

 While at $z \gg 10^3$ photon interactions occur on a timescale much shorter than the Hubble time, it is not the case around and after recombination, at $z \lesssim 10^3$. Therefore, one must study the time-dependent evolution of the photon spectrum, and we do so with a Monte Carlo radiative transport simulation, following the temporal and spatial evolution of sub-10 MeV photons.\par 

For a given time-dependent injected photon spectrum $\Psi(E_\gamma, z)$ (such that $\int d E_\gamma \Psi(E_\gamma, z) = 1$), we run a separate simulation for every injection redshift $z_{i}$. We assume a matter- and radiation-dominated Universe (i.e.~neglect dark energy), with cosmological parameters consistent with the \emph{Planck} 2018 results \cite{planck20b}. We assume a standard ionization history computed with \texttt{HyRec} \cite{YAH_10, yacine11a}, i.e.~we do not account for the feedback of a modified ionization history on the energy deposition efficiency. 

\emph{Initialization --} We initialize the simulation at redshift $z_i$ with $N = 10^6$ photons, all located at the origin of coordinates $r = 0$. The photons energies are distributed according to $d N/dE_\gamma = E_{\rm tot} \Psi(E_\gamma,z_{\rm inj})/E_\gamma$, where $E_{\rm tot} = N \left(\int dE_\gamma \Psi(E_\gamma)/E_\gamma\right)^{-1}$ is the total injected energy.

\emph{Quantities evolved --} In the course of a simulation, we keep track of four phase-space coordinates for each of the $N$ photons, namely their energy $E_\gamma$ and 3-dimensional comoving vector to the origin ${\vec r}_\gamma$. Note that we do not store these quantities as a function of redshift, but simply update them at each timestep. Since the photons do not interact with one another, we may use a different coordinate system for each photon, and choose it such that, at any given time, the photon's direction of propagation is along the $z$ axis.

\emph{Timestep --} We take logarithmic time steps in scale factor $d \ln a$, no larger than 0.0025, and such that the probability of any photon to either Compton scatter or photoionize a hydrogen or helium atom is at most 0.005.

\emph{Free-streaming step -- } We account for cosmological redshifting by updating each photon's energy to $E_\gamma := E_\gamma \rme^{- d \ln a}$. We update each photon's position by freely propagating it from its position at the previous timestep along the current direction of propagation $\hat{z}$, i.e.~$\vec{r}_\gamma := \vec{r}_\gamma + (d \ln a/ aH) ~ \hat{z}$.

\emph{Interaction step -- } For each photon, we compute the probability of photoionizing a hydrogen or helium atom, and of Compton scattering during $d \ln a$. Explicitly, for a photon of energy $E_\gamma$, the probability for each process $X$ is given by 
\barr
P_{X}(E_\gamma) = n_{X} \sigma_X(E_\gamma) \frac{d \ln a}{H(a)},
\earr
where the relevant cross sections are given in Appendix \ref{app:cross_sections}, and $n_X$ is the number density of scatterers relevant to process $X$ -- neutral hydrogen or helium abundance for photoionization, total abundance of free and bound electrons for Compton scattering. 

We draw a first random number for each photon, uniformly distributed in (0, 1). If this number is less than $P_{\rm ion, H}(E_\gamma)$, the photon photoionizes a hydrogen atom, leading to an electron of energy $E_e = E_\gamma - 13.6$ eV. The original photon is then terminated. We reiterate this procedure with the remaining photons for Helium photoionization. If a photon photoionizes a neutral Helium atom, it is terminated and leads to an electron of energy $E_e = E_\gamma - 24.6$ eV.

With the same procedure, we determine whether each remaining photon Compton scatters. If so, we sample the polar angle $\theta$ (with respect to the propagation direction) into which the photon scatters and resulting final energy $E_\gamma'$ from Eqs.~\eqref{eq:kn}~\&~\eqref{eq:Ef}, and uniformly sample the azimuthal angle $\phi$ in $[0, 2 \pi)$. This process results in an electron with energy $E_e = E_\gamma - E_\gamma'$. We then update the photon's energy $E_\gamma := E_\gamma'$, and rotate the photon's coordinate system such that the new direction of propagation is along the $z$ direction; explicitly, we update its spatial coordinates $\vec{r}_{\gamma} := \mathbf{R}(\theta, \phi) \cdot  \vec{r}_{\gamma}$, with the rotation matrix
\begin{align}
\mathbf{R}(\theta, \phi)\equiv
    \begin{pmatrix}
    \cos\theta\cos\phi & \cos\theta\sin\phi & -\sin\theta \\
    -\sin\phi & \cos\phi & 0 \\
    \sin\theta \cos\phi & \sin\theta\sin\phi & \cos\theta 
    \end{pmatrix}.
\end{align}
At the end of each timestep, we thus have an updated table of photon energies and position vectors, for photons that have not been terminated. Moreover, for each photon that interacted during the timestep, we have extracted the energy $E_e$ of the secondary electron produced in the interaction.\par

\emph{Maintaining a large photon sample --} As a simulation progresses, photons lose energy to redshifting and Compton scattering, and the gas is increasingly neutral. As a consequence, photons are increasingly likely to be terminated in photoionization events. In order to maintain low statistical errors, we duplicate the remaining photons every time their number decreases by a factor of 2. This procedure is equivalent to having initialized the simulation with twice the original photon number, and we therefore update $E_{\rm tot} \rightarrow 2 \times E_{\rm tot}$ every time we duplicate photons. Depending on the injection energy and redshift, this duplication can happen up to $\mathcal{O}(10)$ times.\par

\emph{Simulation outputs -- } For a given photon injection spectrum $\Psi$, the end results of each simulation (with initial scale factor $a_i$) is a 2-dimensional table of the injection-to-deposition Green's function $G_{\rm dep}^{\rm inj}(a_d, a_i, r_k)$, in predetermined bins in deposition scale factor $a_d$ and comoving distance from the origin $r_k$. The scale factor bins are logarithmically distributed, in ($6.6\times 10^{-4}$, 0.020) with bin width $\Delta \ln a=0.005$ (this is fixed and not to be confused with the adaptive timestep $d \ln a$). The radial distance bins are logarithmically distributed in (1, $10^3$) Mpc with bin width $\Delta \ln  r = 0.05$; we also include a single bin for $ 0 < r < 1$ Mpc and a single final bin for $r > 10^3$ Mpc. The Green's function table $G_{\rm dep}^{\rm inj}(a_d, a_i, r_k)$ is initialized to zero; at each timestep in the simulation, if the scale factor $a$ falls within the $d$-th bin $(a_d - \Delta \ln a/2, a_d + \Delta \ln a/2)$, the relevant table row is incremented by
\beq
G_{\rm dep}^{\rm inj}(a_d, a_i, r_k) \mathrel{+}= \frac{\sum_{\{ r_\gamma \textrm{ in $k$-th bin} \}} E_e F_{\rm dep}(E_e)}{E_{\rm tot}~ \Delta \ln a ~\Delta \ln r}, \label{eq:G_dep-sim}
\eeq
where the sum goes over all photons that have interacted during the timestep, and the fraction $F_{\rm dep}(E_e)$ of the secondary electron's energy efficiently deposited was described in Sec.~\ref{subsec:elec}. This numerical Green's function matches our formal definition \eqref{eq:GPsi}. This can be checked explicitly by inserting the simulation's input, corresponding to $\epsilon_{\rm inj}(a, \bs{r}) = E_{\rm tot} \delta^3(\bs{r}) \delta(\ln a - \ln a_i)/n_{\rm H}^0$, where $n_{\rm H}^0$ is the comoving baryon density, into Eq.~\eqref{eq:GPsi}.\par

\rev{\emph{Convergence --} To check convergence with respect to the number of injected photons, timestep length, and bin resolution, we have run a Dirac-delta photon injection spectrum simulation with 60 times as many photons ($N_\gamma=60\times 10^6$), half the maximum logarithmic timestep ($d\ln a$ no larger than 0.00125), and double the radial and temporal bin resolution ($\Delta \ln r=0.025$, $\Delta \ln a=0.0025$). We find that, although there is less noise in the real-space output Eq.~\eqref{eq:G_dep-sim}, computing the Fourier transformed Green's function defined in Eq.~\eqref{eq:genG_fourier} produced indistinguishable results. In other words, our simulations produce well-converged large-scale Fourier-space Green's functions, of interest here.}\\

Note that we do not keep track of the energy deposited directly into photoionizations, as it is small relative to the energies of the secondary electron produced in photoionization events, of which we do keep track. This neglect is moreover consistent with our Taylor-expansion approximation for $F_{\rm dep}(E_e)$, which breaks down near the ionization threshold. These approximations become inaccurate for sub-keV photons, for which the neglected atomic binding energy exceeds a percent of the energy deposited.  For injected photons with energies below a few keV, the timescale for photoionization is much shorter than the Hubble time even at $z\sim 100$, and thus these photons could be treated as depositing energy effectively instantaneously. The comoving mean free path for helium photoionization is $\ell_{\rm mfp} \approx 4 ~\textrm{Mpc} (10^2/z)^2 (E_\gamma/\textrm{keV})^{3.3}$, implying that energy deposition can be approximated as spatially on-the-spot at $z \gtrsim 10^2$, for scales larger than a few Mpc. Our treatment would therefore have to be improved if one is interested in sub-keV photon injection at redshifts $z \lesssim 10^2$, relevant e.g.~for 21-cm fluctuations. Our main focus is on photons injected at $z \gg 10^2$, with initial energies well above a keV, and whose vast majority photoionize well before reaching a keV. We thus expect our neglect of photoionization energies to be very accurate for our purposes.

\subsection{Results}

\subsubsection{Spatially averaged Green's function: crosscheck against existing results and analytic solutions}

As a crosscheck of our numerical code, we extract the spatially averaged Green's function for a Dirac spectrum of injected photon energies, i.e.~$\overline{G}^{\rm inj}_{\rm dep} = \overline{G}_{E_\gamma}$. The most sophisticated numerical computations of $\overline{G}_{E_\gamma}$ are provided in Refs.~\cite{slatyer16a,hongwan20a}, and simple approximations are provided in Refs.~\cite{dvorkin13a,yacine17a}. In what follows, we compare the results from our Monte Carlo radiation transport simulations with existing results and a new analytic approximation we develop in Appendix \ref{app:Greens}. 

First, we consider a simplified problem: only accounting for Compton scattering, neglecting photoionizations, and assuming $F_{\rm dep}(E) = 1$, i.e.~full efficiency of secondary electron energy deposition. We derive a simple semi-analytic approximation for the corresponding Green's function in Appendix \ref{app:Greens}, generalizing the result of Ref.~\cite{yacine17a}. The final result is 
\barr
\overline{G}^{\rm analytic}_{E_\gamma}(a_d, a_i) = \frac{\dot{\mathcal{E}}_{\rm C}(a_d, E_{\rm trj}(a_d; E_\gamma, a_i))}{E_\gamma H(a_d)}\label{eq:Ge-analytic2},
\earr
where $\dot{\mathcal{E}}_{\rm C}(a, E)$ is the mean rate of energy loss due to Compton scattering defined explicitly in Eq.~\eqref{eq:dotE_Compt}, and $E_{\rm trj}(a; E_\gamma, a_i)$ is the energy trajectory of a photon at scale factor $a$, subject to redshifting and to the mean rate of energy loss to Compton scattering. \rev{That is, we solve $\dot{E}_{\rm trj} = - H E_{\rm trj} - \dot{\mathcal{E}}_{\rm C}(a, E_{\rm trj})$ with initial conditions $E_{\rm trj}(a_i) = E_\gamma$, where $E_\gamma$ is the photon's initial injection energy. In words, Eq.~\eqref{eq:Ge-analytic2} is the normalized instantaneous energy loss of photons to Compton scattering at scale factor $a_d$, assuming they have followed a ``mean" energy trajectory $E_{\rm trj}$ since their injection at $a_i$.}

\begin{figure}[ht]
\includegraphics[trim={0.7 0 0.4cm 0},width=\columnwidth]{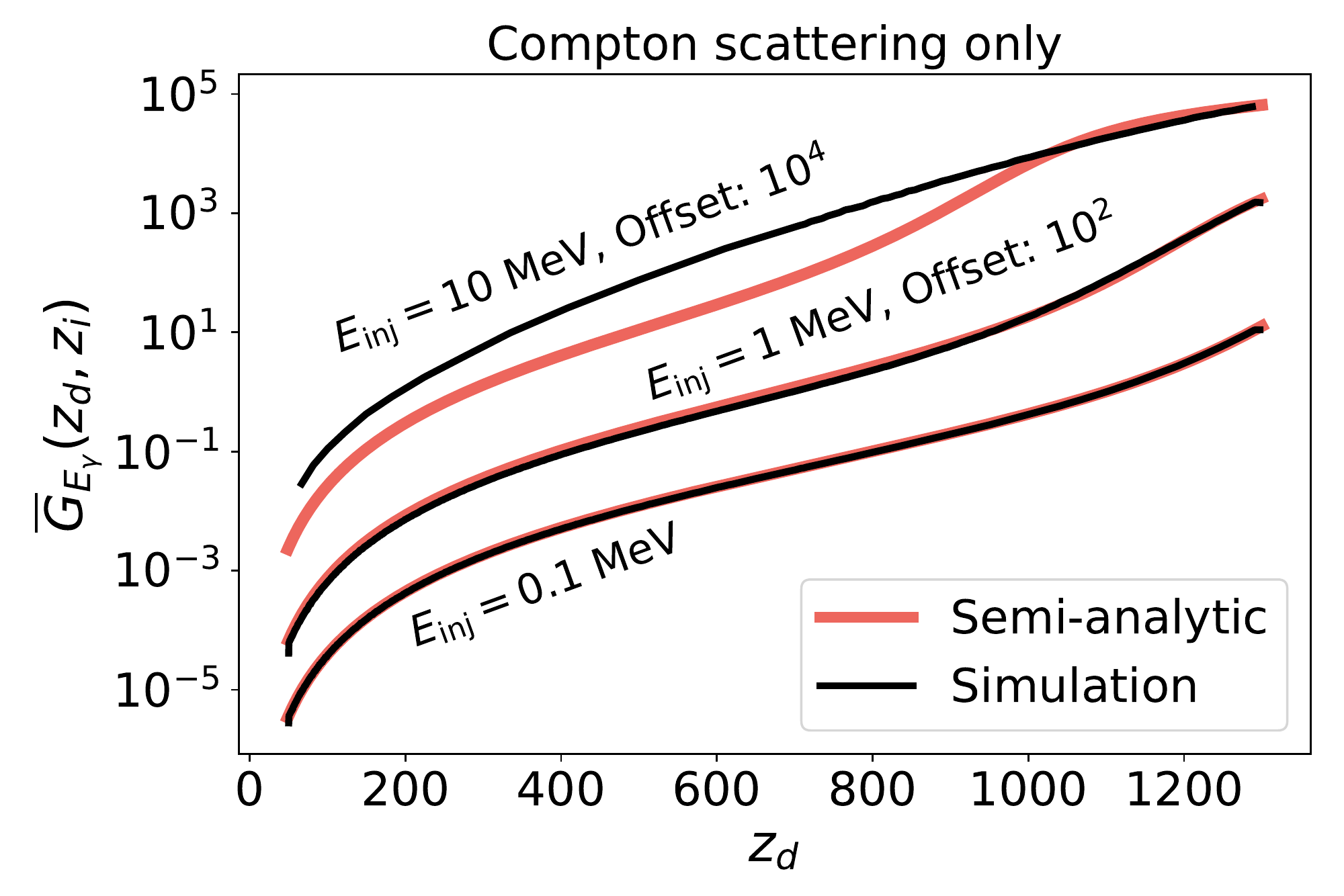}
\caption{\label{fig:comp} Comparison of the spatially averaged Green's function obtained in our simulation (black) with the semi-analytic approximation \eqref{eq:Ge-analytic2} derived in Appendix \ref{app:Greens} (red), when ignoring photoionizations and assuming full efficiency of secondary electron energy deposition. The labeled curves correspond to $E_{\rm inj}=0.1$, 1, and 10 MeV at $z_{i}=1300$, with multiplicative offsets for clarity. The agreement is excellent for low energies, giving us confidence in our simulations' accuracy; it is poor for 10 MeV injected photons, for which the semi-analytic approximation breaks down.}
\end{figure}

\begin{figure*}[ht]
\includegraphics[trim={0.7 0 0.4cm 0},width=\columnwidth]{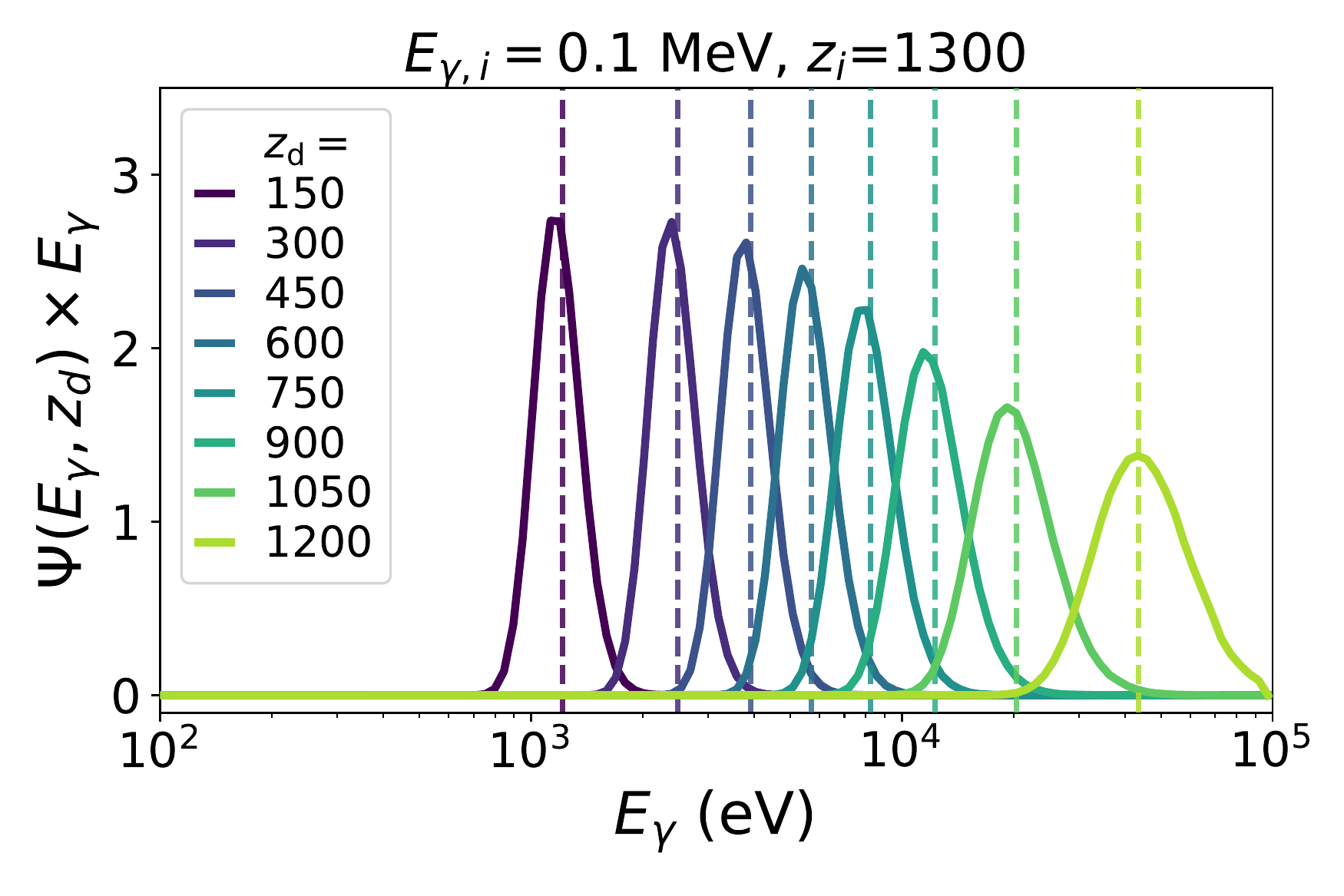}
\includegraphics[trim={0.7 0 0.4cm 0},width=\columnwidth]{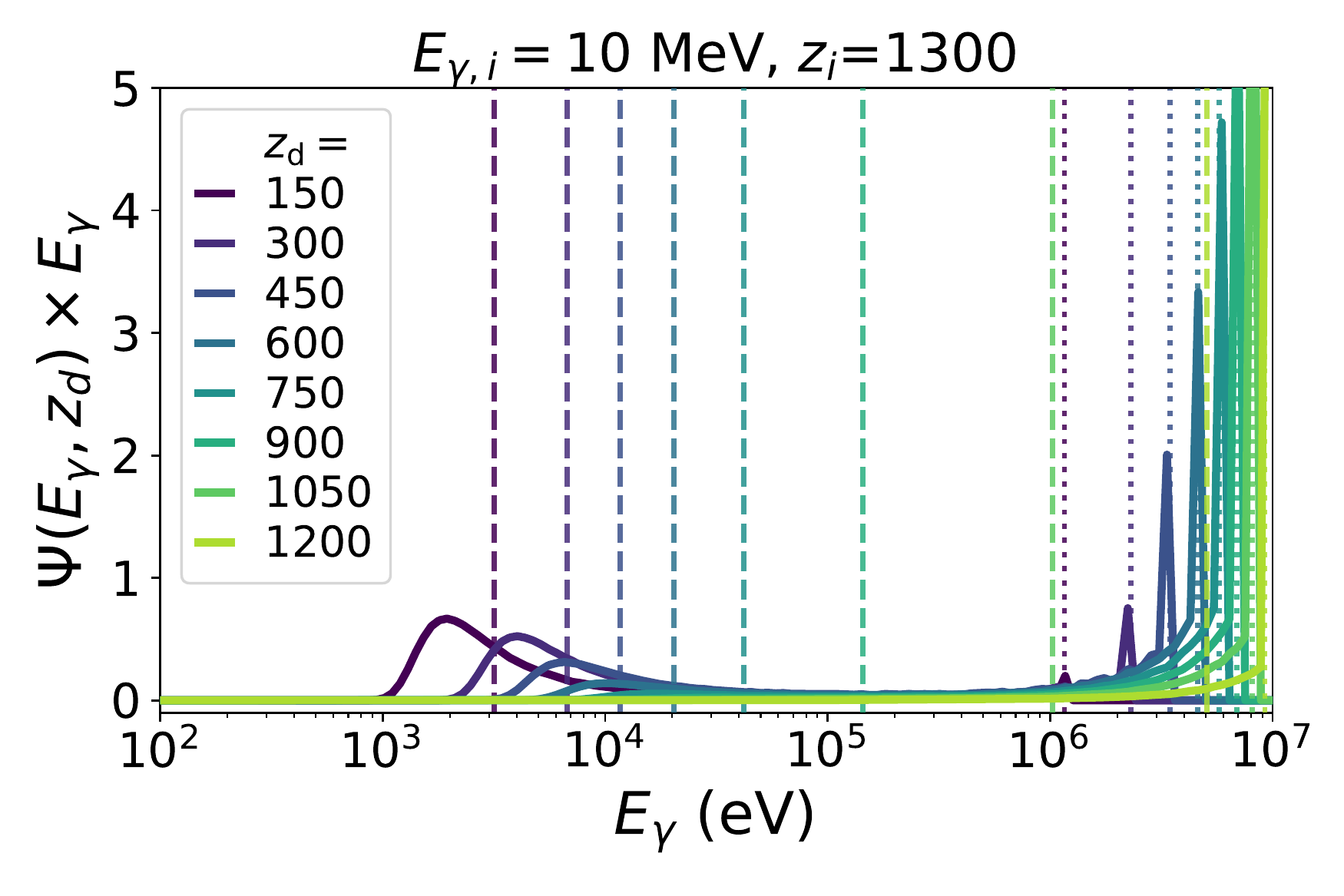}
\caption{\label{fig:Espec}Evolution of the normalized photon energy spectrum with $z_i=1300$ as a function of energy over several deposition redshifts, for a Compton-scattering-only simulation, with a Dirac-delta spectrum of injected photons, at energy $E_{\gamma, i} = 0.1$ MeV (left) and 10 MeV (right). The vertical dashed lines show the mean energy trajectory of photons subject to Compton scattering and Hubble flow discussed in Appendix~\ref{app:Greens}. For 10 MeV photons, the right panel reveals a bimodal distribution that is not well approximated by this mean energy: the high-energy mode is dominated by photons that have not scattered once and hence have energies $E_{\gamma,i}(1+z_d)/(1+z_i)$ (dotted lines in the right panel), and the low-energy mode is photons that have scattered many times.}
\end{figure*}

In Fig.~\ref{fig:comp}, we compare the semi-analytic result \eqref{eq:Ge-analytic2} to our numerical Green's function obtained from Compton-scattering-only simulations, for initial photon energies $E_{\gamma}=0.1$, 1, and 10 MeV injected at $z_{i}=1300$. It can be seen that the two agree remarkably well for injected energies $E_\gamma = 0.1$ and 1 MeV, giving us confidence in the robustness of our simulations. For $E = 10$ MeV, however, the match is rather poor. \rev{This results from the fact that, within our analytic approximation, an initially narrow distribution of injected photons evolves into a narrow distribution centered at $E_{\rm traj}(E_\gamma; a_d, a_i)$.} As can be seen in Fig~\ref{fig:Espec}, this is not accurate at either injected energies, as the time-evolved photon spectrum is broad, as a result of the finite width of the distribution of final photon energies in Compton scattering. Nevertheless, for injection energy $E_\gamma = 0.1$ MeV, the photon spectrum is indeed centered around $E_{\rm trj}(a_d; E_\gamma, a_i)$, shown with dashed lines. In contrast, for $E_\gamma = 10$ MeV, the evolved photon energy distribution is clearly bimodal, with a significant fraction of photons near the free-streaming energy $E_\gamma a_i/a$, due to preferentially forward scattering in the Compton limit. The disagreement of the semi-analytic and numerical Green's functions at $E_\gamma = 10$ MeV is thus due to the breakdown of the assumptions underlying the former. 

\begin{figure}[ht]
\includegraphics[width=\columnwidth]{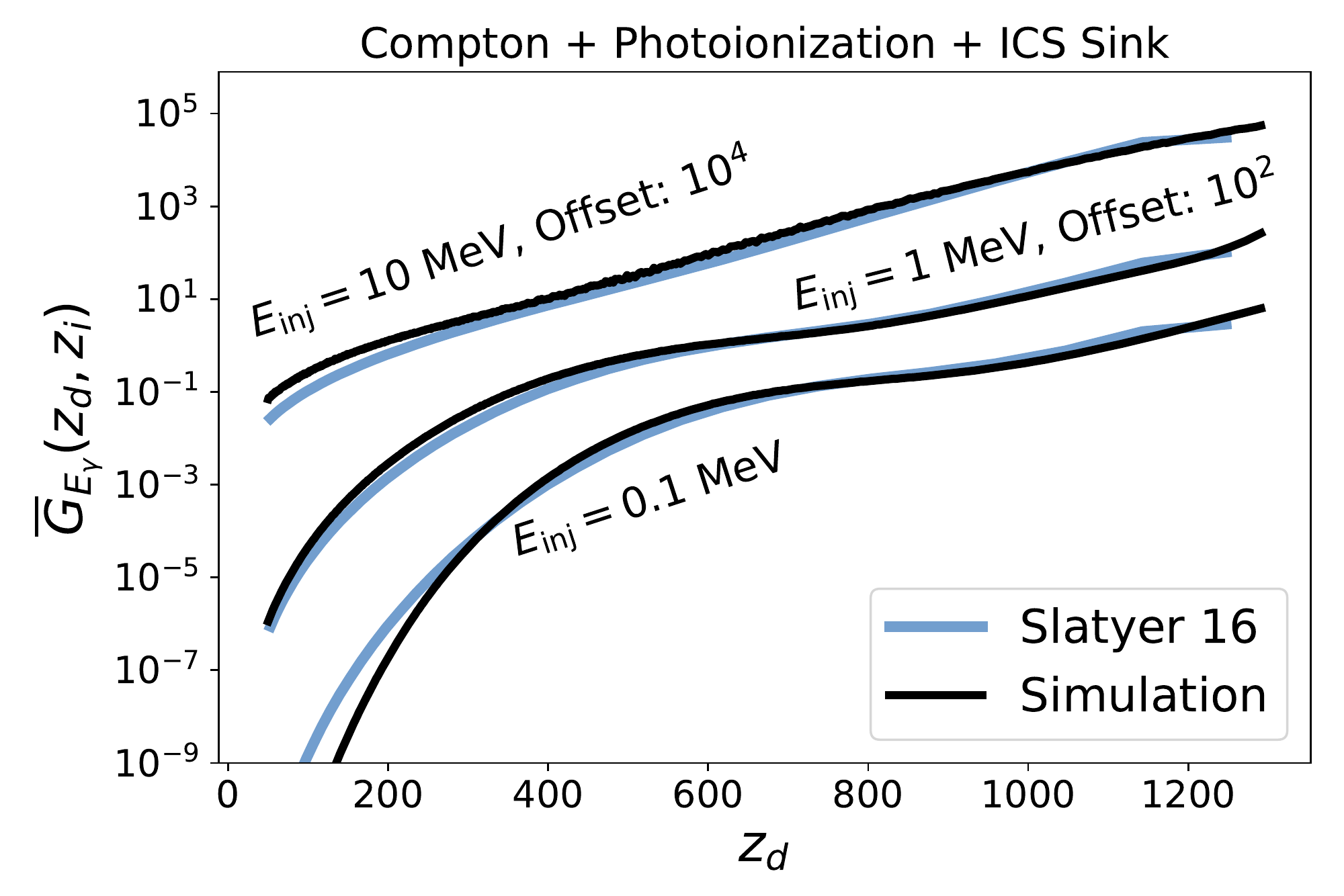}
\caption{\label{fig:slat_sim} Comparison of the spatially averaged Green's function obtained in our simulation (black) with the results from Ref.~\cite{slatyer16a} (blue), including Compton scattering, photoionizations, and secondary electrons' ICS sink. The labeled curves correspond to $E_{\rm inj}=0.1$, 1, and 10 MeV at $z_{i}=1300$, with multiplicative offsets for clarity. The agreement of these curves gives us confidence in the robustness of our simulation.}
\end{figure}

Second, we compare our full-fledged simulations (including photoionizations and the ICS sink, i.e.~$F_{\rm dep}(E) < 1$) to the results of Ref.~\cite{slatyer16a}, where the homogeneous Green's function is computed, accounting for all photon processes at all energy scales. Note that our simulation has a higher resolution in deposition time than that of Ref.~\cite{slatyer16a}. At injected photon energies $E_\gamma \lesssim 10$ MeV, the results of Ref.~\cite{slatyer16a} should match our simulation. This is indeed the case, as shown in Fig.~\ref{fig:slat_sim} for various injection energies. The factor-few disagreement at low redshifts has no observable consequence as it only affects the Green's function in the regime where it is exponentially suppressed.

\subsubsection{Spatial part of the Green's function}\label{subsec:spatial}

\begin{figure*}[hbt!]
\centering
\includegraphics[width=\columnwidth]{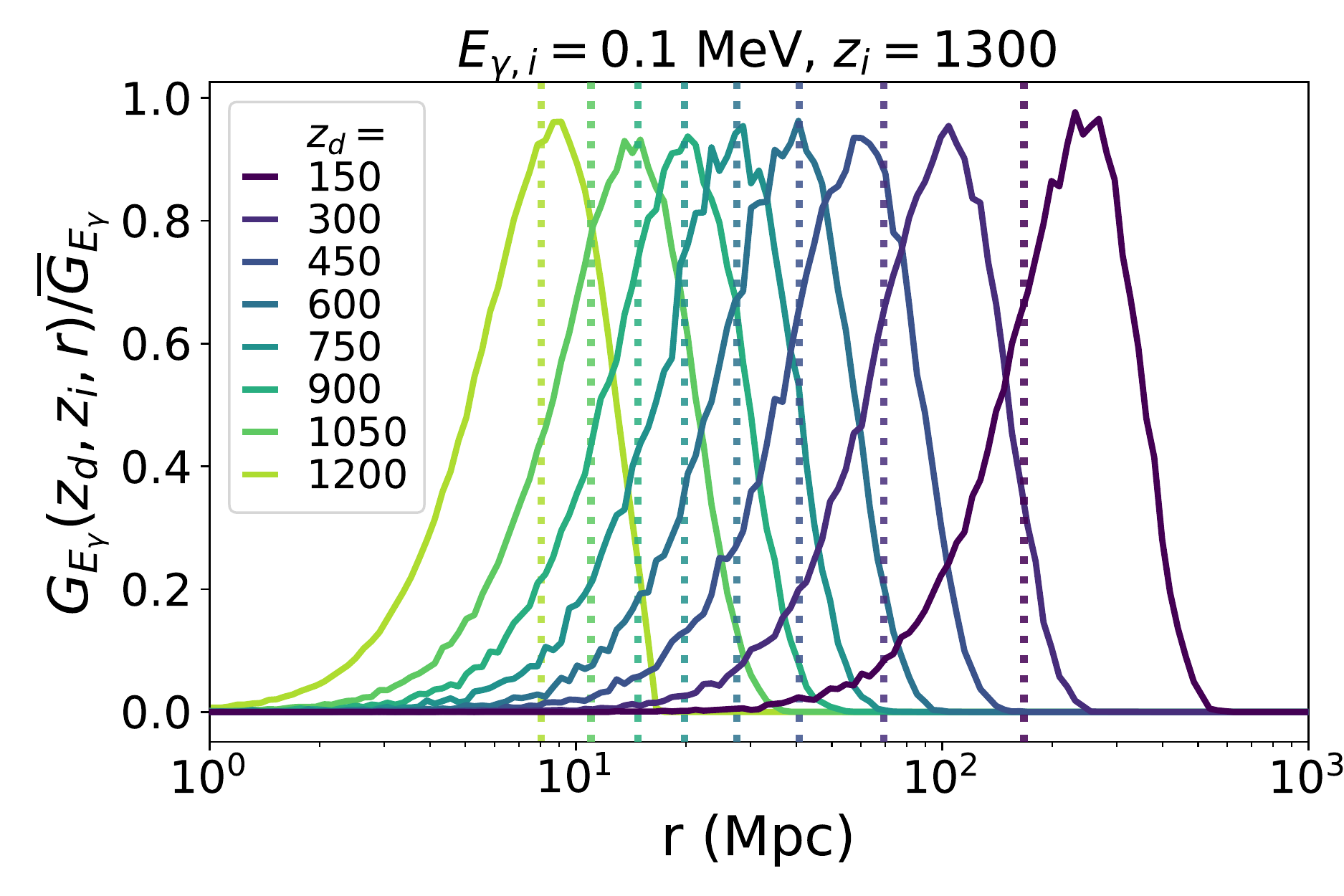}
\includegraphics[width=\columnwidth]{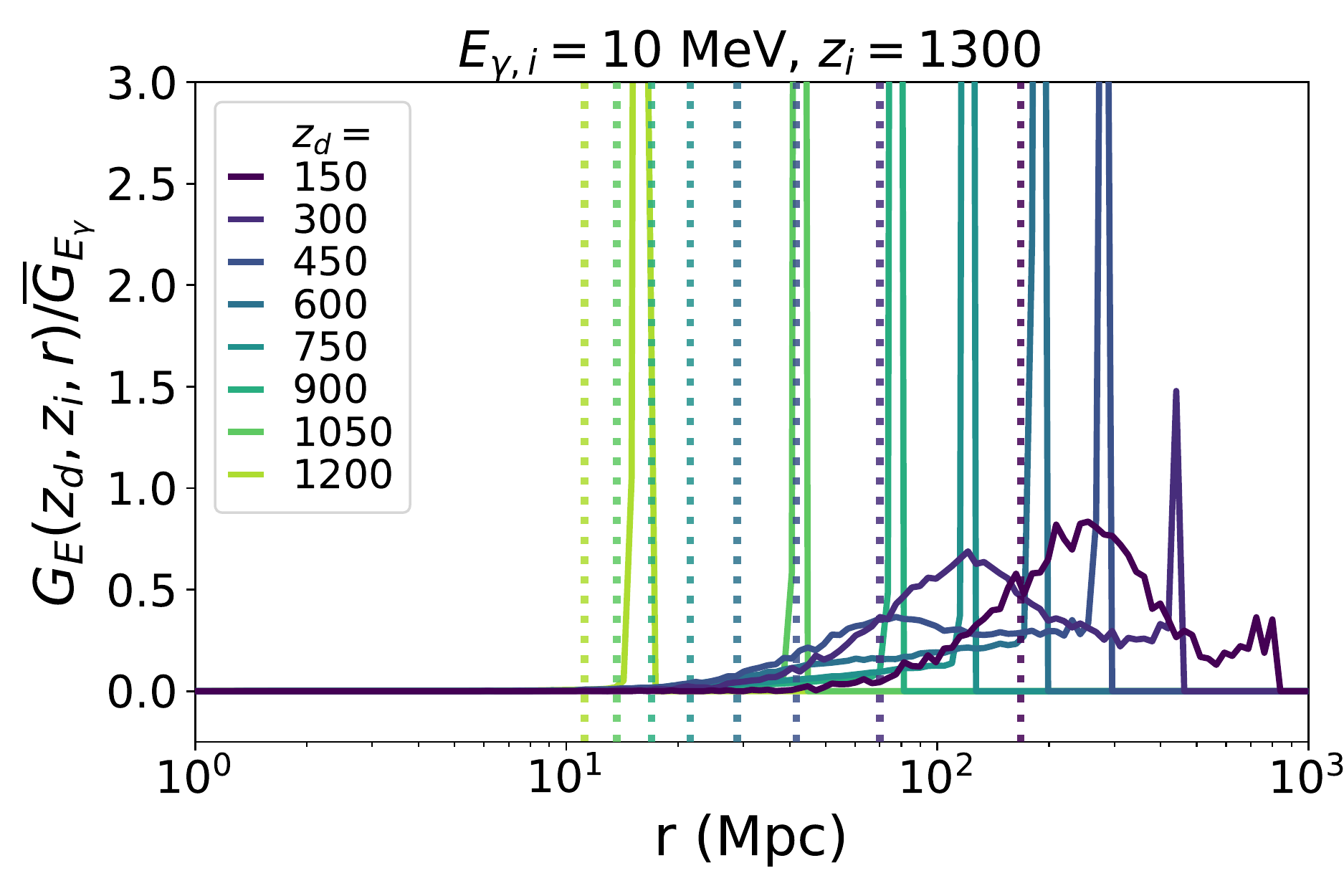}
\caption{\label{fig:z1200spac}Injection-to-deposition Green's function normalized to its spatial average, $G_{E_\gamma}(z_d,z_i,r)/\overline{G}_{E_\gamma}(z_d, z_i)$, for injection redshift $z_i=1300$, and a Dirac spectrum of injected photons, at $E_\gamma=0.1$ MeV (left) and 10 MeV (right). Solid lines are outputs from our full simulation (including photoionization, Compton scattering, and ICS sink), and the vertical dotted lines show the Compton diffusion scale $\lambda_C$ defined in Eq.~\eqref{eq:mfp}. At low energies the spatial distribution has a near-Gaussian shape with a peak near $\lambda_C$. At high energies, the distribution shows a feature near the light horizon radius, in addition to a near-Gaussian shape at low redshifts; see main text for a qualitative explanations of these features. }
\end{figure*}

We now discuss the spatial distribution of energy deposition, for which no other numerical code currently accounts. In Fig.~\ref{fig:z1200spac}, we show $G_{E_\gamma}(z_{d},z_{i}, r)/\overline{G}_{E_\gamma}(z_{d}, z_{i})$ for Dirac-delta photon spectra at energies $E_\gamma = 0.1$ MeV and $10$ MeV injected at $z_{i} = 1300$, including all physical processes (i.e.~Compton scattering, photoionization, and ICS sink). The shapes of the Green's function are significantly different for the low- and high-energy cases: for the former, the Green's function has a broad, Gaussian-like shape, while for the latter, it is initially concentrated at a narrow ring, and eventually develops an additional broad feature. These features can be understood qualitatively and semi-quantitatively, as we show below. 

As long as the Compton scattering timescale is short relative to the Hubble time, and that photon energies are within or near the Thomson regime such that scattering is approximately forward-backward symmetric, photons undergo a random walk. The spatial diffusion scale is then qualitatively similar to the Silk damping scale, except for the fact that at the relevant energies, photons may Compton scatter with both free and bound electrons (this point was missed in the Appendix of Ref.~\cite{dvorkin13a}). Explicitly, we may estimate the Compton diffusion scale as 
\begin{align}\label{eq:mfp}
    \lambda_{\rm C}^{2}(t)\equiv \int_{t_i}^t\frac{\td t'}{a'^2 n_{\rm H}' \sigma_{\rm C}(E_{\rm trj}(a'; E_ \gamma, a_i))},
\end{align}
where the Compton scattering cross section $\sigma_{\rm C}$ is evaluated along the energy trajectory $E_{\rm trj}(a; E_\gamma, a_i)$ defined in the previous section, and again $n_{\rm H}$ is the total abundance of neutral and ionized hydrogen. \rev{If the number of scatterings during a Hubble time is large, and if photon propagation directions are uncorrelated between two scatterings (as is the case in the Thomson limit, with forward-backward symmetry), from the central-limit theorem we expect the photon spatial distribution -- and thus the spatial distribution of energy deposition -- to be a Gaussian with vanishing mean and variance of order $\lambda_{\rm C}^2$.} We overlay the Compton diffusion scale $\lambda_{\rm C}$ on top of the numerically evaluated Green's function in Fig.~\ref{fig:z1200spac}. We see that for injection energy $E_\gamma=0.1$ MeV, the numerical Green's function has an approximately Gaussian shape, whose peak is within a factor $\sim 2$ from $\lambda_C$.

When Compton scattering events are rare, or when they are preferentially forward, as in the case in the limit $E_\gamma \gg m_e$, we do not expect the Green's function's spatial dependence to be Gaussian. Instead, in this regime we expect photons to be mostly located on the light horizon of the injection point, at $r = \int_{t_i}^t dt'/a'$, either because they propagate mostly freely, or because they effectively propagate along straight lines due to preferentially forward scattering. This is confirmed in the right panel of Fig.~\ref{fig:z1200spac}, where the Green's function exhibits a sharp light horizon feature at early times. \rev{We have checked explicitly that these narrow features are robust and independent of the radial and temporal bin resolution, simulation timestep and photon number.} At late times, the 10 MeV-photon Green's function shows a bimodal spatial distribution, consisting of a sharp light cone feature, and a broad Compton diffusion bump. These two sub-distributions correspond to the high-energy photons that have scattered less than a few times and/or preferentially forward, and to those that have lost a significant part of their energy and reached the Thomson regime, respectively. This is a spatial manifestation of the bimodal photon energy distribution shown in Fig.~\ref{fig:Espec}. As time progresses, more and more photons reach the Thomson regime, and the Green's function approaches the Gaussian shape with scale near $\lambda_C$.

\begin{figure*}[hbt!]
\centering
\includegraphics[width=\columnwidth]{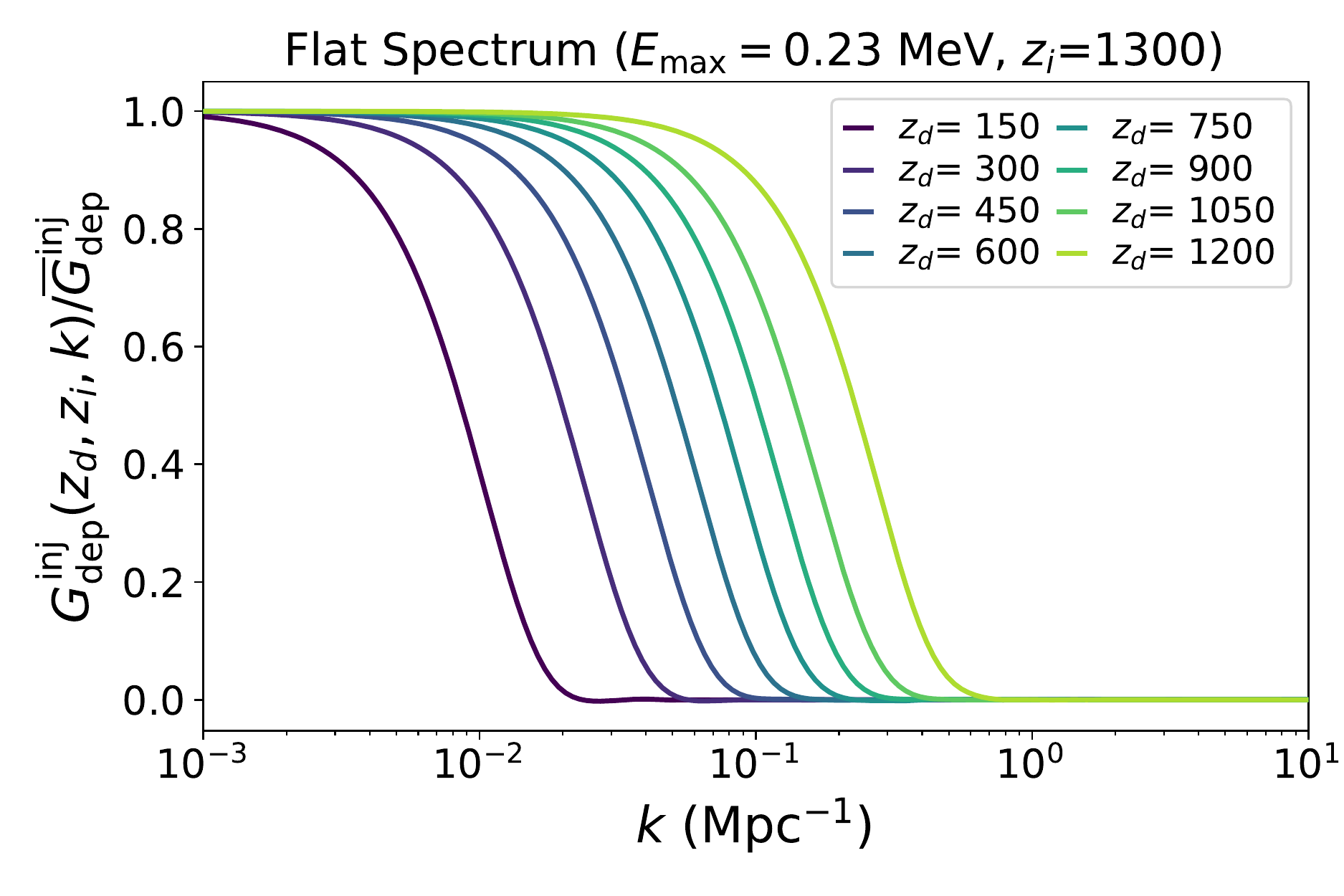}
\includegraphics[width=\columnwidth]{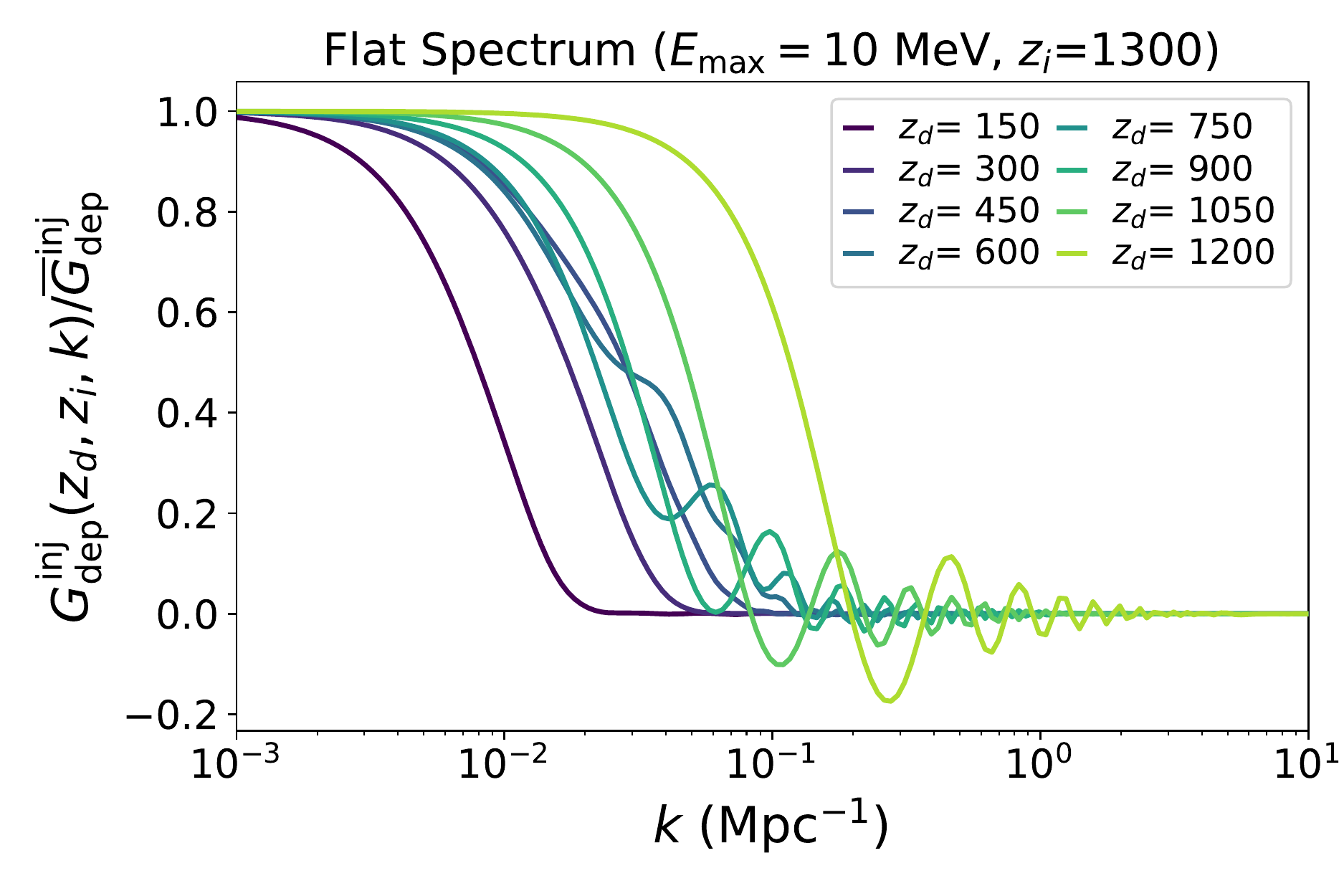}
\caption{Normalized Fourier transform of the injection-to-deposition Green's function $G_{\rm dep}^{\rm inj}(z_d, z_i, k)/\overline{G}_{\rm dep}^{\rm inj}(z_d, z_i)$, for a flat spectrum of injected photons $\Psi(E_\gamma) = \Theta(E_{\max} - E_\gamma)/E_{\max}$, with cutoff $E_{\rm max}=0.23$ MeV (left) and $E_{\rm max}=10$ MeV (right). These spectra correspond to two limiting assumptions for energy injected by accreted PBHs, discussed in Sec.~\ref{sec:PBHs}. At low injected energies, the Green's function and its Fourier transform are approximately Gaussian, but at high injected energies, the sharp light cone feature in the real-space Green's function results in ringing in its Fourier transform. These Green's functions were extracted from our full simulation, including photoionization, Compton scattering, and the ICS sink. }\label{fig:Gk_1300} 
\end{figure*}

As an application of our code beyond a Dirac spectrum, we consider injected photon spectra that are flat up to some cutoff energy, $\Psi(E_\gamma) = \Theta(E_{\max} - E_\gamma)/E_{\max}$, where $\Theta$ is the Heaviside step function. We tabulated the spectrum-averaged Green's function for two values of the cutoff energy $E_{\max} = 0.2$ MeV and $E_{\max} = 10$ MeV, including all physical processes. These flat injected spectra and upper energy cutoffs are relevant to PBH accretion considered in Sec.~\ref{sec:PBHs}. In practice, due to the diverging photon number distribution $dN/dE_{\gamma} \propto 1/E_\gamma$ in this case, we have to impose a numerical lower cutoff $E_{\min}$ in the injected photon energies. We set $E_{\rm min}=0.02$ MeV, and checked explicitly that setting $E_{\rm min}=0.002$ MeV instead leads to no noticeable differences in results for the $E_{\max} = 0.2$ MeV case. We emphasize that this numerical cutoff is only applied to the \emph{injected} spectrum, but that our code follows photons down to arbitrarily low energies. In Fig.~\ref{fig:Gk_1300}, we show the Fourier transforms of the corresponding Green's functions, for $z_i = 1300$. Their qualitative features can be understood in light of the real-space Green's functions for Dirac-function spectra discussed above. At low energies (such that photon energies are always in the Thomson regime), the real-space Green's function resembles a Gaussian, and so does its Fourier transform. At high energies (such that photon energies are in the Compton regime initially), $G_{\rm dep}^{\rm inj}(a_d, a_i, k)$ shows ringing features; these are due to the sharp light-cone feature in the real-space Green's function.

\section{From energy injection to delayed recombination}\label{sec:xe}

\subsection{Definitions: deposition-to-ionization and injection-to-ionization Green's functions}

In this section we study the change to the free-electron fraction $x_e(z)$ for a given rate of energy injection, thus deposition, into the plasma. We make two simplifying approximations to keep the problem tractable. First, we assume that the effect on the ionization history only depends on the \emph{local} (in space) energy deposition rate. This ought to be an excellent approximation at the scales of interest $k \ll 10^3$ Mpc$^{-1}$, much larger than the scales at which Lyman-continuum and Lyman-$\alpha$ transport is relevant \cite{Venumadhav_15}. Second, we assume that perturbations to the standard ionization history are small, i.e.~that $\Delta x_e \ll x_e^0, 1-x_e^0$, where $x_e^0$ is the standard ionization fraction. This is motivated by CMB anisotropy constraints on changes to recombination history near the peak of the visibility function \cite{planck20b}. It is also consistent with our assumption that $x_e = x_e^0$ when computing the energy deposition efficiency. These assumptions allow us to define a purely temporal deposition-to-ionization Green's function $G_{x_e}^{\rm dep}(a, a_d)$, such that
\beq
    \Delta{x_e}(a,{\bm r})\approx \int\td\ln a_{d}~ G^{\rm dep}_{x_e}(a,a_{d})\frac{{\epsilon}_{\rm dep}(a_{d},{\bm r})}{E_I}.\label{eq:Gxe}
\eeq
In this definition, we have normalized $\epsilon_{\rm dep}$ to the ionization energy of hydrogen, $E_I = 13.6$ eV, so that ${{\epsilon}}_{\rm dep}/E_I$ represents the effective number of ionizing photons deposited per baryon, per Hubble time. With this convention, we expect $G^{\rm dep}_{x_e}$ to be of order unity, and the linear approximation to hold as long as $\epsilon_{\rm dep}/E_I \ll 1$.

Combining with Eq.~\eqref{eq:GPsi}, we then obtain (in the case of a homogeneous injection spectrum $\Psi$),
\barr
\Delta x_e(a, \bs{r}') &=& \iint d \ln a_i ~\frac{d^3 r}{4 \pi r^3} \nonumber\\
&&\times  G_{x_e}^{\rm inj}(a, a_i, r|\Psi) \frac{\epsilon_{\rm inj}(a_i, \bs{r}' + \bs{r})}{E_I},\label{eq:Gxe-inj}
\earr
where the injection-to-ionization Green's function $G_{x_e}^{\rm inj}$ is obtained from the temporal convolution of the injection-to-deposition and deposition-to-ionization Green's functions:
\beq
    G^{\rm inj}_{x_e}(a,a_{i},r|\Psi)\equiv \int_{a_{i}}^{a}\td \ln a_{d}~ G^{\rm dep}_{x_e}(a, a_{d})G^{\rm inj}_{\rm dep}( a_{d},a_{i},r|\Psi).\label{eq:W}
\eeq
As in the case of the injection-to-deposition Green's function, we may define the spatially averaged injection-to-ionization Green's function $\overline{G}_{x_e}^{\rm inj}(a, a_i|\Psi)$, as well as its Fourier transform $G_{x_e}^{\rm inj}(a, a_i, k|\Psi)$, both defined as in Eq.~\eqref{eq:W} above, with $G^{\rm inj}_{\rm dep}( a_{d},a_{i},r|\Psi)$ replaced by $\overline{G}^{\rm inj}_{\rm dep}(a_d, a_i|\Psi)$ and $G^{\rm inj}_{\rm dep}( a_{d},a_{i},k|\Psi)$, respectively. With these Green's functions, we obtain the homogeneous part of the perturbation to the free-electron fraction, as well as its Fourier components, as follows: \barr
\overline{\Delta x_e}(a) &=& \int d\ln a_i ~ \overline{G}_{x_e}^{\rm inj}(a, a_i|\Psi) \frac{\overline{\epsilon}_{\rm inj}(a_i)}{E_I},\\
\Delta x_e(a, \bs{k}) &=& \int d\ln a_i ~ G_{x_e}^{\rm inj}(a, a_i, k|\Psi) \frac{\epsilon_{\rm inj}(a_i, \bs{k})}{E_I}. \label{eq:Delta-xe-k}
\earr
We now turn to the numerical evaluation of $G_{x_e}^{\rm dep}$.

\subsection{Calculation of the ionization Green's functions}

\rev{We now describe how we compute the ionization Green's functions. For simplicity, we focus on hydrogen recombination, and neglect the effect of energy deposition on helium recombination (but note that we do account for helium photoionization as a means of converting injected photons into energetic electrons, as described in Sec.~\ref{sec:dep}). This neglect is justified as follows. First, during helium recombination at $z \gtrsim 1800$, the matter temperature is effectively locked to the radiation temperature by Compton scattering, and therefore all the energy deposited in the form of heat has virtually no impact on the thermal history. Second, and most importantly, CMB anisotropies have very little sensitivity to changes to the ionization history at these high redshifts. With that being said, helium atoms can also get ionized at $z \sim 1000$, i.e.~during the epoch of hydrogen recombination. Given the small helium-to-hydrogen number ratio $n_{\rm He}/n_{\rm H} \approx 0.08$, we expect that accounting for this effect would lead to order $\mathcal{O}(10\%)$ corrections to our results, as found in Ref.~\cite{Kawasaki_21} in the context of dark matter annihilation. Our results should therefore be accurate at the $\mathcal{O}(10\%)$ level.}

Computing hydrogen recombination history with sub-percent accuracy requires solving ordinary differential equations (ODEs) for the free-electron fraction $x_e$ and baryon temperature $T_b$, coupled with a partial differential radiative transfer equation for the photon distribution near the Lyman resonances \cite{YAH_10b}, accounting for two-photon emission and absorption \cite{Chluba_05, Hirata_08} and resonant scattering \cite{Chluba_09, Hirata_09}. These equations are solved numerically by the state-of-the-art codes \texttt{HyRec} \cite{YAH_10, yacine11a} and \texttt{CosmoRec} \cite{Chluba_11}. In order to compute the deposition-to-ionization Green's function with these codes, one could in principle include energy deposition source terms with narrow redshift support, approximating Dirac functions. Instead, we use the simpler system of two coupled ODEs solved in \texttt{HyRec-2} \cite{hyrec2}, which allow us to compute the Green's function more robustly. These ODEs are based on the effective 4-level atom model for hydrogen \cite{YAH_10}, with correction functions accounting for detailed radiative transfer, calibrated with \texttt{HyRec}.

The ODEs solved by \texttt{HyRec-2} take the form
\barr
    \dot{x}_e&=&\mathcal{F}(a,x_e,T_b) + S_{x_e}, \label{eq:xedot}\\
    \dot{T}_b &=& -2HT_b+ \Gamma_{\rm C}(x_e) \times (T_\gamma - T_b) + S_{T_b}, \label{eq:Tmdot} 
\earr
where $\mathcal{F}(a, x_e, T_b)$ is the standard rate of change of the free-electron fraction provided explicitly in Eq.~(6) of Ref.~\cite{hyrec2}, $\Gamma_{\rm C}(x_e)$ is the Compton heating rate, which depends on $x_e$ (but not on $T_b$), and can be found, e.g.~in Eq.~(12) of Ref.~\cite{yacine11a}, and $T_\gamma$ is the radiation temperature. The last terms in Eqs.~\eqref{eq:xedot} and \eqref{eq:Tmdot} are source terms due to nonstandard energy deposition in the plasma. Making the approximation that the deposited energy goes into ionization, excitation and heating with fractions $(1-x_e)/3,(1-x_e)/3, (1 + 2x_e)/3$, respectively \cite{chen04a}, and generalizing the results of Ref.~\cite{Giesen_12}, the source terms are
\barr    
S_{x_e}&\equiv& \frac{1-x_e}{3}\left(1+\frac{4}{3}(1-\overline{C})\right)\frac{\dot{\rho}_{\rm dep}}{E_I n_{\rm H}}, \label{eq:Sxe}\\
    S_{T_b}&\equiv& \frac{2}{3 n_{\rm tot}}\frac{1+2 x_e}{3}\dot{\rho}_{\rm dep},\\
    1-\overline{C}&\equiv& \frac{1}{4}(1-C_{2s})+\frac{3}{4}(1-C_{2p}),
\earr
where $n_{\rm tot}$ is the total number density of baryons (electrons, nuclei and atoms), and the coefficient $C_{2s}$ ($C_{2p}$) is the effective probability that an atom in excited state $2s$ ($2p$) reaches the ground state rather being photoionized, generalizing Peebles' $C$ factor \cite{Peebles_68}, and is provided in Eq.~(7) of Ref.~\cite{hyrec2}. The first term in the parenthesis of Eq.~\eqref{eq:Sxe} corresponds to direct ionizations, and the second term corresponds to excitations followed by ionizations.

We now assume that the source terms lead to small perturbations $\Delta x_e, \Delta T_b$ to the ionization fraction and baryon temperature, and linearize Eqs.~\eqref{eq:xedot}, \eqref{eq:Tmdot}:  
\barr
    \Delta\dot{x}_e&=&\p{\mathcal{F}}{x_e}\Delta x_e+\p{\mathcal{F}}{T_b}\Delta T_b+S_{x_e}\label{eq:Dxedot}\\
    \Delta\dot{T}_b&=& - (2 H + \Gamma_C) \Delta T_b + \frac{d \Gamma_C}{d x_e} \Delta x_e +S_{T_b}, \label{eq:DTmdot}
\earr
where the derivatives of $\mathcal{F}$ and $\Gamma_C$, as well as the source terms, are evaluated along the standard ionization and thermal history.

The deposition-to-ionization Green's function is extracted by setting $\dot{\rho}_{\rm dep}/(E_I n_{\rm H}) = H {\epsilon}_{\rm dep}/E_I= H \delta(\ln a - \ln a_d) = \delta(t-t_d)$. Explicitly, for a given deposition scale factor $a_d$, we solve Eqs.~\eqref{eq:Dxedot}-\eqref{eq:DTmdot}, with $S_{x_e}$ and $S_{T_b}$ set to zero, and starting with initial conditions at $a = a_d$
\barr
\Delta x_e(a_d) &=& \frac{1-x_e}{3}\left(1+\frac{4}{3}(1-\overline{C})\right), \label{eq:Dxe-init}\\
\Delta T_b(a_d) &=& \frac{2 n_{\rm H}}{3 n_{\rm tot}} \frac{1 + 2 x_e}{3} E_I,\label{eq:DTm-init}
\earr
with both right-hand sides to be evaluated at $a = a_d$.

\begin{figure}[ht]
    \centering
    \includegraphics[width = \columnwidth]{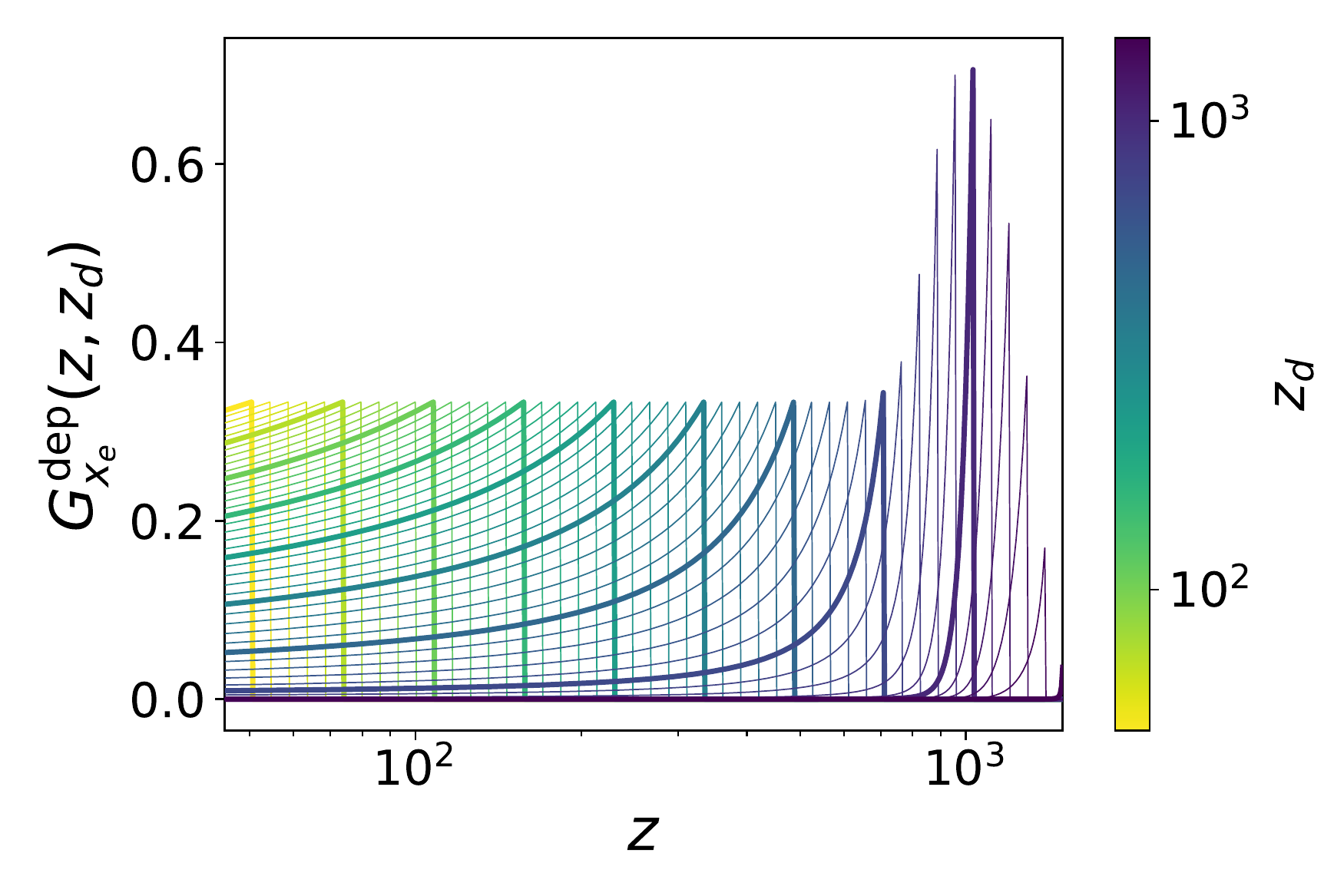}
    \caption{Dimensionless deposition-to-ionization Green's function $G_{x_e}^{\rm dep}(z, z_d)$, as a function of the ionization redshift $z$, for various energy deposition redshifts $z_d$. Different line thicknesses are used only for the purpose of better visualization.}
    \label{fig:gxe}
\end{figure}

We show the deposition-to-ionization Greens function $G_{x_e}^{\rm dep}(a, a_d)$ in Fig.~\ref{fig:gxe}. The envelope of this function follows the initial condition \eqref{eq:Dxe-init}. Much before recombination, when the plasma is fully ionized ($x_e \rightarrow 1$), the impact of energy deposition on the ionization history is suppressed. Once the plasma becomes significantly neutral, the envelope is approximately $(1 + 4/3 (1 - \overline{C}))/3$; the effective Peebles $C$-factor $\overline{C}$ is initially small, and reaches unity for $z \lesssim 800$ (see e.g.~Ref.~\cite{YAH_thesis}), which translates to the bump in the Green's function envelope around $z \sim 1000$, and the 1/3 plateau at $z \lesssim 800$. Note that, although we simultaneously solve for the temperature evolution for completeness, at $z \gtrsim 200$, $\Gamma_C \gg H$, implying an exponential damping of temperature perturbations. The evolution of the ionization Green's function is thus mostly controlled by the term proportional to $\p{\mathcal{F}}{x_e}$. As time progresses, $\p{\mathcal{F}}{x_e}/H$ decreases (the recombination process slows down), leading to a more extended tail for the Green's function at low redshifts.

\begin{figure*}[tbp]
    \centering
    \includegraphics[width = \columnwidth]{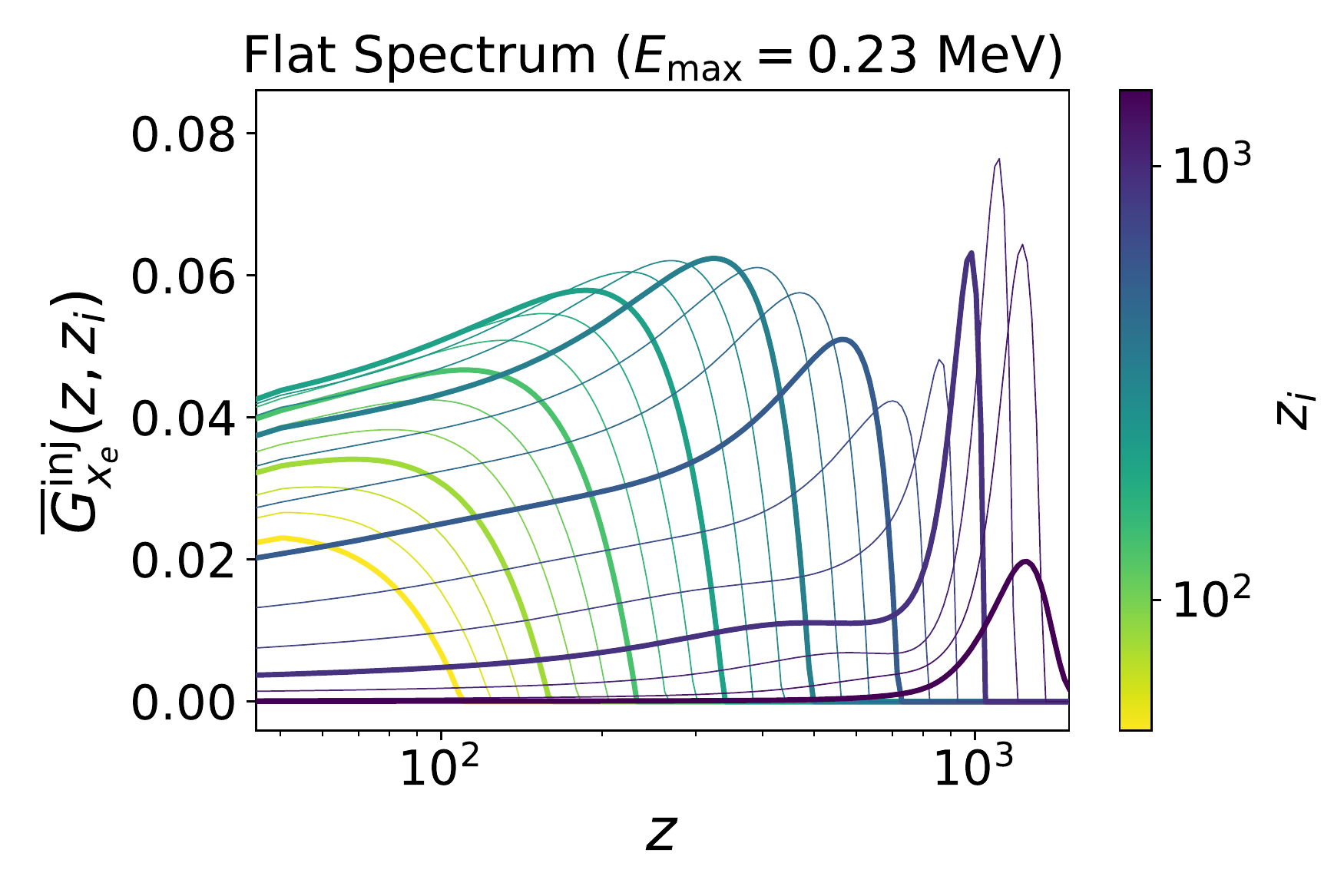}
    \includegraphics[width=\columnwidth]{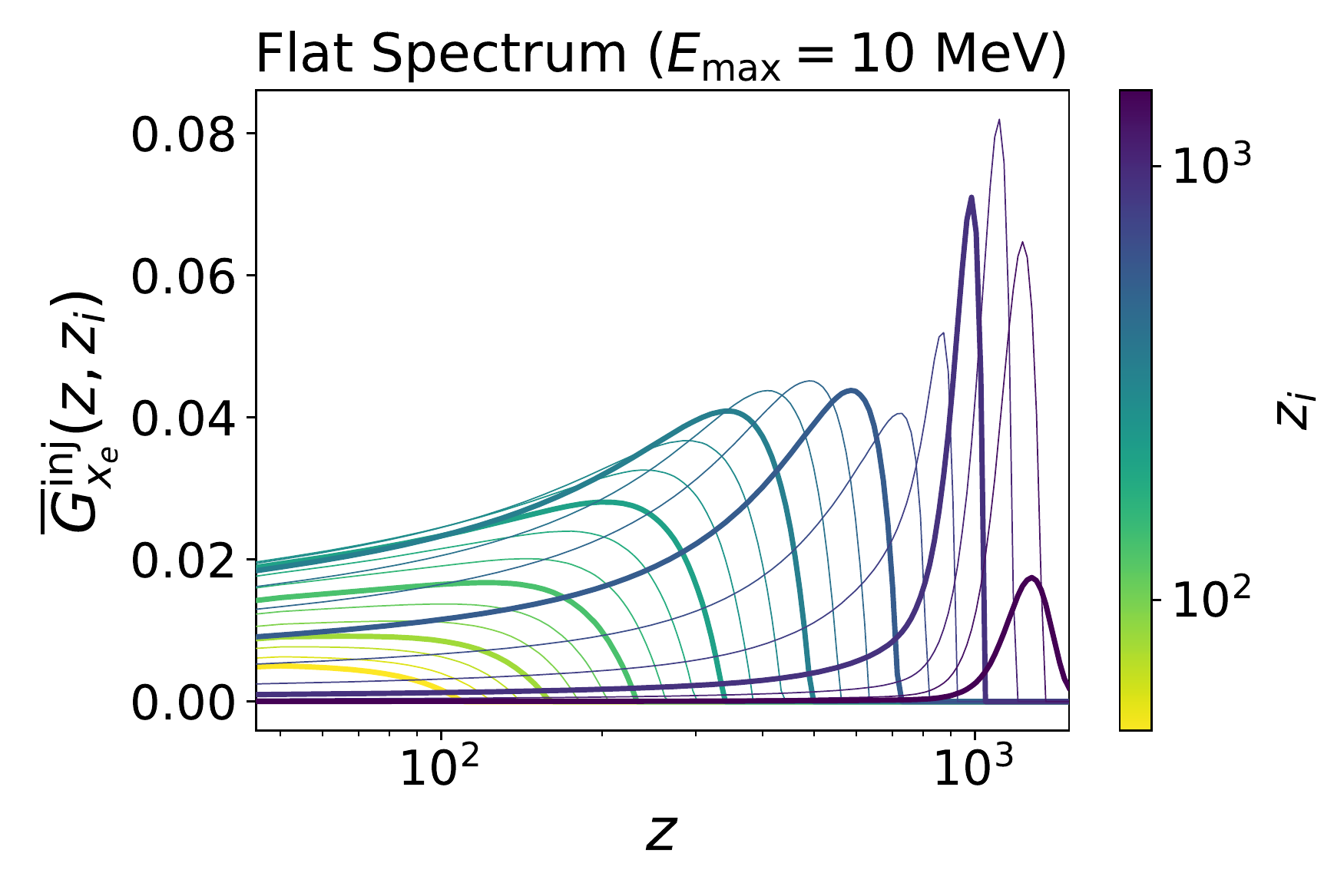}
    \caption{Homogeneous injection-to-ionization Green's function (defined in Eq.~\eqref{eq:W}) for a flat photon spectrum with cutoff $E_{\rm max}=0.23$ MeV (left) and  $E_{\rm max}=10$ MeV (right), as a function of ionization redshift $z$. Different line colors correspond to different injection redshifts $z_i$; note that we vary the thickness of lines only for clarity.}
    \label{fig:Gxe-inj-mean}
\end{figure*}

\begin{figure*}[btp]
\centering
\includegraphics[width=\columnwidth]{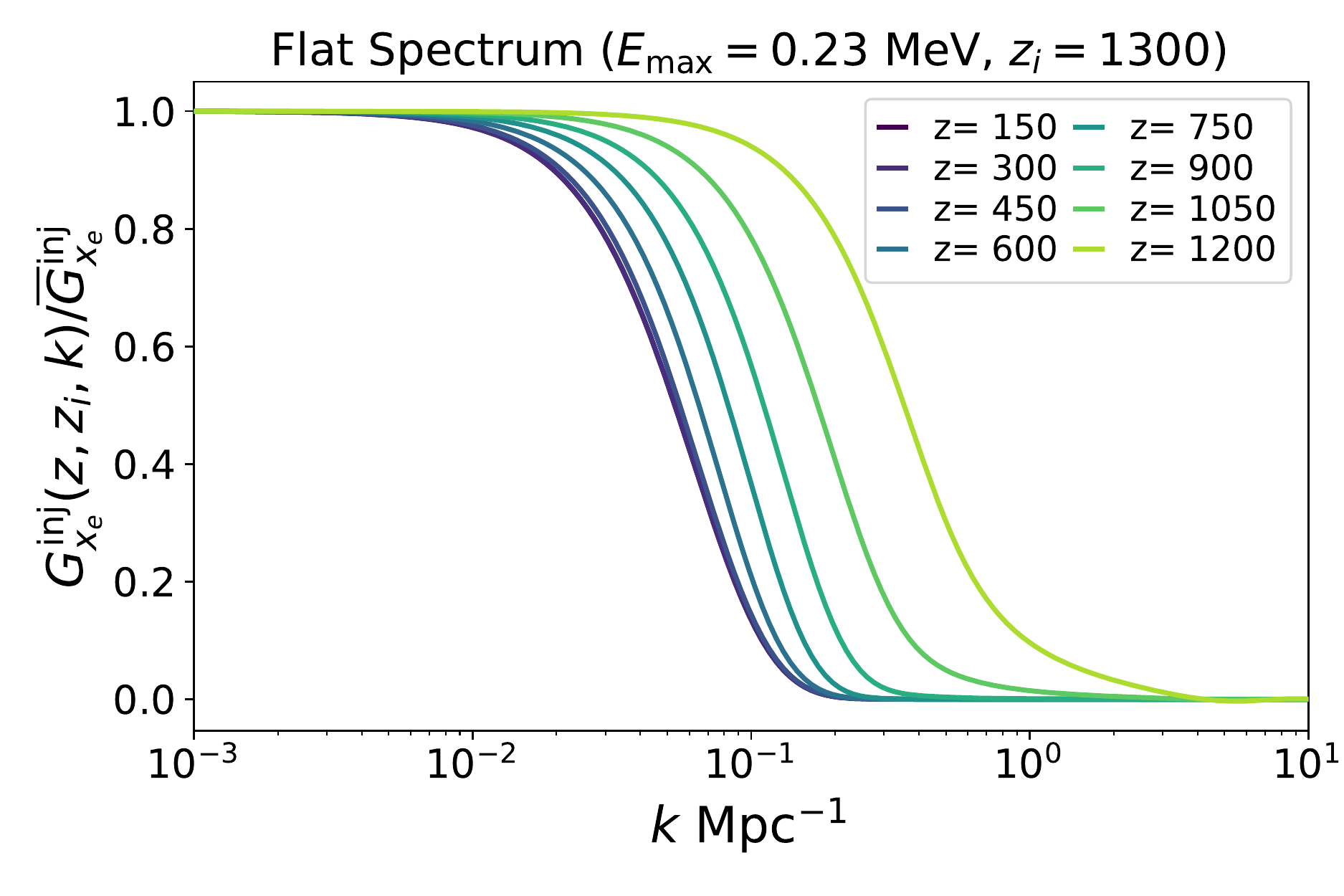}
\includegraphics[width=\columnwidth]{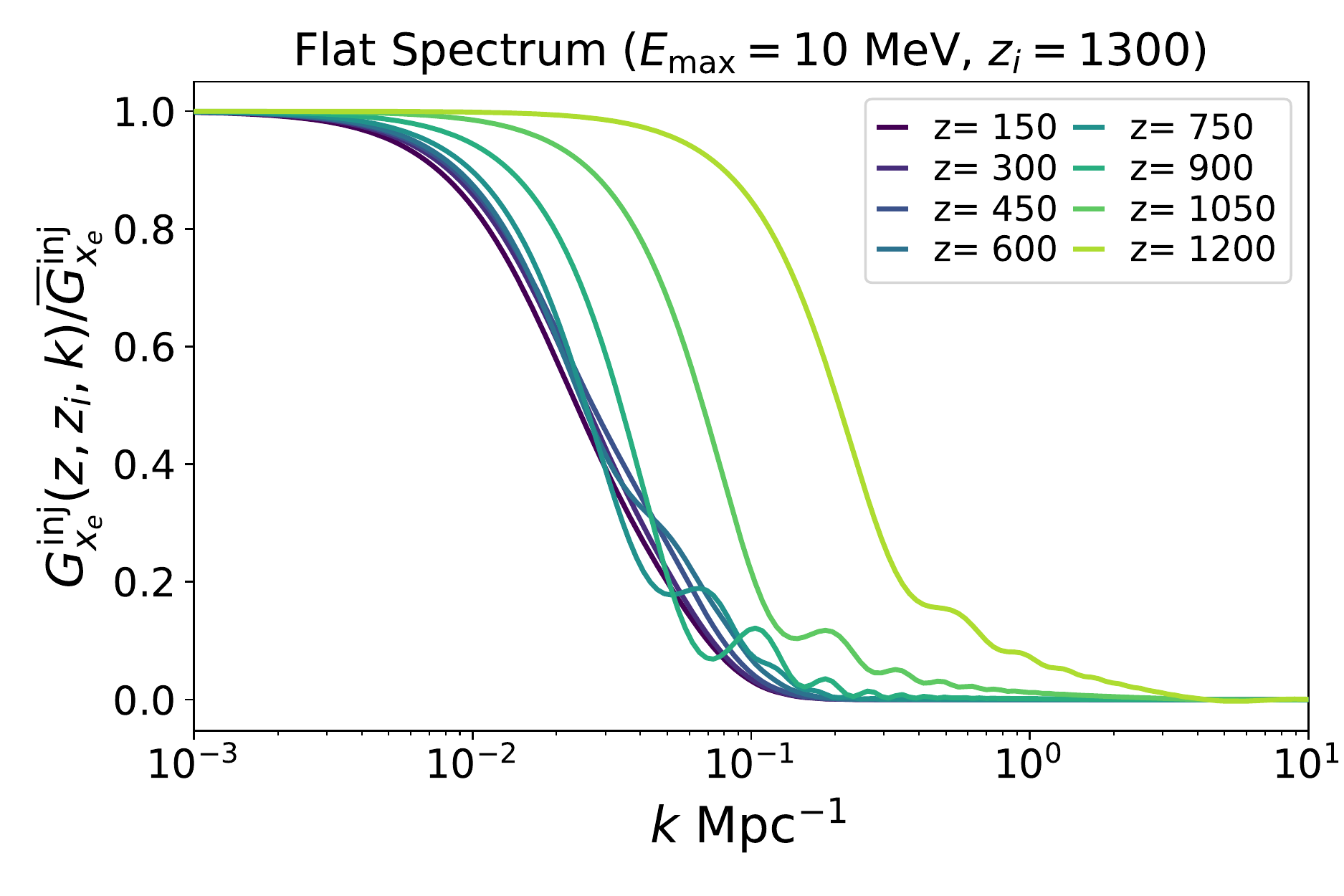}
\caption{\label{fig:W_100} Normalized Fourier transform of the injection-to-ionization Green's function $G_{x_e}^{\rm inj}(z,z_i,k)/\overline{G}_{x_e}^{\rm inj}(z,z_i)$, for a flat energy spectrum injected at $z_i=1300$, as a function of wavenumber. The left panel is with a high-energy cutoff of $E_{\rm max}=0.23$ MeV, and the right is for $E_{\rm max}=10$ MeV, showing a more pronounced small-scale suppression due to the propagation of higher-energy photons, as well as ringing features due to the light-cone feature in the injection-to-deposition Green's function.}
\end{figure*}

Given $G_{x_e}^{\rm dep}$ and $G_{\rm dep}^{\rm inj}$ for a given injection spectrum $\Psi$, we convolve them to obtain the injection-to-deposition Green's function $G_{x_e}^{\rm inj}$ defined in Eq.~\eqref{eq:W}. In Fig.~\ref{fig:Gxe-inj-mean}, we show the spatially averaged Green's function, for a flat injected spectrum, up to cutoff energies $E_{\max} = 0.2$ MeV and 10 MeV, respectively. For the latter energy cutoff, we see that, after recombination, changes in the free electron fraction are suppressed in comparison with the lower energy cutoff photon injection spectrum. This is likely due to a combination of two effects: one, the higher-energy photons can result in secondary electrons ($E_e\sim $ MeV) that are mostly inefficient in depositing energy into the plasma (c.f. Fig.\ref{fig:rat}); and two, higher-energy photons have a lower probability of scattering and thus disperse their energy deposition over later redshifts. In Fig.~\ref{fig:W_100}, we show the normalized Fourier transform $G_{x_e}^{\rm inj}(z, z_i = 1300, k)$, as a function of wavenumber and ionization redshift, for each of the two injected spectra. In both cases, fluctuations in the ionization fraction are suppressed for $k \gtrsim$ few times $10^{-1}$ Mpc$^{-1}$, with a suppression lengthscale increasing as time progresses, and typically larger for higher-energy photons, which can propagate over larger distances.

\section{Application to accreting PBHs}\label{sec:PBHs}

\subsection{PBH accretion and radiation model}\label{subsec:accmod}

We now apply the formalism developed above to energy injection by accreting PBHs. Despite the relatively simple physical conditions in the early Universe, the problem of accretion on PBHs remains complex, and existing estimates of their accretion rate and luminosity vary by orders of magnitude and are highly uncertain \cite{ricotti08a, poulin17a, yacine17a}. One of the major uncertainties is the geometry of the accretion flow: if an accretion disk forms, the radiative efficiency and overall luminosity is expected to be significantly larger than for a quasi-spherical flow \cite{poulin17a}.   
Lacking a definitive proof that an accretion disk must form, in order to make headway while remaining conservative, in this work we adopt the quasi-spherical accretion model of Ref.~\cite{yacine17a} (hereafter AK17). We emphasize that our main point should remain qualitatively valid regardless of the details of the accretion flow: the luminosity of accreting PBHs ought to depend on the local supersonic relative velocities of accreted baryons, leading to a spatially modulated energy injection rate. \rev{Our final results, applied to the spherical accretion model of AK17, should therefore be understood as an example application, and can be generalized to more sophisticated accretion models as they become available.}

\rev{We begin by briefly reviewing the idealized, spherically symmetric accretion model. Consider a stationary, nonrotating black hole with mass $M$ embedded in a homogeneous plasma with an average baryon mass density $\overline{\rho}_b$, temperature $\overline{T}_b$, and mean ionization fraction $\overline{x}_e$ (all defined far from the black hole). The rate of accretion, $\dot{M}$, was computed by Bondi \citep{bondi52a} by solving for the physical solution to the steady-state continuity and Euler equations. Namely, by enforcing the monotonic increase of the fluid's radial velocity approaching the black hole and avoiding singularities in the flow outside the event horizon, the unique solution can be found to be
\beq
\dot{M} = \lambda\times 4\pi \overline{\rho}_b\frac{M^2}{v_{\rm B}^3},
\eeq
where $\lambda $ is a dimensionless parameter and $v_{\rm B} \equiv \sqrt{\overline{P}_b/\overline{\rho}_b}$ is a characteristic velocity, where $\overline{P}_b \equiv \overline{\rho}_b (1 + \overline{x}_e) \overline{T}_b/m_p$ is the average gas pressure. With this definition, $v_{\rm B}$ is equal to the isothermal sound speed (but note that the gas is not isothermal in general).}

\rev{In Bondi's original calculation, the only non-gravitational force was assumed to be pressure, and the gas was assumed to be barotropic. In the cosmological context of interest, accreting PBHs are embedded in an intense photon bath. In light of this, AK17 generalize this accretion model as well as the calculations of Refs.~\cite{ricotti07a, ricotti08a}, accounting for Compton drag and Compton cooling by CMB photons. In practice, AK17 compute $\lambda$ numerically by solving the steady-state fluid equations, and provide an analytic approximation as a function of the ratios of the accretion timescale $t_{\rm B} \equiv M/v_{\rm B}^3$ to the Compton drag and Compton cooling timescales. The parameter $\lambda(v_{\rm B})$ is thus a function of redshift as well as of $v_{\rm B}$. When Compton drag is negligible (typically after recombination), $\lambda$ lies somewhere between the adiabatic and isothermal limits, $\lambda_{\rm ad} \approx 0.12$ and $\lambda_{\rm iso} \approx 1.12$, respectively, depending on the strength of Compton cooling. When the Compton drag timescale is short relative to the accretion timescale (typically before recombination), accretion is suppressed by Compton drag, i.e.~$\lambda \ll 1$.}

\rev{It was shown in Ref.~\cite{ricotti07a} that the steady-state approximation is valid for black hole masses $M \lesssim 3 \times 10^4$ $M_\odot$. Additionally, AK17 found that local thermal feedback (i.e.~Compton heating of the accreting gas by the radiated gamma rays, which we describe shortly) is also negligible in this mass range. We therefore limit ourselves to PBH masses $M \lesssim 10^4 ~M_{\odot}$, for which the simple spherical accretion model is self consistent. }

As the accreted gas falls towards the black hole's horizon, it gets compressed and heated up, and eventually fully ionized. At minimum, we thus expect the accreted plasma to produce free-free radiation. The total free-free luminosity is dominated by emission near the black hole's horizon. It scales as the local gas density squared, thus as the accretion rate squared, and is a function of the gas temperature $T_s$ near the horizon. The temperature $T_s$ determines not only the overall PBH luminosity, but also the radiation spectrum, as the differential free-free spectrum $d L/dE_\gamma$ is approximately constant up to photon energies $E_\gamma = E_{\max} \sim T_s$. In other words, the assumed PBH luminosity per photon energy interval takes the form 
\beq
\frac{d L}{d E_\gamma} \approx \frac{L}{E_{\max}} \Theta(E_{\max} - E_\gamma), \ \ \ \ \ \ \ L \equiv \mathcal{L}(T_s) \dot{M}^2,
\eeq
where $\Theta$ is the Heaviside step function, and $\mathcal{L}(T_s)$ is a known function of temperature \cite{svensson82a}. 

\begin{figure}[ht]
\includegraphics[trim={0 0 0 0},width=\columnwidth]{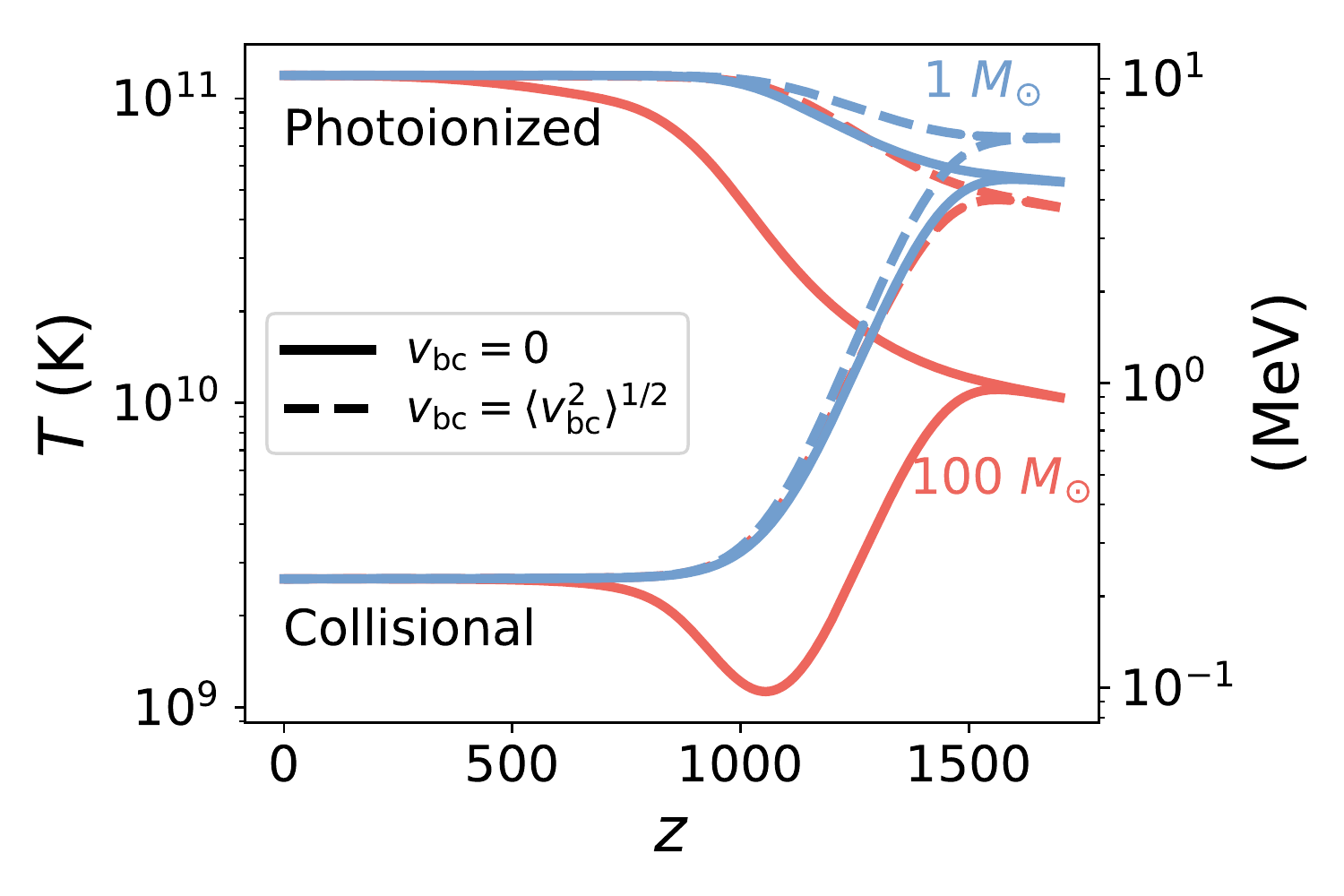}
\caption{\label{fig:T_v_comp} Temperature near the Schwarzschild radius derived in AK17 for black hole masses $M= 1$, 100 $M_\odot$ as a function of redshift, evaluated for two different values of the relative velocity $v_{\rm bc}=0$ (solid) and $v_{\rm bc}=\langle v_{\rm bc}^2\rangle ^{1/2}$ (dashed). We show both accretion scenarios where the accreting plasma is either collisionally ionized or photoionized by the black hole radiation itself. This temperature governs the overall PBH luminosity as well as the energy cutoff of the approximately flat radiated spectrum.}
\end{figure}

AK17 approximately determine the gas temperature $T_s$ near the black hole's horizon in two limiting regimes. In the first case, they assume the accreting gas is \textit{photoionized} by the black hole's radiation; in this case, the gas temperature near the horizon can reach up to $T_s \sim 10^{11}$ K $ \sim 10$ MeV \cite{Shapiro_73} once Compton cooling is negligible. The other limiting regime is that of \emph{collisional ionization} of the accreting gas. This scenario would apply if the radiation produced by the PBH is not strong enough to photoionize the infalling gas. In this case, the gas remains mostly isothermal throughout a collisional ionization region \cite{Shapiro_73}, and can only heat up to $T_s \sim 3 \times 10^9$ K $\sim 0.2$ MeV near the horizon. In both cases, $T_s$ can be significantly lower than the quoted maximum values before recombination, when Compton cooling inhibits significant heating of the infalling gas, as can be seen in Fig.~\ref{fig:T_v_comp}. Importantly, AK17 showed that neither of these two limiting regimes is self-consistent, as in both cases the PBH luminosity is neither large enough to fully photoionize the gas nor small enough to be entirely negligible. The PBH luminosity can vary by up to two orders of magnitude between these two extreme limits, which highlights the broad theoretical uncertainty remaining in this problem. Still, the collisionally ionized case should provide a very conservative, minimum-plausible estimate of the luminosity of accreting PBHs.

Let us remark that, prior to recombination, the temperature $T_s$ depends on the ratio of the Compton cooling and accretion timescales. Just like the accretion rate $\dot{M}$, it is thus a function of redshift and of the characteristic velocity $v_{\rm B}$, and so is the PBH luminosity $L$. 

Given the differential luminosity $dL/dE_\gamma$ of each PBH, one can finally infer the volumetric rate of energy injection by accreting PBHs, per photon energy interval:
\beq
\frac{d\dot{\rho}_{\rm inj}}{d E_\gamma} = f_{\rm pbh} ~\frac{\rho_{\rm cdm}}{M} \frac{d L}{d E_\gamma},
\eeq
where $f_{\rm pbh}$ is the fraction of PBHs that comprise dark matter. This equation holds for a single PBH mass and can be trivially generalized to an extended mass distribution. \par

\subsection{Relative velocity of accreted gas}\label{subsec:relvel}

The simple accretion model described above applies to a spherically symmetric flow. In practice, the accreted gas has significant velocity at infinity in the PBH rest-frame. In this subsection we describe the properties of these velocities, and describe their estimated impact on PBH accretion in Sec.~\ref{subsec:relvel-effect}.
 
It is now well known that, with adiabatic initial conditions, CDM and baryons have supersonic relative velocities $v_{\rm bc}$ around and after recombination \cite{Tseliakhovich_10}. Indeed, before recombination baryons and photons are tightly coupled and undergo acoustic oscillations, while the CDM free-falls in gravitational potentials. These different dynamics result in large-scale relative velocities, fluctuating on $\sim 100$ Mpc scales, and with a rms reaching about five times the baryon sound speed at recombination. After baryons kinematically decouple from photons at $z \sim 1000$, they behave as a cold fluid on scales larger than their Jeans length, and the baryon-CDM relative velocities then decay as $1/a$, as long as baryons and CDM perturbations remain linear. As the gas cools and the sound speed decreases too, relative velocities remain supersonic until baryons heat up again at reionization. 

If PBHs make up a significant part of the CDM, they must have adiabatic initial conditions on large scales in order to satisfy observational constraints \cite{Planck_18_inflation}. Even if they make up a subdominant part of the CDM, if PBHs are produced through the gravitational collapse of rare overdense regions upon horizon entry, their large-scale distribution is expected to follow that of the radiation, in the absence of significant primordial non-Gaussianity \cite{Suyama_19, Young_20}. In other words, in this formation scenario, we also expect PBHs initial density perturbations to be adiabatic on large scales. We thus expect the velocity of baryons relative to PBHs to follow the baryon-CDM relative velocity on large scales, regardless of PBH abundance. 

The picture is notably different on small scales, at which PBH (thus CDM) density perturbations get enhanced by unavoidable Poisson fluctuations \cite{Afshordi_03, Ali-Haimoud_18, Desjacques_18}, leading to early formation of nonlinear structures \cite{Kashlinsky_16, YAH_17}. This results in additional relative velocities on small scales, with magnitude similar to the virial velocities of collapsed halos. In their mixed PBH and particle-like CDM simulations, Ref.~\cite{inman19} found these virial velocities to be smaller than the background baryon sound speed for $z\gtrsim 300$ and smaller than the rms large-scale relative velocity for $z \gtrsim 100$. These results indicate that small-scale nonlinear relative velocities can be neglected in the PBH accretion problem, at least for $z \gtrsim 100$. Note that even if PBHs make a very small fraction of CDM, and Poisson fluctuations do not significantly affect structure formation, nonlinear structures eventually form at $z \lesssim 30$, leading to small-scale relative velocities \cite{poulin17a}. 

Given that CMB anisotropies have little sensitivity to the ionization history at $z \lesssim 100$, in this work, following previous studies, we shall only consider the large-scale, linear component of baryon-CDM relative velocities $v_{\rm bc}$, with a Gaussian distribution entirely characterized by its power spectrum, which can be extracted from linear Boltzmann codes, and with rms $\langle v_{\rm bc}^2\rangle^{1/2} \approx 30$ km/s $\approx 5~v_{\rm B}$ at $z \approx 10^3$ \cite{Tseliakhovich_10}.

\subsection{Effect of relative velocities on PBH luminosity and radiated spectrum}\label{subsec:relvel-effect}

Generalizing Bondi's prescription \cite{bondi52a}, we account approximately for relative velocities by making the following substitution in $\lambda, \dot{M}$ and $T_s$:
\beq
v_{\rm B} \rightarrow \sqrt{v_{\rm B}^2 + v_{\rm bc}^2}. \label{eq:vbc-sub}
\eeq
Modulo the weak dependence of $v_{\rm B}$ in the dimensionless accretion parameter $\lambda(v_{\rm B})$, simulations have shown the functional form of Eq.~\eqref{eq:vbc-sub} to be accurate in accretion rates within several tens of percents \cite{shima85a,ruffert96a,lee14a}.\par
\begin{figure}[ht]
\includegraphics[width=\columnwidth]{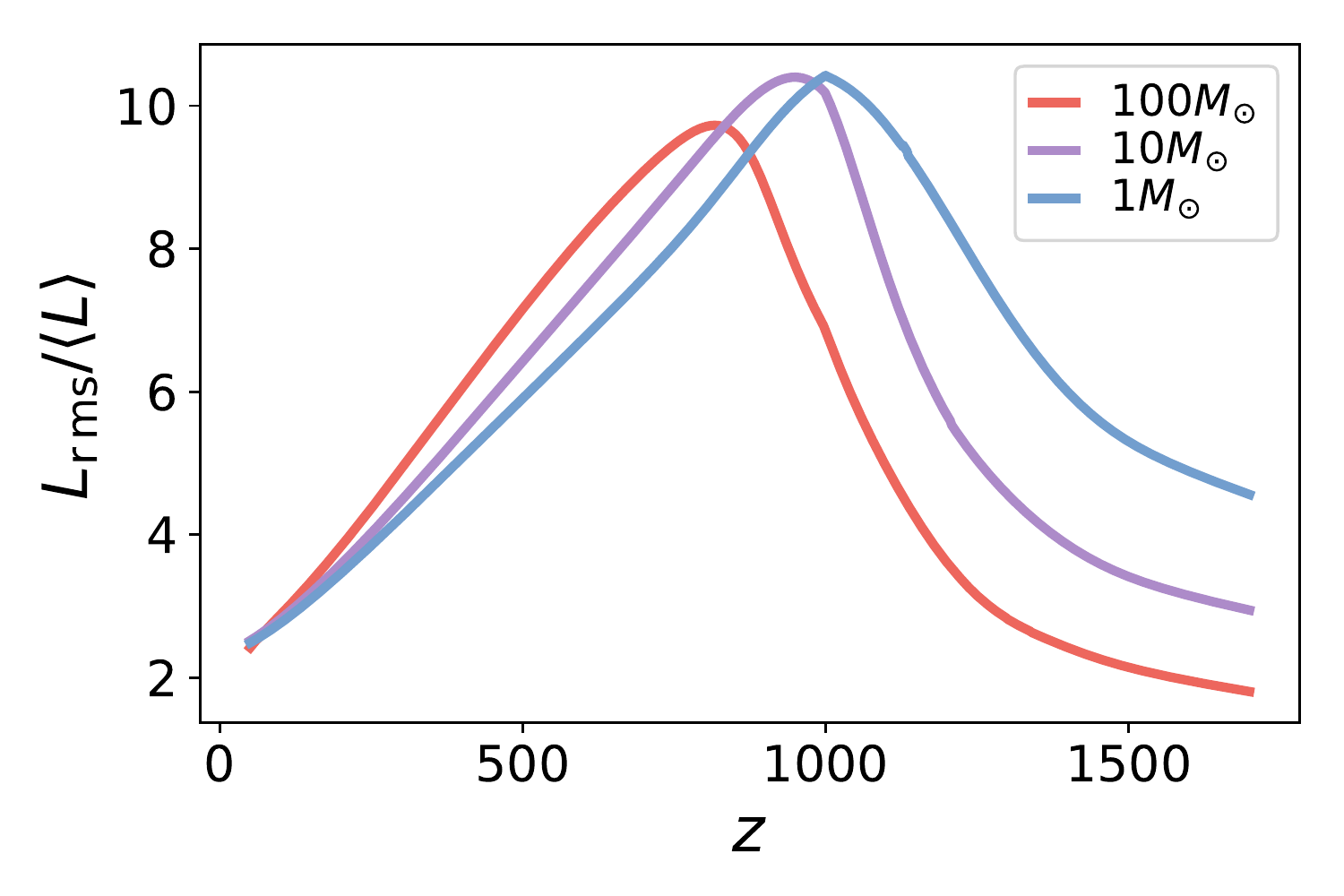}
\caption{\label{fig:Lrms_L} The ratio of $\sqrt{\braket{L^2}-\braket{L}^2}$ to $\braket{L}$ (where average here means over realizations of relative velocity between cold dark matter and baryons) as a function of redshift for several different black hole mass, as computed in AK17. These curves are obtained in the collisional-ionization accretion scenario.}
\end{figure}

With the prescription \eqref{eq:vbc-sub}, relative velocities affect most significantly the accretion rate, which scales approximately as $\dot{M} \propto 1/v_{\rm B}^3$. To a lesser extent, they also affect the ratio of the Compton drag and cooling timescales to the accretion timescale, thus the accretion constant $\lambda(v_{\rm B})$, as well as the gas temperature $T_s$ near the horizon. The first effect dominates the PBH luminosity, which scales as $L \propto \dot{M}^2 \propto 1/(v_{\rm B}^2 + v_{\rm bc}^2)^3$. The luminosity is thus dominated by the rare regions where the relative velocity is subsonic, which occupy a volume fraction of order $(v_{\rm B}/\langle v_{\rm bc}^2\rangle^{1/2})^3 \sim 10^{-2}$ at $z \sim 10^3$. In the remaining vast majority of the volume, the accretion luminosity is strongly suppressed by the supersonic relative motions. This physical picture is illustrated in Fig.~\ref{fig:relv_L}, showing the function $v_{\rm B}^6/(v_{\rm B}^2 + v_{\rm bc}^2)^3$ computed for a 3-dimensional realization of the relative velocity field. This figure shows that the PBH luminosity, thus energy injection rate, is concentrated in small and rare islands with subsonic relative velocities, surrounded by a mostly quiet sea of regions with supersonic velocities. It moreover indicates that the distribution of PBH luminosities is highly skewed, and that its average value is not at all a representative luminosity. We illustrate this in a different way in Fig.~\ref{fig:Lrms_L}, where we show that the rms luminosity can be up to an order of magnitude greater than its mean.

Relative velocities also modulate the gas temperature $T_s$ near the PBH horizon, thus the cutoff $E_{\max} \sim T_s$ of the free-free emission spectrum. In practice, for a given accretion scenario (collisional ionization or photoionization of the accreted gas), $T_s$ varies by no more than a factor-of-a-few when $v_{\rm bc}$ is varied from $0$ to $\langle v_{\rm bc}^2 \rangle^{1/2}$, as can be seen in Fig~\ref{fig:T_v_comp}. Moreover, these variations are mostly limited to $z \gtrsim 10^3$. Given that our simple accretion model is certainly not accurate to that degree, in order to simplify computations, we shall neglect the variations of $E_{\max}$ with relative velocity. For consistency, we shall also neglect its variations with PBH mass and redshift, which are of comparable magnitude. For each accretion scenario, we therefore assume a flat spectrum up to a \emph{constant cutoff} $E_{\max}$, independent of redshift, mass and relative velocity. We take $E_{\rm max}\approx 0.2$ MeV for the collisional-ionization scenario, and $E_{\rm max}\approx 10$ MeV for the photoionization scenario. We checked explicitly that restoring variations of $E_{\rm max}$ with respect to mass and redshift per accretion scenario does not significantly affect our final results (the rms fluctuations of $x_e$). It is important to note that this approximation only concerns the photon spectrum cutoff, not the overall amplitude of luminosity $L(T_s)$, in which we include the full dependence on PBH mass, redshift and relative velocity.

\subsection{Spatially perturbed recombination due to accreting PBHs}\label{subsec:delxe_power}

We now have all the ingredients to compute the perturbations to the free-electron fraction $\Delta x_e$ due to inhomogeneously accreting PBHs. For a given accretion scenario (photoionization or collisional ionization of the accreted gas), we assume a uniform injected photon spectrum $\Psi(E_\gamma) = \Theta(E_{\max} - E_\gamma)/E_{\max}$, with constant $E_{\max}$.

We compute the injection-to-deposition and injection-to-ionization Green's functions for this spectrum, as described in Secs.~\ref{sec:dep} and \ref{sec:xe}. We then obtain $\Delta x_e$ from Eq.~\eqref{eq:Gxe-inj}, with
\beq
\epsilon_{\rm inj}(\bs{r}) = \overline{\epsilon}_{\rm inj} \left(1 + \delta_L(\bs{r})\right), \ \ \ \ \ \overline{\epsilon}_{\rm inj} \equiv \frac{f_{\rm pbh} \rho_{\rm cdm}}{M n_{\rm H} H} \overline{L}, \label{eq:eps_inj_L}
\eeq
where $\delta_L(\bs{r}) \equiv L(\bs{r})/\overline{L} -1$ is the relative fluctuation of PBH luminosity. 

The new quantities computed in this work are the spatial perturbations to the free-electron fraction $\Delta x_e(a, \bs{k})$, obtained from Eq.~\eqref{eq:Delta-xe-k}, with $\epsilon_{\rm inj}(\bs{k}) = \overline{\epsilon}_{\rm inj} \delta_L(\bs{k})$. These perturbations are not directly observable, but leave an imprint on CMB anisotropy power spectra and higher-order correlations, the computation of which we defer to upcoming works. Importantly, to lowest order in $f_{\rm pbh}$, these CMB observables depend on the part of $\Delta x_e$ that is correlated with terms quadratic in unperturbed CMB anisotropies. In order to obtain a proxy for this correlated part, we split PBH luminosity perturbations in a piece tracing fluctuations in $v_{\rm bc}^2$ (superscript c), and a piece that is uncorrelated with them (superscript unc):
\barr
\delta_L(\bs{r}) &=& \delta_L^{\rm c}(v_{\rm bc}(\bs{r})) + \delta_L^{\rm unc}(v_{\rm bc}(\bs{r})), \label{eq:delta_L-decomp}\\
\delta_L^{\rm c}(v_{\rm bc}(\bs{r})) &=& b \left(\frac{v_{\rm bc}^2(\bs{r})}{\braket{v_{\rm bc}^2}} - 1 \right), \ \ \ \ b \equiv \frac32 \frac{\braket{v_{\rm bc}^2 \delta_L}}{\braket{v_{\rm bc}^2}}, \label{eq:b-def}\\
\braket{v_{\rm bc}^2 \delta_L^{\rm unc}} &=& 0,
\earr
where $\langle ... \rangle$ represents averaging over the Gaussian distribution of relative velocities, and all quantities implicitly depend on scale factor. Note that the two-point correlation function of $\delta_L^{\rm c}$ matches the large-scale limit of the two-point correlation function of $\delta_L$ \cite{dalal10a, Ali-Haimoud_14}. Numerically, we find that the power spectrum of $\delta_L^{\rm c}$ reproduces well the power spectrum of $\delta_L$ for $k \lesssim 0.1$ Mpc$^{-1}$, but significantly underestimates it for smaller scales. Using $\delta_L^{\rm c}$ in lieu of $\delta_L$ should therefore provide a conservative estimate of the spatial perturbations of $\Delta x_e$, which, in any case, are suppressed at $k \gtrsim 0.1$ Mpc$^{-1}$ due to nonlocal energy deposition, as can be seen in Fig.~\ref{fig:W_100}.
 
We thus define $\Delta x_e^{\rm c}(a, \bs{k})$ as the free-electron perturbation resulting from the $v_{\rm bc}^2-$correlated part of the injected energy, $\epsilon_{\rm inj}^{\rm c} = \overline{\epsilon}_{\rm inj} \delta_L^{\rm c}(\bs{k})$, inserted in Eq.~\eqref{eq:Delta-xe-k}. This ``correlated part" of the free-electron fluctuations serves as a proxy for the quantity relevant to CMB anisotropies.

For any field $X(a, \bs{k})$ we define the dimensionless unequal-time power spectrum $\Delta^2_X(a, a', k)$ through
\beq
\braket{X(a, \bs{k}) X^*(a', \bs{k}')} = \frac{16 \pi^5}{k^3} \Delta^2_X(a, a', k)\delta^3(\bs{k}' - \bs{k}),
\eeq
and for short denote $\Delta^2(a, k) \equiv \Delta^2(a, a, k)$, which gives the variance of $X(a, \bs{r})$ per logarithmic $k$ interval:
\beq
\textrm{var}(X(a)) \equiv \braket{X(a, \bs{r})^2}= \int d \ln k ~ \Delta_X^2(a, k). \label{eq:var}
\eeq 

\begin{figure}[ht]
\includegraphics[trim={0 0 0 0},width=\columnwidth]{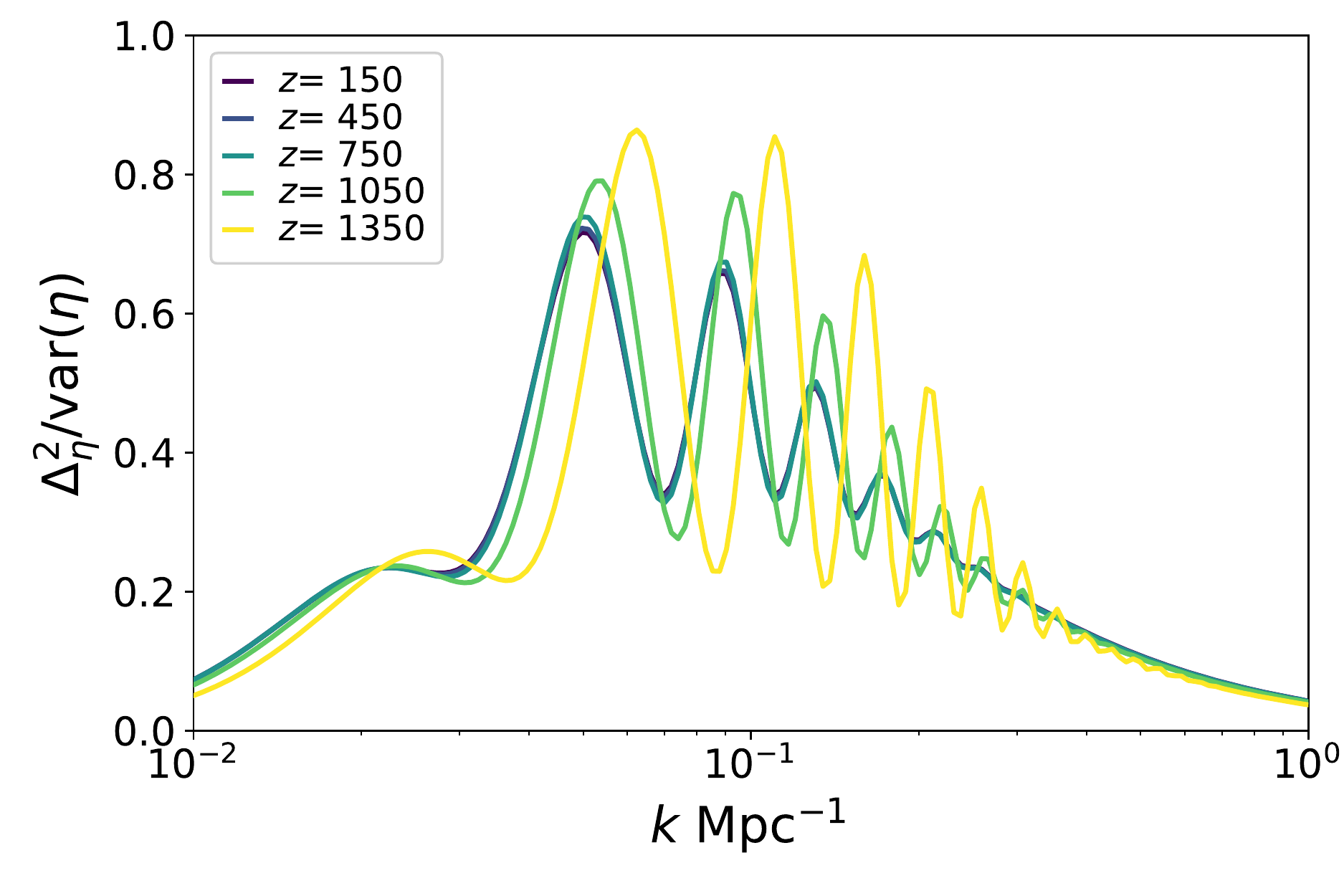}
\caption{\label{fig:eta_power} Dimensionless power spectrum of $\eta(z,\bs{r}) \equiv v_{\rm bc}^2(z, \bs{r})/\braket{v_{\rm bc}^2(z)} -1$ at several redshifts, normalized by its variance $\textrm{var}(\eta) = 2/3$. The shape is approximately constant at $z\lesssim 1000$, after kinematic decoupling \cite{Tseliakhovich_10}.}
\end{figure}

From Eq.~\eqref{eq:Delta-xe-k}, with $\epsilon_{\rm inj} \rightarrow \overline{\epsilon}_{\rm inj} \delta_L^{\rm c}$, we obtain 
\barr
\Delta^2_{\Delta x_e^c}(a, k) = \iint^a d \ln a_i d \ln a_i' G_{x_e}^{\rm inj}(a, a_i, k)G_{x_e}^{\rm inj}(a, a_i', k) \nonumber\\
\times b(a_i) \overline{\epsilon}_{\rm inj}(a_i) b(a_i')  \overline{\epsilon}_{\rm inj}(a_i') \Delta^2_{\eta}(a_i, a_i', k),~~~~~ \label{eq:Pow-Dxe}
\earr
where $b(a)$ is defined in Eq.~\eqref{eq:b-def} and $\eta(a,\bs{r}) \equiv v_{\rm bc}^2(a, \bs{r})/\braket{v_{\rm bc}^2(a)} -1$. The power spectrum of $\eta$ is easliy computed, and can be expressed as an integral quadratic in the power spectrum of $v_{\rm bc}$, given explicitly in Ref.~\cite{ferraro12a}; we show it for several redshifts in Fig.~\ref{fig:eta_power}.

\begin{figure}[ht]
\includegraphics[trim={0 0 0 0},width=\columnwidth]{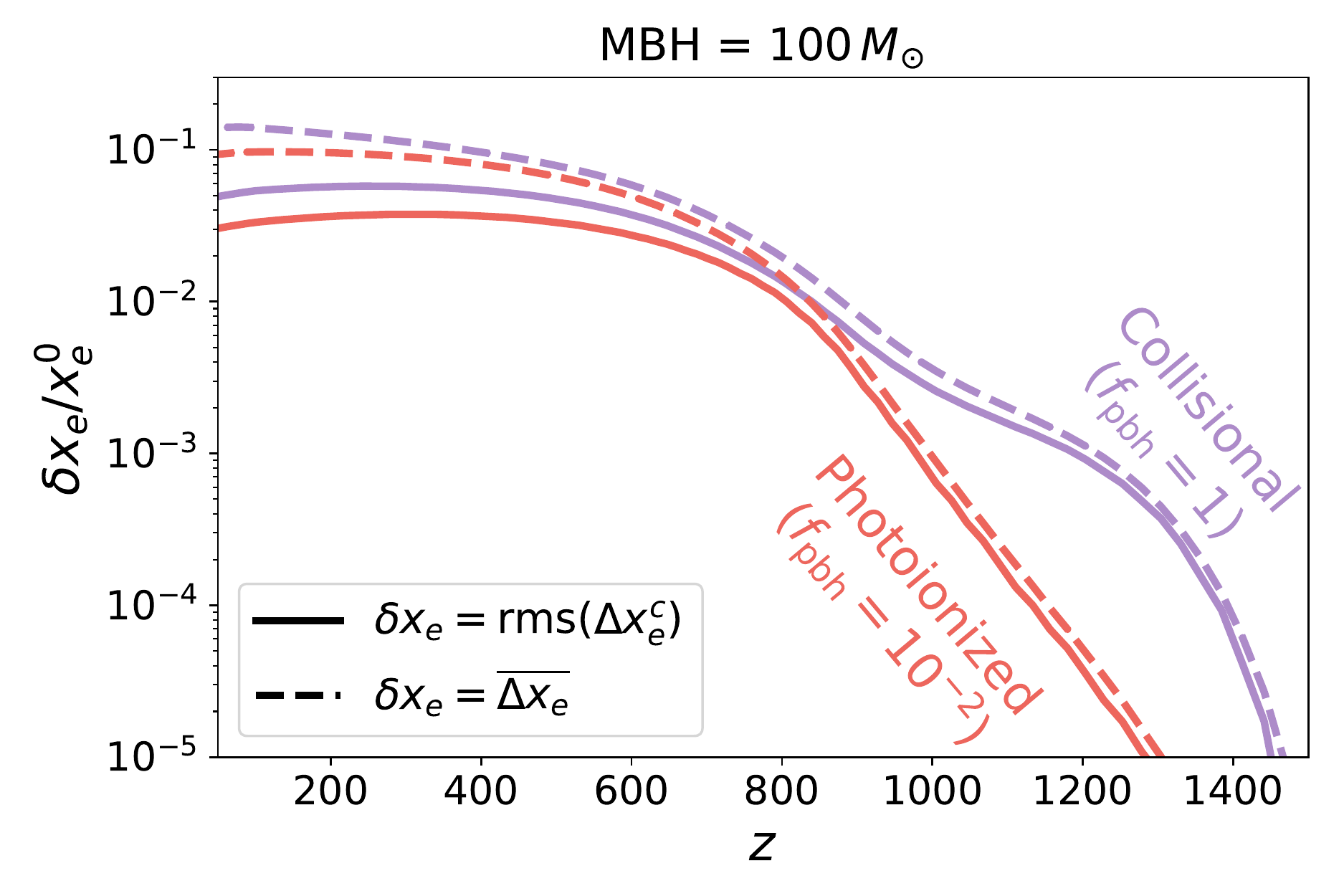}
\caption{\label{fig:xe_rms} The rms of spatial fluctuations of the ($v_{\rm bc}^2$-correlated part of) free-electron fraction perturbations induced by accreting PBHs (solid lines), compared to the mean change in the free-electron fraction (dashed lines), both normalized to the standard ionization history $x_e^0$. The two colors correspond to the two accretion scenarios discussed in Sec.~\ref{subsec:accmod}. In both cases, we consider 100-$M_{\odot}$ PBHs, whose abundances roughly saturate AK17's limits: $f_{\rm pbh}=1$, $10^{-2}$ for the collisional ionized and photoionized cases respectively.}
\end{figure}

In Fig.~\ref{fig:xe_rms}, we show $\textrm{rms}(\Delta x_e^c)/x_e^0 \equiv \sqrt{\textrm{var}(\Delta x_e^{\rm c})}/x_e^0$, as a function of redshift, for both accretion scenarios; this quantity is obtained from $\Delta_{\Delta x_e^{\rm c}}^2(k)$ through Eq.~\eqref{eq:var}. For reference, we also show the mean relative change $\overline{\Delta x_e}/x_e^0$, obtained through our Green's function. We see that $\textrm{rms}(\Delta x_e^c)$ is comparable to $\overline{\Delta x_e}$, within a factor of order unity. Note that our estimate of $\overline{\Delta x_e}$ is somewhat different from that of AK17. Our treatment is more accurate in some respects, as AK17 have a simplified treatment of energy deposition and do not account for photoionization nor the ICS energy sink. On the other hand, our Green's function approach does not capture nonperturbative changes to the free-electron fraction, for which AK17 do solve. As we show in Appendix \ref{app:xe-checks}, our Green's function approach is accurate as long as $\overline{\Delta x_e}/x_e^0 \ll 1$, in which case our treatment is overall more accurate.
 
\begin{figure*}[htbp]
\centering
\includegraphics[trim={0 0 0 0},width=\columnwidth]{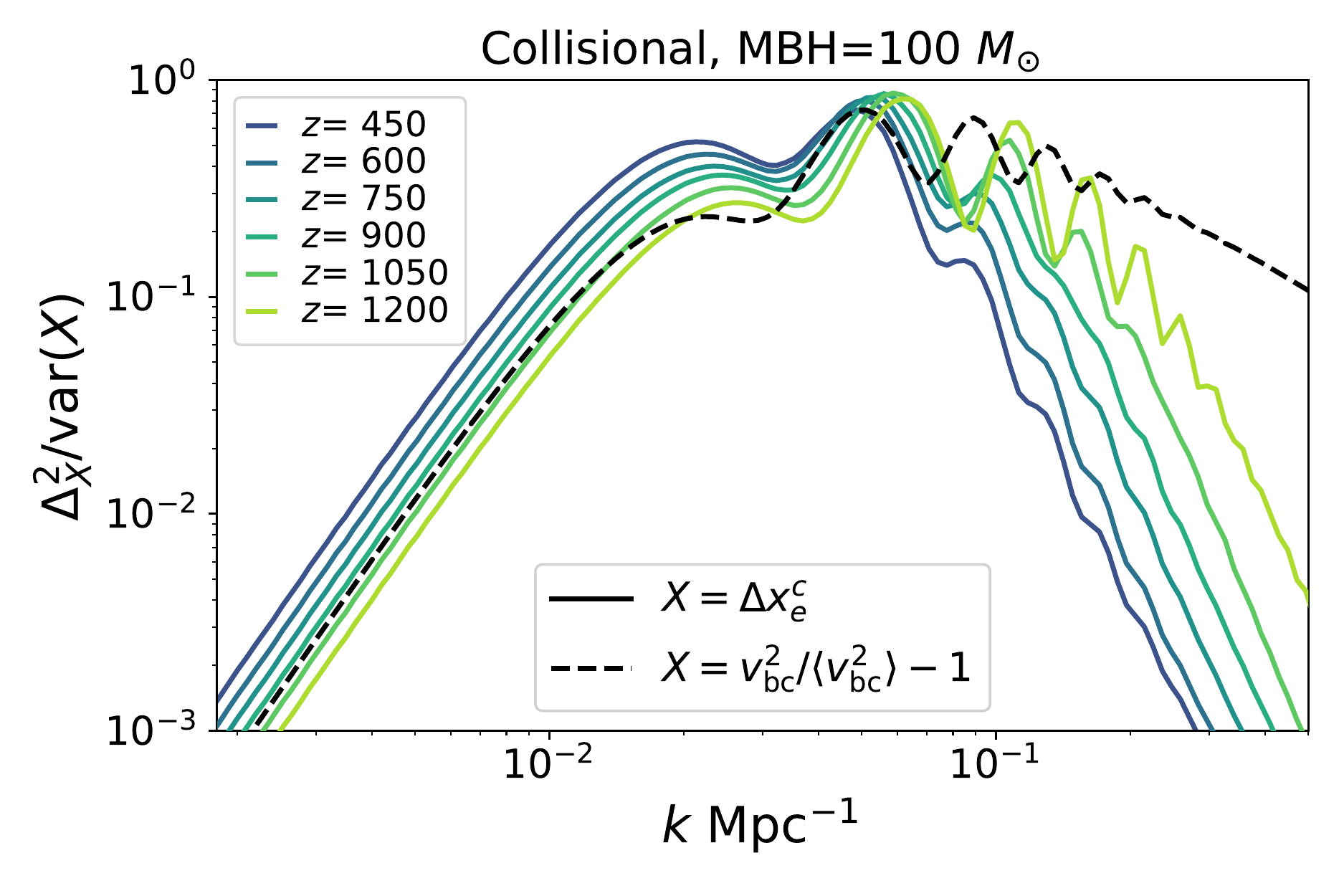}
\includegraphics[trim={0 0 0 0},width=\columnwidth]{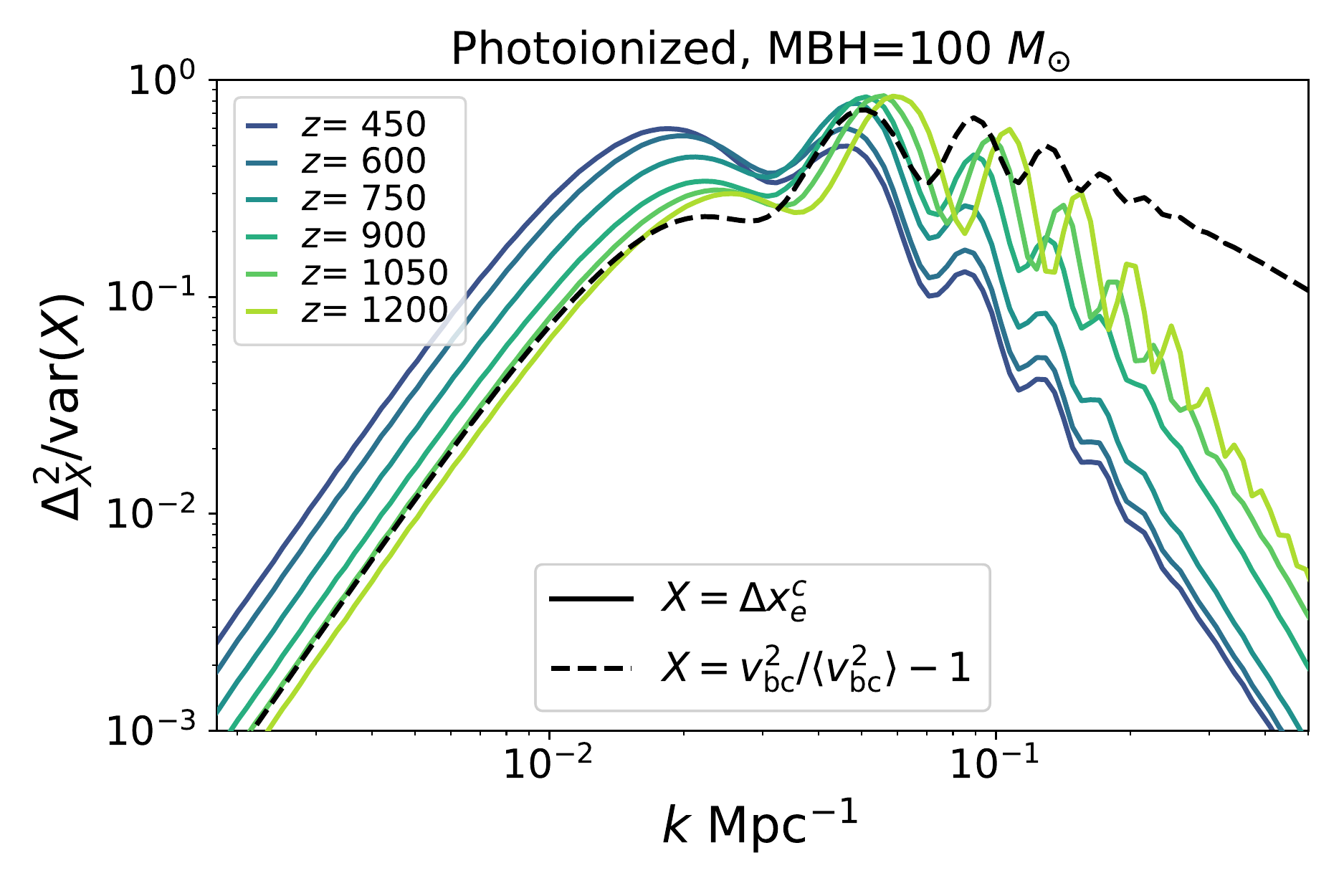}
\caption{\label{fig:Delxe} Dimensionless power spectrum of the ($v_{\rm bc}^2$-correlated part of the) free-electron fraction $\Delta x_e^{\rm c}$ at several redshifts, obtained from Eq.~\eqref{eq:Pow-Dxe}, normalized by its variance. Each panel corresponds to a PBH accretion scenario: collisionally ionized (left) and photoionized (right), see Sec.~\ref{subsec:accmod} for details. For reference, the black dashed line shows the normalized power spectrum of $\eta(z,\bs{r}) \equiv v_{\rm bc}^2(z, \bs{r})/\braket{v_{\rm bc}^2(z)} -1$, at $z = 1000$ (this quantity is independent of redshift for $z \lesssim 1000$). These plots reveal the small-scale suppression of free-electron perturbations due to nonlocal energy deposition, which becomes more pronounced for lower redshifts, as photons have propagated farther from their injection point. Still, free-electron perturbations retain significant spatial fluctuations up to $k \sim 0.1$ Mpc$^{-1}$.}
\end{figure*}

We show $\Delta^2_{\Delta x_e^{\rm c}}(z, k)/\textrm{var}(\Delta x_e^c(z))$ at several redshifts in Fig.~\ref{fig:Delxe}, for accretion onto 100-$M_{\odot}$ PBHs in either accretion scenario. This quantity represents the overall shape of the relative fluctuations in the ionization fraction $\Delta x_e^{\rm c}$ per logarithmic $k$-interval. For contrast, we overlay the normalized power spectrum $\Delta^2_{\eta}(k)/\textrm{var}(\eta)$ at $z = 1000$, i.e.~after kinematic decoupling, for which $\eta$ is time invariant \cite{Tseliakhovich_10}. This comparison reveals the small-scale suppression of power in $\Delta x_e^{\rm c}$ due to nonlocal energy deposition that is more stark for later redshifts, as we would expect from photon propagation. Additionally, for the photoionized accretion scenario, the higher-energy injected photons have larger propagation distances, resulting in a more pronounced suppression of small-scale power relative to large-scale power (this can be seen by examining the ratios of the amplitudes of the second and first peaks). Note that, besides the redshift-dependent small-scale suppression, the power spectrum of $\Delta x_e$ does not quite have the universal shape of low-redshift velocity-induced acoustic oscillations \cite{munoz19a, munoz19b} shown as a black dashed line, which is relevant to large-scale structure \cite{dalal10a} or 21-cm fluctuations \cite{Ali-Haimoud_14, munoz15a}. Indeed, free-electron fraction perturbations at a given redshift are affected by energy injection at all prior epochs, including before kinematic decoupling, during which the scale dependence of relative velocities is time dependent, as seen in Fig.~\ref{fig:eta_power}.

Figs.~\ref{fig:xe_rms} and \ref{fig:Delxe} constitute the main results of this study. Fig.~\ref{fig:xe_rms} shows that the spatial modulations of the free-electron fraction perturbations sourced by accreting PBHs are comparable in amplitude to their mean. Moreover, Fig.~\ref{fig:Delxe} shows that these perturbations have support on large scales $k \sim 0.01-0.1$ Mpc$^{-1}$, similar to the scales at which CMB anisotropies are maximal. \rev{In past studies, the mean perturbation to recombination $\overline{\Delta x_e}$ has been the sole quantity considered when estimating the impact of accreting PBHs on CMB anisotropies. Qualitatively, a homogeneous increase to the free-electron fraction affects CMB-anisotropy power spectra similarly to an increase in the reionization optical depth: it damps anisotropies on small angular scales and enhances polarization on large angular scales. The order-unity spatial perturbations in $\Delta x_e$ shown in Figs.~\ref{fig:xe_rms} and \ref{fig:Delxe} ought to impact CMB anisotropies in two different ways. First, they should lead to order-unity modifications to the perturbation to CMB power spectra, with an angular dependence qualitatively different from that resulting from the homogeneous $\overline{\Delta x_e}$. This means that this additional perturbation to CMB power spectra should have different degeneracies with standard cosmological parameters, in particular the reionization optical depth, and should therefore help improve constraints on accreting PBHs. Secondly, the large-scale perturbations to $\Delta x_e$ should source higher-order correlations functions in CMB anisotropies, beyond the power spectra. We will study and quantify these effects in upcoming publications.}

\section{Discussion and Conclusion}\label{sec:conc}

In this work, we have developed a set of analytic and numerical tools to compute spatial perturbations to recombination resulting from inhomogeneous injection of sub-10 MeV photons, and applied them to the specific case of accreting PBHs. \\
The first step was to translate energy \textit{injection} to a time- and space-dependent energy \textit{deposition} rate. To that end, we developed a Monte Carlo radiation transport code, incorporating all relevant plasma interactions with up-to-date cross sections. This code follows the evolution of an injected photon spectrum, accounting for photoionization and Compton scattering, and tabulates the energy deposited into secondary electrons as a function of time and distance from the injection point. While secondary electrons dissipate their energy almost instantaneously through rapid interactions, a non-negligible part of this energy is lost to upscattering CMB photons to sub-10.2 eV energies, at which they do not interact efficiently with the plasma. We computed the fraction of electron energy that is lost to this sink with a novel analytic integral expression \eqref{eq:elec_app}, matching existing numerical results remarkably accurately. The final output of our radiation transport code is a time- and space-dependent injection-to-deposition Green’s function, self-consistently accounting for this energy sink. \\
The second step was to convert the energy deposited in the primordial plasma to a perturbation of its ionization and thermal history. We extracted a deposition-to-ionization Green's function by linearizing the effective 4-level atom differential equations solved by \texttt{HyRec-2}, which provide a highly accurate approximation of the exact numerical radiative transfer calculation of \texttt{HyRec}. By convolving the two Green’s functions, we obtained the injection-to-ionization Green’s function, which directly connects an energy injection rate to a time-dependent inhomogeneous free-electron fraction. We find that recombination inhomogeneities are typically washed out for scales $k \gtrsim 0.1$ Mpc$^{-1}$, due to the finite propagation of injected photons. The Green’s function we compute allows us to quantify this suppression in detail, as a function of time, injected spectrum, and comoving scale.

We applied these new tools and methods to inspect, for the first time, the imprint on cosmological recombination of inhomogeneous photon injection by accreting PBHs. The physical origin of this inhomogeneity is the dependence of the accretion rate on the velocities of accreted baryons relative to dark matter, thus PBHs. Importantly, these relative velocities are typically supersonic, and therefore ought to have a strong, nonperturbative effect on PBH luminosities. Fluctuations of relative velocities on $\sim$100 Mpc scales thus translate to a large-scale spatial modulation of the PBH accretion rate and luminosity, thus energy injection rate. To quantify this effect, we adopted the accretion model of Ref.~\cite{yacine17a}, which was used to derive conservative upper limits to the PBH abundance from CMB-anisotropy power spectra. Within this model, the PBH luminosity is \emph{highly} inhomogeneous, concentrated in small islands with subsonic relative velocities (see Fig.~\ref{fig:relv_L}). Conservatively, we extracted the free-electron variations resulting from the component of luminosity fluctuations that is correlated with relative velocities squared, as we expect those to give the dominant contributions to observable effects in CMB anisotropies. We found that spatial perturbations to the free-electron fraction $\Delta x_e$ peak at $k \sim 10^{-2}$ Mpc$^{-1}$, and are only partially washed out by the finite propagation distance of high-energy photons. Importantly, we found that the rms of $\Delta x_e$ is comparable to its mean, which was the only quantity evaluated in previous studies. 

\rev{While we focused on accreting PBHs in this work, the tools we developed ought to be useful to study other sources of non-standard energy injection, and their effect beyond the CMB. For instance, it may be useful to extend our study to annihilating or decaying DM particles in the cosmic dark ages, during which the DM density distribution is significantly inhomogeneous. The resulting energy injection should heat the gas inhomogeneously, and therefore leave unique signatures on the high-redshift 21-cm signal \citep{furlanetto06a,valdes07a,natarajan09a,cumberbatch10a}.}

This work presents the first detailed calculation of the spatial aspect of energy deposition and ionization perturbations, and it is worth mentioning several aspects in which it could be improved or expanded upon. First, it would be interesting to generalize our spatial injection-to-ionization Green’s function to arbitrary photon energies, and to arbitrary types of injected particles. This would allow this formalism to be applied, e.g. to inhomogeneously distributed annihilating or decaying dark matter particles \cite{dvorkin13a}. Second, we derived a novel and accurate analytic expression for the fraction of electron energy resulting in sub-10.2 eV inverse-Compton-scattered (ICS) photons. This result can be extended to the branching ratios of electron energy deposition into ionization, excitation, heating and ICS, as we outline in the main text. \rev{In particular, our analytic approximation can be generalized to derive the full spectrum of sub-10.2 eV photons produced in ICS; this will be useful to quantify the impact of sub-10.2 eV photons on cosmological recombination, and thus check the standard assumption that their effect is entirely negligible.} Third, when computing the deposition-to-ionization Green’s function, we assumed that the deposited energy is shared equally between ionization, excitation and heating (for a fully neutral gas) \cite{chen04a}. Our Green’s function could be made more accurate by computing these branching ratios explicitly. Lastly, for the specific problem of energy injection by accreting PBHs, we adopted the extended-Bondi accretion model of Ref.~\cite{yacine17a}, which provides a clear prescription for the effect of relative velocities on PBH luminosities. It would be interesting to consider the case of disk-like accretion \cite{poulin17a}, for which the dependence of luminosity on relative velocities has not yet been studied. 

To conclude, we have laid the groundwork for studying the imprints of inhomogeneous energy injection in the early Universe on CMB anisotropies. In the context of accreting PBHs, our preliminary result shows that spatial fluctuations in ionization perturbations are as large as their mean, which suggests novel signatures in CMB anisotropies. First, these fluctuations should source additional contributions to CMB temperature and polarization power spectra, comparable in magnitude to the effect of the mean perturbation to the free-electron fraction. Importantly, these additional contributions should have very different shapes than the previously computed perturbations to CMB power spectra, thus could help break parameter degeneracies and probe lower PBH abundances. Second, we expect these spatial perturbations to recombination to imprint non-Gaussian signatures in CMB anisotropies — specifically, a nonzero trispectrum at lowest order in PBH abundance. This novel qualitative effect should provide a sensitive window into PBHs, given the tight limits on CMB non-Gaussianities \citep{planck20c}. To quantify these auspicious signatures requires perturbing the photon Boltzmann-Einstein system. We tackle this challenging problem in upcoming companion papers.

\begin{acknowledgments}
We are grateful to Nanoom Lee, Hongwan Liu, Juli\'an Mu\~{n}oz and Tracy Slatyer for useful comments on the draft of this paper, and thank Juli\'an Mu\~{n}oz for providing Fig.~\ref{fig:relv_L}. This material is based upon work supported by the National Science Foundation Graduate
Research Fellowship Program under Grant No. DGE1839302, and the NSF grant 1820861. Any opinions,
findings, and conclusions or recommendations expressed in this material are those of the
author(s) and do not necessarily reflect the views of the National Science Foundation.
\end{acknowledgments}

\appendix

\section{cross sections and energy loss rates}\label{app:cross_sections}

\subsection{Sub-10 MeV photons}
\subsubsection{Compton Scattering}
The Compton cross section $\sigma_{\rm C}(E)$ is the integral of the Klein-Nishina differential cross section computed from tree-level quantum electrodynamics. Namely, for a given initial photon energy $E$,
\beq
    \deriv{\sigma_{\rm C}}{\cos\theta}(E)=\frac{3}{8}\sigma_T\left(\frac{E'}{E}\right)^2\left(\frac{E}{E'}+\frac{E'}{E}-1 +\cos^2 \theta\right),\label{eq:kn}
\eeq
where $\sigma_T$ is the Thomson cross section and the outgoing photon energy, $E'$, is a function of $\cos\theta$ and $E$,
\begin{align}\label{eq:Ef}
    \frac{E'}{E} =\left(1+ \frac{E}{m_e}(1-\cos\theta)\right)^{-1}.
\end{align}

\subsubsection{Photoionization}
We adopt the following cross sections for photoionization of hydrogen and neutral helium \citep{zdziarski89a},
\barr
    \sigma_{\rm H} (E)&=&\frac{64\pi}{\alpha^3}\sigma_T(E_I/E)^4\frac{\exp(-4\eta\arctan(1/\eta))}{1-\exp(-2\pi\eta)},~~\\
    \eta&=&\frac{1}{\sqrt{E/E_I-1}},\\
    \sigma_{\rm {HeI}}(E)&=&-12\sigma_{\rm H}(E) \nonumber\\
    &+&5.1\times 10^{-20}\textrm{ cm}^2\left(\frac{250\textrm{ eV}}{E}\right)^{\Gamma(E)},
\earr
where $\alpha$ is the fine structure constant, and $E_I=13.6$ eV is the ionization energy of hydrogen. The exponent in the Helium cross section, $\Gamma(E)$, is a broken power-law fit \citep{zdziarski89a}: $\Gamma= 3.30$ for $E> 250$~eV, and $\Gamma=2.65$ for 50~eV$<E<$~250~eV, but a posteriori our injected photons do not reach this latter limit even accounting for Hubble expansion and Compton scattering.
For comparison to Thomson scattering, when $E\gg E_I$, 
\beq    
    \sigma_{\rm H}(E)\approx 0.24\,\sigma_T \left(\frac{\alpha m_e}{E}\right)^{7/2}.
\eeq

\subsection{Electrons}

\subsubsection{Inverse Compton Scattering (ICS)}

For ICS we use the general spectrum with no assumptions on the energy regime originally from \cite{fargion97a}, which we checked matches the asymptotic low- and high-energy approximations (see the appendix of \cite{hongwan20a} for discussion). 

In the situation of interest, electrons interact with CMB photons, with a blackbody spectrum: the number density of photons per energy interval is
\begin{align}
    n_{\rm BB}(\epsilon,T)\equiv \frac{1}{\pi^2\hbar^3}\frac{\epsilon^2}{\exp(\epsilon/k_B T)-1}.
\end{align}
Given an electron with initial energy $E$, the doubly differential ICS rate, per initial CMB photon energy $\epsilon$ and final photon energy $\epsilon_1$, is then 
\begin{align}
    \frac{d^2 \Gamma_{\rm ICS}}{d \epsilon d\epsilon_1} &= \frac{3\sigma_T  n_{\rm BB}(\epsilon)}{32\beta^6\gamma^2\epsilon}\left\{\frac{1}{\gamma^4}\left(\frac{\epsilon}{\epsilon_1}-\frac{\epsilon_1^2}{\epsilon^2}\right)\right.\nonumber\\
    &+(1+\beta)\left[\beta(\beta^2+3)+\frac{1}{\gamma^2}(9-4\beta^2)\right]\nonumber\\
    &+(1-\beta)\left[\beta(\beta^2+3)-\frac{1}{\gamma^2}(9-4\beta^2)\right]\frac{\epsilon_1}{\epsilon}\nonumber\\
    &\left.-\frac{2}{\gamma^2}(3-\beta^2)\left(1+\frac{\epsilon_1}{\epsilon}\right)\log\left(\frac{1+\beta}{1-\beta}\frac{\epsilon}{\epsilon_1}\right)\right\},
\end{align}
where $\gamma \equiv E/m_e$ and $\beta \equiv \sqrt{1-1/\gamma^2}$ are the Lorentz factor and velocity of the incoming electron, respectively. The expression above holds for $(1-\beta)\epsilon_1/(1+\beta)<\epsilon<\epsilon_1$. In addition, for $\epsilon_1 < \epsilon <(1+\beta)\epsilon_1/(1-\beta)$,
\begin{align}
    \frac{d^2 \Gamma_{\rm ICS}}{d \epsilon d\epsilon_1}(\epsilon_1< \epsilon; \beta) =  -\frac{d^2 \Gamma_{\rm ICS}}{d \epsilon d\epsilon_1}(\epsilon_1 > \epsilon; -\beta) 
\end{align}
All other values of the incoming photon energy are kinematically forbidden. The energy lost by the electron per scattering is $\Delta E = E - E'  =  \epsilon_1 - \epsilon$. The quantity of interest to us is the rate of electron energy loss to sub-10.2eV photons,
\barr
\dot{\mathcal{E}}_{\rm sink}(E) = \int d \epsilon \int^{E_{\rm exc}} d \epsilon_1 (\epsilon_1 - \epsilon) \frac{d^2 \Gamma_{\rm ICS}}{d \epsilon d\epsilon_1}. 
\earr

\subsubsection{Ionization}
We use the relativistic binary-encounter-dipole model from Ref.~\cite{yong-ki00a} for an incident electron ionizing ground-state hydrogen and neutral helium. Taking the liberated electron's outgoing energy as $W=E-E'-E_I$, where $E_I$ is the binding energy of the target atom, the differential ionization cross section per atomic orbital is,
\begin{align}
    &\frac{d \sigma_{\rm ion}(E)}{d E'}=\frac{3\sigma_T N m_e}{(\beta^2_E+\beta^2_U+\beta^2_{E_I})4E_I^2}\nonumber\\
    &\quad\times\left\{\frac{(N_i/N)-2}{t+1}\left(\frac{1}{w+1}+\frac{1}{t-w}\right)\frac{1+2t'}{(1+t'/2)^2}\right.\nonumber\\
    &\quad+[2-(N_i/N)]\left[\frac{1}{(w+1)^2}+\frac{1}{(t-w)^2}+\frac{(E_I/m_e)^2}{(1+t'/2)^2}\right]\nonumber\\
    &\quad\left.+\frac{1}{N(w+1)}\deriv{f}{w}\left[\ln \left(\frac{\beta_E^2}{1-\beta_E^2}\right)-\beta_E^2-\ln\left(\frac{2 E_I}{m_e}\right)\right]\right\},
\end{align}
where $N$ is the orbital electron occupation number for the relevant atom, $t\equiv E/E_I$, $t'\equiv E/m_e$, $w\equiv W/E_I$, and $u\equiv U/E_I$ where $U=\braket{p^2/2m}$ is the average orbital kinetic energy of the target electron. $\beta_i$ here are the velocities computed for energies $i\in\{E,U,E_I\}$. $N_i\equiv\int_0^\infty (\td f/\td w)\td w$, where $\td f/\td w$ is the differential dipole oscillator strength, and is taken from Ref.~\cite{kim94a} as a fitted power series
\begin{align}
   \deriv{f}{w}=Ay^2+By^3+Cy^4+Dy^5+Fy^6,
\end{align}
with $y\equiv E_I/(W+E_I) = 1/(1+w)$. The coefficients are given by:
\begin{itemize}
    \item Hydrogen: $E_I=13.6$, $U=13.6$, $N=1$, $A=0$, $B=12.2$, $C=-29.6$, $D=31.3$, and $F=-12.2$.
    \item Helium: $E_I=24.6$, $U=39.5$, $N=2$, $A=-0.0225$, $B=1.18$, $C=-0.463$, $D=0.0891$, and $F=0$.
\end{itemize}
The relevant energy loss is then,
\begin{align}
    \dot{\mathcal{E}}_{\rm ion}(E) = v_E \sum_{i = \rm H, He} n_i \int\td E'~ \frac{d \sigma_{{\rm ion},i}(E)}{d E'}(E-E'),
\end{align}
where $v_E = \sqrt{1 - m_e^2/E^2}$ is the velocity of the incoming electron.

\subsubsection{Excitation}
We only consider excitations from the ground state to the first excited. We use the fitting functions in Ref.~\cite{stone02a},
\begin{align}
    \sigma_{\rm exc}(E)&=\nonumber\\
    &\frac{3\sigma_T R}{\alpha^4 2\pi(E+E_I+E_{\rm exc})}\left(A\ln\left(\frac{E}{R}\right)+B+C\frac{R}{E}\right)
\end{align}
where $R\approx 13.6$ eV is the Rydberg energy. The coefficients are given by,
\begin{itemize}
    \item Hydrogen: $E_{\rm exc}=10.204$, $E_I=13.6$,  $A=0.5555$, $B=0.2718$, and $C=0.0001$.
    \item Helium: $E_{\rm exc}=21.218$, $E_I=24.6$, $A=0.1656$, $B=-0.07694$, and $C=0.03331$.
\end{itemize}
We assume the energy lost by the electron is simply the excitation energy. We therefore obtain the following rate of energy loss through excitation
\beq
\dot{\mathcal{E}}_{\rm exc}(E) = v_E \sum_{i = \rm H, He} n_i E_{\rm exc,i} \sigma_{\rm exc,i}(E).
\eeq

\subsubsection{Heating}
An energetic electron propagating in a plasma shares its kinetic energy with ambient electrons, thus heats the plasma. 
Because we eventually compute the rate of energy loss, we simply use from Ref.~\cite{furlanetto10a}, 
\begin{align}
  \dot{\mathcal{E}}_{\rm heat}(E) =\frac{4\pi(\alpha\hbar c)^2n_{\rm H} x_e\ln \Lambda}{m_e v_E},
\end{align}
where $v_E$ is the electron velocity, and the Coulomb logarithm is taken as,
\begin{align}
    \ln \Lambda=\ln\left(\frac{4E}{\zeta_e}\right),\ \ 
    \zeta_e\equiv 2\hbar\left(\frac{4\pi n_{\rm H} x_e\alpha\hbar c}{m_e}\right)^{1/2}.
\end{align}

\section{Semi-analytic approximation for the spatially averaged Green's function with Compton scattering only}\label{app:Greens}

In this appendix we focus on the temporal part of the Green's function $\overline{G}_E(a_d, a_i)$ defined in Eq.~\eqref{eq:G-av}. The most detailed numerical computation of $\overline{G}_E$ is given in Refs.~\cite{slatyer16a,hongwan20a}. Here we provide a simple semi-analytic approximation, holding when photoionizations are negligible, and assuming secondary electrons efficiently deposit all their energy (i.e.~$F_{\rm sink} = 0$). This approximation provides a consistency check for our simulations.

We define $\mathcal{N}_E$ as the photon number density per energy interval. In the homogeneous limit, and neglecting photoionizations, the collisional Boltzmann equation satisfied by $\mathcal{N}_E$ takes the form 
\barr
a^{-2} \frac{d}{dt}(a^2 \mathcal{N}_E) &=& \frac{1}{E} \frac{d \dot{\rho}_{\rm inj}}{dE}  +\mathcal{C}_{\rm C}[\mathcal{N}_E]. \label{eq:Boltzmann-NE}
\earr
In the left-hand-side, $d/dt \equiv \partial_t - H E \partial_E$ is the total derivative along free photon trajectories. The first term in the right-hand-side of Eq.~\eqref{eq:Boltzmann-NE} accounts for energy injection in the form of photons, and the second term is the Compton collision operator, which is a linear integral operator. 

Given the photon number density $\mathcal{N}_E(t)$, the (homogeneous) volumetric power deposited in secondary electrons is then simply
\beq
\dot{\rho}_{e}(t) = \int dE ~\dot{\mathcal{E}}_{\rm C}(E) \mathcal{N}_E(t), \label{eq:dot_rho_e}
\eeq
where 
\beq
   \dot{\mathcal{E}}_{\rm C}(E) \equiv n_{\rm H}  \int_{E_{\rm min}}^E\td E' \deriv{\sigma_{\rm C}(E)}{E'}(E-E')\label{eq:dotE_Compt}
\eeq
is the rate of photon energy loss through Compton scattering -- it can be expressed analytically, but the expression is not particularly enlightening and we do not provide it here. Note that in this equation (just like our simulations) the ionization energy is neglected relative to the initial photon energy.

With the exact Compton collision operator, the Boltzmann equation \eqref{eq:Boltzmann-NE} has to be solved numerically. Following Ref.~\cite{yacine17a}, we shall approximate $\mathcal{C}_{\rm C}[\mathcal{N}_E]$ by a simple number-conserving divergence term, reproducing the exact energy loss rate for a given $\mathcal{N}_E$:
\beq
\mathcal{C}_{\rm C}[\mathcal{N}_E] \approx \frac{\partial}{\partial E} \left(\dot{\mathcal{E}}_{\rm C}(E) \mathcal{N}_E\right). \label{eq:C_C}
\eeq
This approximation ought to be accurate when the Compton scattering kernel is narrowly peaked at final photon energies close to the initial photon energy. In particular, we expect it to be accurate for $E \lesssim m_e$.

With this approximation, the Boltzmann equation for $\mathcal{N}_E$ becomes
\barr
\partial_t (a^2 \mathcal{N}_E) - \left( H E + \dot{\mathcal{E}}_{\rm C}(E)\right) \partial_E (a^2 \mathcal{N}_E) \nonumber\\
 - \frac{\partial \dot{\mathcal{E}}_{\rm C}}{\partial E} a^2 \mathcal{N}_E = \frac{a^2}{E} \frac{d \dot{\rho}_{\rm inj}}{dE}. \label{eq:Boltz-approx}
\earr
We now derive an explicit semi-analytic solution for this approximate photon Boltzmann equation.
 
\begin{figure}[ht]
\includegraphics[width=\columnwidth]{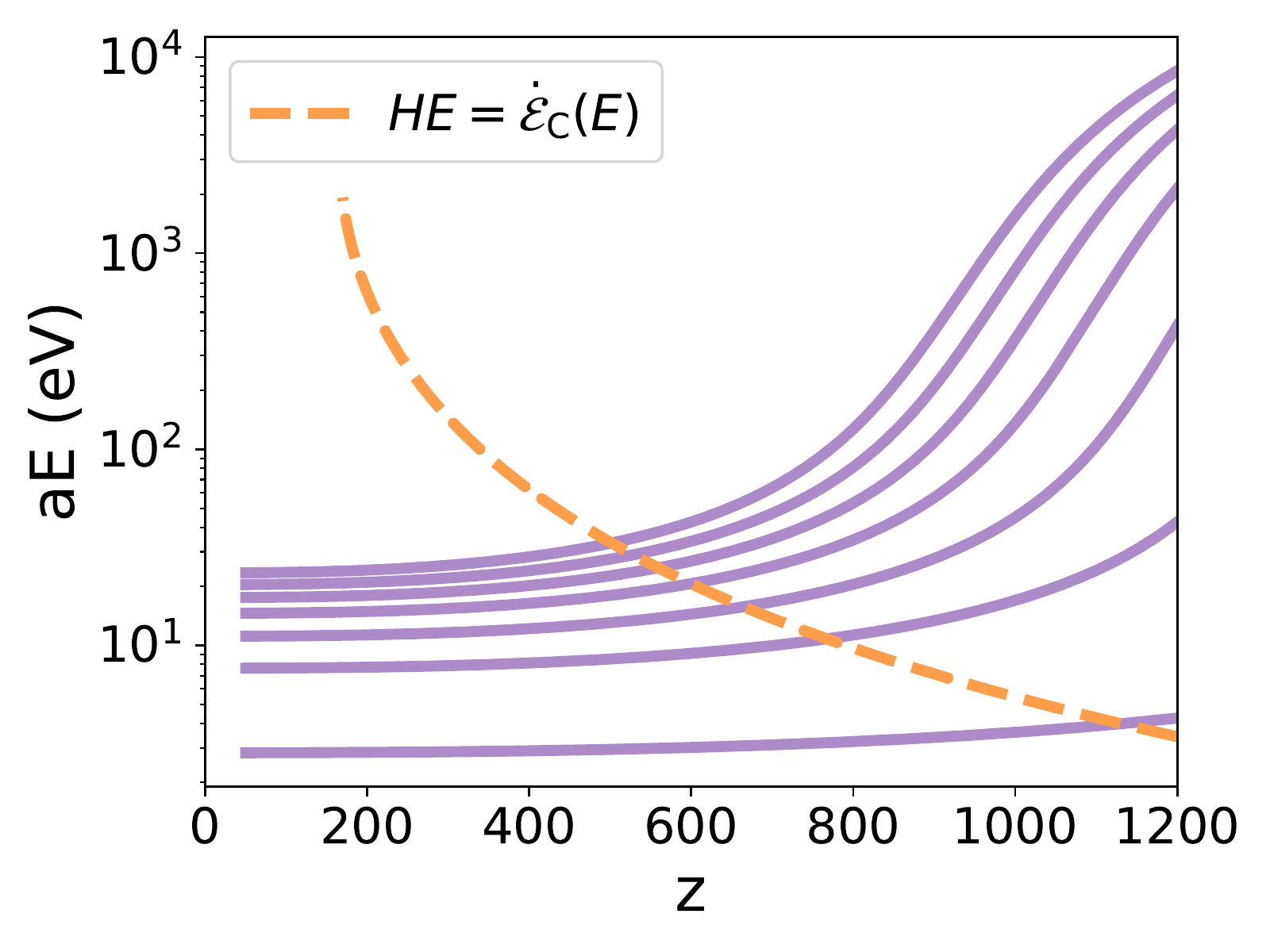}
\caption{\label{fig:E(t)} Trajectories of a photon's energy subject to Compton scattering and Hubble flow with $z_{\rm inj}=1200$ and energies of $E_{\rm inj}=(20,\,15,\,10,\,5,\,1,\,0.1,\,0.01)\times m_e$. The orange dashed line marks the redshift below which redshifting becomes dominant over energy loss through Compton scattering. Note the $y$-axis is scaled by $a$.}
\end{figure}

Given an initial photon energy $E_{t'}$ at time $t'$, we define $E_t( E_{t'})$ to be the solution of the differential equation
\barr   
\deriv{E_t}{t} &=& -H E_t - \dot{\mathcal{E}}_{\rm C}(E_t),\label{eq:Etraj}\\
E_t(t = t') &=& E_{t'},
\earr
In words, $E_{t}$ is the evolution of the mean energy of photons injected with energy $E_{t'}$ at time $t'$, subject to energy loss through cosmological redshifting and Compton scattering. The solution $E_t(E_{t'})$ is unique, and one can therefore uniquely define $E_{t'}(E_t)$, the initial energy at some time $t'$ evolving to $E_t$ at a later time $t$. We will refer to the solutions $E_t$ as the ``Compton trajectories". We show a few trajectories in Fig.~\ref{fig:E(t)}. Since $\dot{\mathcal{E}}_{\rm C} \propto n_{\rm H} \propto a^{-3}$ and $H \propto a^{-3/2}$ during matter domination, energy loss through Compton scattering dominates at early times, and cosmological expansion dominates at late times. 

The first two terms in Eq.~\eqref{eq:Boltz-approx} have a simple interpretation: they represent the total time derivative along Compton trajectories, $d/dt|_{\rm C}$ (in contrast with $d/dt$, the total derivative along free trajectories). In other words, we may rewrite Eq.~\eqref{eq:Boltz-approx} as
\barr
&&\frac{d}{dt}\Big{|}_{\rm C}(a^2 \mathcal{N}_E)+ \gamma(E) a^2 \mathcal{N}_E = \frac{a^2}{E} \frac{d \dot{\rho}_{\rm inj}}{dE}, \\
&&\gamma(E) \equiv  - \frac{\partial \dot{\mathcal{E}}_{\rm C}}{\partial E} 
\earr
This is now a simple first-order ODE along Compton trajectories, which has an explicit integral solution: 
\barr
a^2 \mathcal{N}_{E_t} &=& \int_0^t dt' \exp\left[- \int_{t'}^t dt'' \gamma(E_{t''}(E_t))\right] \nonumber\\
&&\times \left[\frac{a^2}{E} \frac{d \dot{\rho}_{\rm inj}}{dE}\right](t', E_{t'}(E_t)). \label{eq:NE-sol}
\earr
To find $\dot{\rho}_e(t)$, we insert this solution into Eq.~\eqref{eq:dot_rho_e}. This involves an integral over $E_t$, the photon energy at time $t$. We are, instead, interested in expressing $\dot{\rho}_e(t)$ as an integral over the \emph{initial} energies at injection, $E_{t'}$. The two integrals are related by the Jacobian
\beq
J(t, t') \equiv \frac{\partial E_t}{\partial E_{t'}}.
\eeq
Differentiating Eq.~\eqref{eq:Etraj} with respect to $E_{t'}$, we find that the Jacobian satisfies the ODE 
\beq
\frac{\partial J}{\partial t} = - \left(H + \frac{\partial \dot{\mathcal{E}}_{\rm C}}{\partial E}\right) J , \ \ \ \ \ \ 
J(t',t') = 1,
\eeq
with solution 
\beq
J(t, t') = \frac{a'}{a} \exp\left[-\int_{t'}^t dt'' \frac{\partial \dot{\mathcal{E}}_{\rm C}}{\partial E}\right],
\eeq
where the integrand is to be expressed at $E_{t''}(E_{t'})$.

Inserting Eq.~\eqref{eq:NE-sol} into \eqref{eq:dot_rho_e}, switching the order of integration and replacing $\int... dE_t = \int... J dE_{t'}$, we finally arrive at
\barr
\dot{\rho}_e(t) = \int dt' \int dE_{t'} \frac{{a'}^3}{a^3}\dot{\mathcal{E}}_{\rm C}(t,E_t(E_{t'}) \frac1{E_{t'}}\frac{d\dot{\rho}_{\rm inj}}{dE_{t'}}.~~~~
\earr
From this expression, we may finally read off the Green's function for energy deposition into secondary electrons (defined with the same convention as $\overline{G}_E$), 
\barr
\overline{G}^e_{E_{t'}}(a, a') = \frac{\dot{\mathcal{E}}_{\rm C}(t, E_t(E_{t'}))}{E_{t'} H(a)}.\label{eq:Ge-analytic}
\earr
This result can be understood rather intuitively. It represents the rate of photon energy deposition through Compton scattering at time $t$, accounting for the fact that photons lost energy between $t'$ and $t$, so that their energy at $t$ is $E_t(E_{t'})$. Let us note that AK17 derived an analytic solution for $\dot{\rho}_{\rm dep}$ given $\dot{\rho}_{\rm inj}$, neglecting photoionizations, assuming that all the secondary electron's energy is efficiently deposited, and moreover approximating $\dot{\mathcal{E}}_{\rm C}(E) \approx 0.1~n_{\rm H} \sigma_{\rm T} E$. It is straightforward to check that our more general solution recovers that of AK17 when making the same approximations.

\section{Validity of the linear approximation for recombination perturbations} \label{app:xe-checks}

In this paper we have computed the effect of energy deposition on ionization perturbations $\Delta x_e$ by linearizing the recombination equations. In this appendix, we check the validity of this linear approximation for the specific case of energy injection from accreting PBHs.

\begin{figure}[!htb]
    \centering
    \includegraphics[width = \columnwidth]{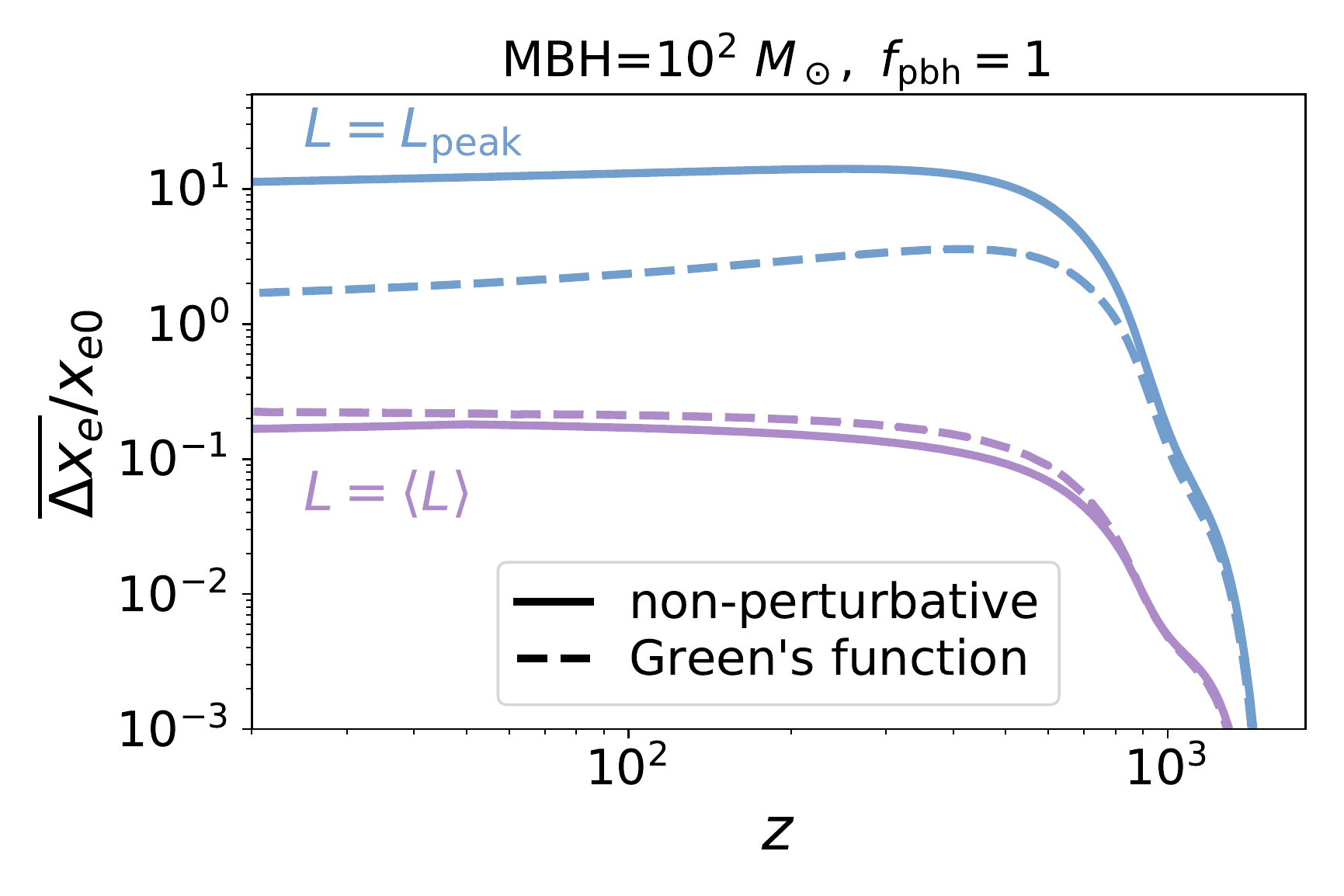}
    \caption{Comparison of the fractional change in ionization history from a homogeneous perturbation to recombination $\overline{\Delta x_e}/x_{e0}$, from PBH's with masses of $10^2$ $M_{\odot}$ and abundance approximately saturating the CMB anisotropy limit, in the collisional-ionization case. We assume that every PBH has an identical luminosity in two limiting cases: luminosity averaged over the distribution of relative velocity ($\langle L \rangle$, purple or lower) as done in AK17; or the luminosity at the peak of the PBH population's luminosity contribution ($L_{\rm peak}$, blue or upper). We expect the actual effect to be in between these limiting cases. Solid lines are computed from our deposition-to-ionization Green's function, relying on the linearization of the recombination equations, and dashed lines show the non-perturbative result from a modified version of \texttt{HyRec} \cite{yacine17a}.}
    \label{fig:delxe_comp}
\end{figure}

The assumption that $\Delta x_e \ll x_e^0$ is not necessarily always justified in the case of inhomogeneously accreting PBHs. To illustrate this point, we have computed the fractional change to the ionization history with the modified version of \texttt{HyRec} \citep{yacine11a} used in AK17. While AK17 use an approximate energy deposition efficiency, this allows them to solve for $x_e$ and $\dot{\rho}_{\rm dep}$ simultaneously and self-consistently, without assuming that perturbations to the ionization history are small. In Fig.~\ref{fig:delxe_comp}, we show the changes to the free-electron fraction in two limiting cases for 100 $M_{\odot}$ PBHs.\par
On the one hand, if one assumes that the energy deposition is fully smeared out spatially, then the relevant energy injection is coming from a relative velocity-averaged luminosity as was done in AK17 (purple lines). In this case, the change in ionization fraction is indeed small at all times and our linearized approximation agrees with the nonperturbative result of AK17 within $\sim 30\%$.\par
On the other hand, if one instead assumes a spatially local energy deposition, the effect on the recombination history is significantly enhanced in regions where the baryon-PBH relative velocity is subsonic. The majority of the PBH population's luminosity then arises from black holes with relative velocity $v_{\rm bc}\approx c_{s,\infty}/\sqrt{2}$, where $c_{s,\infty}$ is the speed of sound of baryons far away from the accreting mass. In this limit, we obtain changes to the free electron fraction as large as $\Delta x_e \sim 10~x_e^0$ at $z \lesssim 800$ (blue lines), which our linearized approximation fails to accurately reproduce. The actual effect lies somewhere in between these two limits, and we therefore expect the linear approximation to be reasonably accurate, especially around the peak of the visibility function.\par

\FloatBarrier

\bibliography{pbh_cmb_ng}

\end{document}